\documentclass[a4paper,11pt]{article}
\usepackage{jheppub}
\usepackage[T1]{fontenc}
\usepackage{diagbox}
\usepackage{adjustbox}
\usepackage{multirow}
\usepackage{pifont}
\usepackage{array}
\usepackage{mathtools}
\usepackage{booktabs}
\usepackage[dvipsnames]{xcolor}
\usepackage{tcolorbox}
\usepackage{float}
\usepackage{placeins}

\usepackage{verbatim}
\usepackage{ulem}

\usepackage{amsmath}
\usepackage{subfig}
\usepackage[compat=1.1.0]{tikz-feynman}
\usepackage{mathtools}

\newcommand{\GeV}{{\text{GeV}}}

\newcommand{\SU}{{\text{SU}}}

\title{\boldmath Multi-step phase transition and gravitational wave from general $\mathbb{Z}_2$ scalar extensions}

\author[a,b]{Qing-Hong~Cao,}
\author[c]{Katsuya~Hashino,}
\author[d]{Xu-Xiang Li,}
\author[e,f,g]{and Jiang-Hao Yu,}

\affiliation[a]{School of Physics, Peking University, Beijing 100871, China}
\affiliation[b]{Center for High Energy Physics, Peking University, Beijing 100871, China}
\affiliation[c]{National Institute of Technology, Fukushima College, Iwaki, Fukushima 970-8034, Japan}
\affiliation[d]{Department of Physics and Astronomy, University of Utah, Salt Lake City, UT 84112, USA}
\affiliation[e]{CAS Key Laboratory of Theoretical Physics, Institute of Theoretical Physics, Chinese Academy of Sciences,
Beijing 100190, China}
\affiliation[f]{School of Physical Sciences, University of Chinese Academy of Sciences, Beijing 100049, P.\ R.\ China}
\affiliation[g]{School of Fundamental Physics and Mathematical Sciences, Hangzhou Institute for Advanced
Study, UCAS, Hangzhou 310024, China}

\emailAdd{qinghongcao@pku.edu.cn}
\emailAdd{hashino@rs.tus.ac.jp}
\emailAdd{xuxiangli@pku.edu.cn}
\emailAdd{jhyu@itp.ac.cn}

\abstract{Multi-step phase transition provides a paradigm in which a broken symmetry during phase transition can be restored, enriching the phenomena of both dark matter and baryon asymmetry.
We study the dynamics of the multi-step phase transition in the standard model extension with additional isospin $N$-plet scalar field $\Phi_2$ under a discrete $\mathbb{Z}_2$ symmetry. 
We find that the multi-step phase transition could be triggered if there is a moderately large coupling between the Higgs and the $\Phi_2$ and this coupling is required to be larger as the mass of the $\Phi_2$ and/or isospin increase. The first-order phase transition at the first (second) step can be realized by the thermal loop (tree-level barrier) effects. Thus it is more likely that a detectable spectrum of gravitational waves can be produced at the second step of the phase transition.
}

\begin{document} 
\maketitle
\flushbottom

%%%%%%%%%%%%%%%%%%%%%%%%%%%%%%%%%%%%%%%%%%%%%%%%%%%%%%%%%%%%
%%%%%%%%%%%%%%%%%%%%%%%%%%%%%%%%%%%%%%%%%%%%%%%%%%%%%%%%%%%%
%%%%%%%%%%%%%%%%%%%%%%%%%%%%%%%%%%%%%%%%%%%%%%%%%%%%%%%%%%%%
\section{Introduction}
%%%%%%%%%%%%%%%%%%%%%%%%%%%%%%%%%%%%%%%%%%%%%%%%%%%%%%%%%%%%
%%%%%%%%%%%%%%%%%%%%%%%%%%%%%%%%%%%%%%%%%%%%%%%%%%%%%%%%%%%%
%%%%%%%%%%%%%%%%%%%%%%%%%%%%%%%%%%%%%%%%%%%%%%%%%%%%%%%%%%%%

Although the Standard Model (SM) of particle physics has been enormously successful in predicting a wide range of phenomena, it fails to explain several observable facts, such as the origin of matter-antimatter asymmetry, the existence of dark matter, etc. Various new physics models have been proposed to address these puzzles; among them, the popular scenarios include the weakly interacting massive particle (WIMP) dark matter and the electroweak baryogenesis, all of which happen around the electroweak scale. Therefore it is quite appealing that both the origin of matter and dark matter are solved at the electroweak scale.

In the early Universe, particles in the SM obtain their masses from the vacuum expectation value (VEV) of the Higgs boson during the electroweak phase transition (EWPT). In this epoch, the baryon asymmetry of the Universe could also be generated via the scenario of electroweak baryogenesis~\cite{Kuzmin:1985mm}, which requires a strongly first-order EWPT to satisfy the Sakharov conditions~\cite{Sakharov:1967dj}.
However, it is well known that the pattern of the EWPT in the SM is only a crossover according to the lattice simulation~\cite{Dine:1992vs}.
Therefore, to realize the first-order EWPT, it is necessary to introduce new physics beyond the SM.

To address the two puzzles, the minimal extension of the SM is adding the WIMP dark matter to realize the first-order phase transition. Although there are thermal loop corrections from the gauge bosons in the SM, the value of the Higgs boson mass is too large to realize the first-order phase transition.  An additional bosonic degree of freedom would increase the thermal loop correction and thus help to realize the first-order phase transition. The scalar multiplet with the discrete $\mathbb{Z}_2$ symmetry would serve as the WIMP dark matter candidate. In this model setup, the scalar multiplet also enhances the thermal loop contribution of the Higgs potential and thus realizes the first-order phase transition.

An interesting signature of the first-order phase transition is the stochastic gravitational wave (GW) from the collision of the nucleated bubbles. The GW spectrum from the phase transition can be typically characterized by the following parameters, the released latent heat $\alpha$, the inverse of the duration of phase transition $\beta$, and transition temperature $T_t$. These parameters can be determined from the shape of the scalar potential for electroweak symmetry breaking; thus, it is likely that new physics models with first-order phase transition can be explored by GW observations. In particular, the frequency of the GW spectrum produced by the first-order EWPT is typical $10^{-3}$--$10$ Hz. Such a GW spectrum with sub-Hz frequency should be detected by the space-based interferometers in the future, such as the Laser Interferometer Space Antenna (LISA)~\cite{LISA}, DECIGO~\cite{Seto:2001qf} and BBO~\cite{BBO}.

Because of the discrete $\mathbb{Z}_2$ symmetry, the scalar multiplet dark matter model exhibits an interesting feature: multi-step phase transition could happen due to a multi-degenerate minimum.  Usually the multi-step happens due to parameter choice, there is no symmetry to guarantee it happens~\cite{Apreda:2001tj,Apreda:2001us,Grojean:2006bp,Huber:2007vva,Espinosa:2008kw,Ashoorioon:2009nf,Kang:2009rd,Jarvinen:2009mh,Konstandin:2010cd,No:2011fi,Wainwright:2011qy,Barger:2011vm,Leitao:2012tx,Dorsch:2014qpa,Kozaczuk:2014kva,Schwaller:2015tja,Xiao:2015tja,Kakizaki:2015wua,Jinno:2015doa,Huber:2015znp,Leitao:2015fmj,Huang:2016odd,Garcia-Pepin:2016hvs,Jaeckel:2016jlh,Dev:2016feu,Hashino:2016rvx,Jinno:2016knw,Barenboim:2016mjm,Kobakhidze:2016mch,Hashino:2016xoj,Artymowski:2016tme,Kubo:2016kpb,Balazs:2016tbi,Vaskonen:2016yiu,Dorsch:2016nrg,Huang:2017laj,Baldes:2017rcu,Chao:2017vrq,Beniwal:2017eik,Addazi:2017gpt,Kobakhidze:2017mru,Tsumura:2017knk,Marzola:2017jzl,Bian:2017wfv,Huang:2017rzf,Iso:2017uuu,Addazi:2017oge,Kang:2017mkl,Cai:2017tmh,Chao:2017ilw,Aoki:2017aws,Huang:2017kzu,Demidov:2017lzf,Chen:2017cyc,Chala:2018ari,Hashino:2018zsi,Morais:2018uou,Croon:2018new,Bruggisser:2018mus,Imtiaz:2018dfn,Huang:2018aja,Bruggisser:2018mrt,FitzAxen:2018vdt,Megias:2018sxv,Alves:2018oct,Baldes:2018emh,Ahriche:2018rao,Prokopec:2018tnq,Fujikura:2018duw,Beniwal:2018hyi,Brdar:2018num,Mazumdar:2018dfl,Addazi:2018nzm,Shajiee:2018jdq,Breitbach:2018ddu,Marzo:2018nov,Megias:2018dki,Croon:2018kqn,Angelescu:2018dkk,Alves:2018jsw,Abedi:2019msi,Fairbairn:2019xog,Kainulainen:2019kyp,Hasegawa:2019amx,Helmboldt:2019pan,Dev:2019njv,Ellis:2019flb,Cutting:2019zws,Bian:2019bsn,Kannike:2019mzk,Bian:2019szo,Dunsky:2019upk,Paul:2019pgt,Brdar:2019fur,Wang:2019pet,Alves:2019igs,DeCurtis:2019rxl,Hall:2019ank,Alanne:2019bsm,Zhou:2019uzq,Morais:2019fnm,Greljo:2019xan,Aoki:2019mlt,Archer-Smith:2019gzq,Croon:2019rqu,Haba:2019qol,Carena:2019une,DelleRose:2019pgi,VonHarling:2019rgb,Chiang:2019oms,Barman:2019oda,Zhou:2020xqi,Ellis:2020awk,Wang:2020jrd,Pandey:2020hoq,Zhou:2020stj,Blasi:2020wpy,Lewicki:2020jiv,Hindmarsh:2020hop,Croon:2020cgk,Nakai:2020oit,Eichhorn:2020upj,Paul:2020wbz,Han:2020ekm,Ares:2020lbt,Wang:2020wrk,Ghosh:2020ipy,Huang:2020crf,Chao:2020adk,Di:2020ivg,Liu:2021jyc,Zhang:2021alu,Cline:2021iff,Cao:2021yau,Niemi:2021qvp,Funatsu:2021gnh,Zhou:2021cfu,Liu:2021mhn,Borah:2021ocu,Aoki:2021oez,Lewicki:2021xku,Dong:2021cxn,Baldes:2021vyz,Marfatia:2021twj,Gould:2021ccf,Lerambert-Potin:2021ohy,Freitas:2021yng,Goncalves:2021egx,Reichert:2021cvs,Borah:2021ftr,Ares:2021ntv,Bai:2021ibt,Ares:2021nap,Lewicki:2021pgr,Bandyopadhyay:2021ipw,Hashino:2021qoq,Demidov:2021lyo,Graf:2021xku,Kanemura:2022ozv,Hamada:2022soj,Blasi:2022woz,Benincasa:2022elt,Biekotter:2022kgf,Azatov:2022tii,Phong:2022xpo}. On the other hand, a discrete symmetry could guarantee the potential form with a multi-degenerate vacuum. This provides room for a multi-step phase transition to happen.  More interestingly, the GW spectra could have one or multi peaks through the multi-step phase transition if several phase transitions in the multi-steps are first order. Therefore, it is possible to use the observation of the multi-peak GW spectra to probe the new physics models with the multi-step transition.

\begin{figure}[htb]
  \begin{center}
\includegraphics[width=0.6\textwidth]{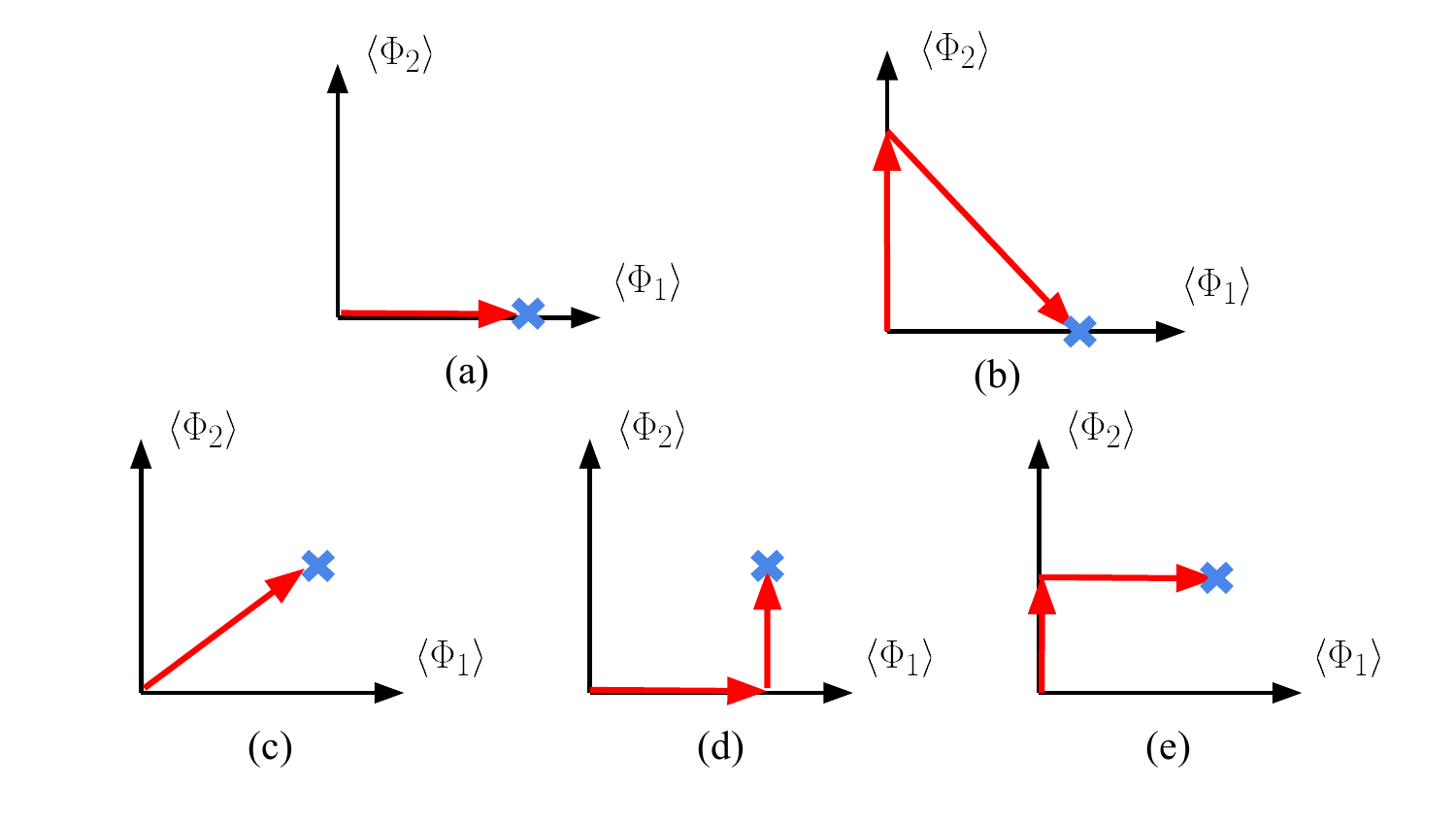}
\caption{The path of phase transitions in the model with two classical fields $ \left\langle \Phi_1 \right\rangle$ and $\left\langle \Phi_2 \right\rangle$. 
The blue cross marks represent the global minimum point of the potential at zero temperature.
The value of $\left\langle \Phi_2 \right\rangle$ is zero in (a) and (b) phase transitions, while it is nonzero in (c)--(e) phase transitions, respectively.
}
\label{RM}
  \end{center}
\end{figure}

In this work, we consider the general isospin $N$-plet scalar model with discrete $\mathbb{Z}_2$ symmetry, which helps to realize the two-step phase transition. The model with the scalar sector extended with a general isospin $N$-plet has been discussed in our previous work~\cite{Cai:2022bcf} where the possibility of a one-step EWPT and the consequent GW spectrum are examined. The general potential of the Higgs and the new $N$-plet scalar accommodates fruitful physics beyond the one-step EWPT discussed in the previous work. Fig.~\ref{RM} shows parts of the possible paths of the EWPT that can be realized.
%%%%%%%%%%%%%%%%%%%%%%%%%%%%%%%%%%%%
%%%%%%%%%%%%%%%%%%%%%%%%%%%%%%%%%%%%
%%%%%%%%%%%%%%%%%%%%%%%%%%%%%%%%%%%%
%%%%%%%%%%%%%%%%%%%%%%%%%%%%%%%%%%%%
In the phase transitions of Figs.~\ref{RM}-(a) and~\ref{RM}-(b), the classical field of the additional scalar field $\left\langle \Phi_2 \right\rangle$ at the blue cross marks is zero after the phase transition, while this value in Figs.~\ref{RM}s(c)--\ref{RM}-(e) is nonzero. The one-step phase transition in Fig.~\ref{RM}-(a) is discussed in the previous work~\cite{Cai:2022bcf} and in this work we will focus on the two-step phase transition pattern shown in Fig.~\ref{RM}-(b) where a broken global symmetry is restored after electroweak symmetry breaking (EWSB).

According to this phase transition path in Fig.~\ref{RM}-(b), a global symmetry in the potential is broken during the phase transition. After the phase transition, the broken symmetry can be restored in the current Universe. Such a multi-step phase transition is significant to explain the phenomena beyond the Standard Model: dark matter existence~\cite{FileviezPerez:2008bj,Cui:2011qe,Cline:2012hg,Patel:2012pi,Fairbairn:2013uta,Alanne:2014bra,Blinov:2015sna,Baker:2016xzo,Grzadkowski:2018nbc,Baker:2018vos,Bian:2018bxr,Ghorbani:2019itr,Robens:2019kga,Chen:2019ebq,Chiang:2019oms,Bell:2020gug,Chiang:2020yym} and baryon asymmetry of the Universe~\cite{Profumo:2007wc,Espinosa:2011ax,Espinosa:2011eu,Profumo:2014opa,Inoue:2015pza,Tenkanen:2016idg,Vaskonen:2016yiu,Kurup:2017dzf,Chen:2017qcz,Ramsey-Musolf:2017tgh,Chala:2018opy,Gould:2019qek,Caprini:2019egz,Kozaczuk:2019pet,Ramsey-Musolf:2019lsf,Senaha:2020mop}.
Furthermore, the phase transition in the second path of Fig.~\ref{RM}-(b) can be strong because the barrier shows up in the potential at the zero temperature by tree-level effects, like Ref.~\cite{Espinosa:2011ax}. We will show the conditions in which the two-step phase transition happens in this general model. For the EWPT in Fig.~\ref{RM}-(b), the first (second) step relies on the thermal loop (tree-level) effects, and thus the sizable barrier could be generated in the second step by the tree-level effects. The detectable GW spectrum is mainly produced by the second step since the tree-level effect is the dominant contribution to this phase transition path. On the other hand, the first-order EWPT in Fig.~\ref{RM}-(a) can be realized by finite thermal effects only. Since the sources of the barrier are different between the one-step and two-step phase transitions, the temperature for starting the EWPT of the Fig.~\ref{RM}-(a) is higher than that of the Fig.~\ref{RM}-(b).
Therefore, we may distinguish the patterns of the EWPT in Figs.~\ref{RM}-(a) and -(b) by the GW observation.

In the following, we introduce the model's detail with additional isospin $N$-plet scalar field $\Phi_2$ with discrete $\mathbb{Z}_2$ symmetry in Secs.~\ref{sec:model} and~\ref{sec:potential}.
Furthermore, we use the high-temperature approximation to discuss the condition of the multi-step phase transition analytically in Sec.~\ref{sec:twostep}. 
In Sec.~\ref{sec:num}, we numerically discuss the dynamics of the EWPT being able to restore the $\mathbb{Z}_2$ symmetry after the EWPT and show detectability of the GW spectrum produced from the first-order EWPT of Figs.~\ref{RM}-(a) and -(b).
The conclusion is drawn in Sec.~\ref{sec:summary}.

%%%%%%%%%%%%%%%%%%%%%%%%%%%%%%%%%%%%%%%%%%%%%%%%%%%%%%%%%%%%
%%%%%%%%%%%%%%%%%%%%%%%%%%%%%%%%%%%%%%%%%%%%%%%%%%%%%%%%%%%%
%%%%%%%%%%%%%%%%%%%%%%%%%%%%%%%%%%%%%%%%%%%%%%%%%%%%%%%%%%%%
\section{General form of $\mathbb{Z}_2$ $N$-plet scalar model}
\label{sec:model}
%%%%%%%%%%%%%%%%%%%%%%%%%%%%%%%%%%%%%%%%%%%%%%%%%%%%%%%%%%%%
%%%%%%%%%%%%%%%%%%%%%%%%%%%%%%%%%%%%%%%%%%%%%%%%%%%%%%%%%%%%
%%%%%%%%%%%%%%%%%%%%%%%%%%%%%%%%%%%%%%%%%%%%%%%%%%%%%%%%%%%%

Before going into details of the evaluation of the phase transition, we first present a general form of the potential under the model extended with an $N$-plet scalar $\Phi_2$, which transforms as the $(I_2, Y_2)$ representation under $SU(2)_L \times U(1)_Y$ gauge group. In order to have a complete analysis, in this section, we adopt a slightly different notation compared to Ref.\,\cite{Ramsey-Musolf:2021ldh}. The $N$-components field
\begin{align}
    \Phi_2 = 
        \left(\begin{array}{crcccccc} 
          \Phi_2^{I_2} \\
          \Phi_2^{I_2 - 1} \\
        \vdots  \\
          \Phi_2^{-I_2} 
        \end{array} 
        \right)
\end{align}
is replaced by a field $(\Phi_2)_{i_1\dots i_{2I_2}}$ with $2I_2 = N-1$ indices all of which take the value of $1$ or $2$. $(\Phi_2)_{i_1\dots i_{2I_2}}$ is fully symmetric in all its $2I_2$ indices and therefore only has $N$ independent components. The two notations can be related by the generalized Clebsch-Gordan coefficients $C^{\mathcal{I}}_{i_1\dots i_{2I_2}}$ as 
\begin{align}\label{eq:defPhi2}
    (\Phi_2)_{i_1\dots i_{2I_2}} = C^{\mathcal{I}}_{i_1\dots i_{2I_2}}\Phi_2^{\mathcal{I}}, \quad C^{\mathcal{I}}_{i_1\dots i_{2I_2}} = \sqrt{\frac{(I_2 + m)!(I_2 - m)!}{(2I_2)!}} \delta_{{\mathcal{I}},\sum_{n=1}^{2I_2} (3/2 - i_n)}.
\end{align}
Note that if $\Phi_2$ is an $SU(2)_L$ doublet, $I_2 = 1/2$, then our notations are the same. It is more convenient to construct gauge-invariant operators using $(\Phi_2)_{i_1\dots i_{2I_2}}$ as building blocks as all indices can be regarded as indices of the fundamental representation of the $SU(2)_L$ group. The conjugated field, $(\Phi_2^*)^{i_1\dots i_{2I_2}} \equiv [(\Phi_2)_{i_1\dots i_{2I_2}}]^*$, is also replaced by\footnote{$(\Phi_2^*)_{i_1\dots i_{2I_2}}$ and the associated conjugate in Ref.\,\cite{Ramsey-Musolf:2021ldh} are related also by
$(\Phi_2^*)_{i_1\dots i_{2I_2}} = C^{\mathcal{I}}_{i_1\dots i_{2I_2}}\overline{\Phi}_2^{\mathcal{I}}$.
}
\begin{align}
    (\Phi_2^*)_{i_1\dots i_{2I_2}} = (\Phi_2^*)^{j_1\dots j_{2I_2}} \prod_{n = 1}^{2I_2} \epsilon_{i_nj_n}
\end{align}
which transforms in the same way as $(\Phi_2)_{i_1\dots i_{2I_2}}$.
The lower and upper indices denote the indices of fundamental and antifundamental representations and $\epsilon_{12} = -\epsilon_{21} = 1,~ \epsilon_{11} = \epsilon_{22} = 0$. As an example, for a complex $SU(2)_L$ triplet field $\Phi_2$,
\begin{align}
    (\Phi_2)_{i_1i_2} = \left(\begin{array}{cc}
        \Phi_2^{{\mathcal{I}} = +1} & \frac{1}{\sqrt{2}}\Phi_2^{{\mathcal{I}} = 0} \\
        \frac{1}{\sqrt{2}}\Phi_2^{{\mathcal{I}} = 0} & \Phi_2^{{\mathcal{I}} = -1}
    \end{array}\right), \quad
    (\Phi_2^*)_{i_1i_2} = \left(\begin{array}{cc}
        (\Phi_2^{{\mathcal{I}} = -1})^* & \frac{1}{\sqrt{2}}(\Phi_2^{{\mathcal{I}} = 0})^* \\
        \frac{1}{\sqrt{2}}(\Phi_2^{{\mathcal{I}} = 0})^* & (\Phi_2^{{\mathcal{I}} = -1})^*
    \end{array}\right).
\end{align}
The SM-like Higgs is
\begin{align}
    (\Phi_1)_i = \left(\begin{array}{crcccccc} 
        G^{+}\\ \frac{1}{\sqrt{2}}\left(h + iG^0\right) \end{array} 
    \right), 
    \quad (\Phi_1^*)_i = \epsilon_{ij} (\Phi_1^*)^{j} = \left(\begin{array}{crcccccc} 
        \frac{1}{\sqrt{2}}\left(h - iG^0\right)\\ -G^{-}  \end{array} 
    \right).
\end{align}
Under this notation, the Lagrangian of the scalar sector is
\begin{align}
    {\cal L}_{\text{Scalar}} = (D_\mu \Phi_1)^\dagger (D^\mu \Phi_1) + (D^\mu \Phi_2^*)_{i_1 \dots i_{2I_2}} (D_\mu \Phi_2)_{j_1 \dots j_{2I_2}} \prod_{n = 1}^{2I_2} \epsilon^{i_nj_n} - V_0(\Phi_1, \Phi_2),
\end{align}
where $D_\mu = \partial_\mu - i g T^a W_\mu^a - ig' Y B_\mu$. Here $g$ and $g'$ are $\SU(2)_L$ and $U(1)_Y$ gauge couplings respectively, and $T^a$ is the generator of $\SU(2)_L$ for corresponding representation. Also note that in principle, the factor in the production should be fully symmetrized over all $i_n$ and $j_n$ respectively, and here we leave it implicit since these indices are symmetric in the fields.

The general form of the potential can be obtained by the Young tableau method. The mass terms or the quadratic terms read as
\begin{align}\label{eq:masspot}
    \Delta V_0 = -\mu_1^2 (\Phi_1^*)^{i} (\Phi_1)_i - \mu_2^2 (\Phi_2^*)_{i_1 \dots i_{2I_2}} (\Phi_2)_{j_1 \dots j_{2I_2}} \prod_{n = 1}^{2I_2} \epsilon^{i_n j_n}.
\end{align}
with the symmetrization over every group of indices implicit. We have imposed a $\mathbb{Z}_2$ symmetry over the scalars such that $\Phi_1 \to \Phi_1$ and $\Phi_2 \to -\Phi_2$ and all other fields are all $\mathbb{Z}_2$ even.
The interaction terms thus are given as
\begin{align}\label{eq:intpot}
    \Delta V_{0} & = (\Phi_1^*)_i (\Phi_1)_j (\Phi_2^*)_{k_1 \dots k_{2I_2}} (\Phi_2)_{l_1 \dots l_{2I_2}} \times \left[ a_0 K_0^{ijk_1\dots l_{2I_2}} + a_1 K_1^{ijk_1\dots l_{2I_2}} \right] \nonumber \\
    & \phantom{=} + (\Phi_1)_i (\Phi_1)_j (\Phi_1^*)_k (\Phi_1^*)_l \times \lambda_1 \epsilon^{ik} \epsilon^{jl} \nonumber \\
    & \phantom{=} + (\Phi_2)_{i_1 \dots i_{2I_2}} (\Phi_2)_{j_1 \dots j_{2I_2}} (\Phi_2^*)_{k_1 \dots k_{2I_2}} (\Phi_2^*)_{l_1 \dots l_{2I_2}} \times \sum_{m = 0}^{\lfloor I_2 \rfloor} b_{m} C_m^{i_1 \dots l_{2I_2}},
\end{align}
where the factors are
\begin{align}
    K_0^{ijk_1\dots l_{2I_2}} & = \frac{1}{[(2I_2)!]^2} \left[ \epsilon^{ij} \prod_{n = 1}^{2I_2} \epsilon^{k_n l_n} + (\text{all perms of } k, l) \right], \\
    K_1^{ijk_1\dots l_{2I_2}} & = \frac{1}{[(2I_2)!]^2} \left[ \epsilon^{ik_1}\epsilon^{jl_1} \prod_{n = 1}^{2I_2 - 1} \epsilon^{k_n l_n} + (\text{all perms of } k, l) \right], \\
    C_m^{i_1 \dots l_{2I_2}} & = \frac{1}{[(2I_2)!]^4} \left[ \prod_{n = 1}^{2m} \epsilon^{i_n j_n} \epsilon^{k_n l_n} \prod_{n = 2m+1}^{2I_2} \epsilon^{i_n k_n} \epsilon^{j_n l_n} + (\text{all perms of }i, j, k, l) \right],
\end{align}
and $\lfloor n \rfloor$ means the rounding down of $n$. The explicit symmetrization of indices is important to calculate Feynman diagrams correctly. Also, note that the terms proportional to $a_1$ do not exist when $I_2 = 0$.

When the isospin and hypercharge of $\Phi_2$ take some special values, there will be additional gauge-invariant terms.
For complex $\Phi_2$ with even $2I_2$ and $Y_2 = 0$, one can add these terms:
\begin{align}\label{eq:extrapoteven}
    \Delta V_{0} & = (\Phi_1^*)_i (\Phi_1)_j (\Phi_2)_{k_1 \dots k_{2I_2}} (\Phi_2)_{l_1 \dots l_{2I_2}} \times a_0' K_0^{ijk_1\dots l_{2I_2}} \nonumber \\
    & \phantom{=} + (\Phi_2)_{i_1 \dots i_{2I_2}} (\Phi_2)_{j_1 \dots j_{2I_2}} (\Phi_2)_{k_1 \dots k_{2I_2}} (\Phi_2)_{l_1 \dots l_{2I_2}} \times \sum_{m = 0}^{\lfloor I_2 / 2 \rfloor} b_{m}' C_m^{i_1 \dots l_{2I_2}} \nonumber\\
    & \phantom{=} + (\Phi_2)_{i_1 \dots i_{2I_2}} (\Phi_2)_{j_1 \dots j_{2I_2}} (\Phi_2)_{k_1 \dots k_{2I_2}} (\Phi_2^*)_{l_1 \dots l_{2I_2}} \times \sum_{m = 0}^{\lfloor I_2 / 2 \rfloor} b_{m}'' C_m^{i_1 \dots l_{2I_2}} + \mathrm{h.c.}.
\end{align}
However, as the complex $\Phi_2$ is in a real representation under this case, it can be divided into two real components with different $CP$ properties. For odd $2I_2$ with $Y_2 = 0$ and $\pm 1/2$,
\begin{align}\label{eq:extrapotodd}
    \Delta V_0^{(Y_2 = 0)} & = (\Phi_1^*)_i (\Phi_1)_j (\Phi_2)_{k_1 \dots k_{2I_2}} (\Phi_2)_{l_1 \dots l_{2I_2}} \times a_1' K_1^{ijk_1\dots l_{2I_2}} + \mathrm{h.c.}, \\
    \Delta V_0^{(Y_2 = 1/2)} & = (\Phi_1)_i (\Phi_1)_j (\Phi_2^*)_{k_1 \dots k_{2I_2}} (\Phi_2^*)_{l_1 \dots l_{2I_2}} \times a_1' K_1^{ijk_1\dots l_{2I_2}} + \mathrm{h.c.}.
\end{align}

The discussion above on the general potential is valid under the case that the new scalar $\Phi_2$ is complex. The potential for a real one will be a bit simpler due to the flavor symmetry. Only the $a_0$- and $b_m$- with $m$ up to $\lfloor I_2 / 2 \rfloor$ terms in Eq.~\eqref{eq:intpot} are nonredundant. Also, there should be a $1/2$ factor before $\mu_2^2$ in Eq.~\eqref{eq:masspot}.

It should be noted that high order $SU(2)_L$ multiplets receive theoretical constraints such as perturbative unitarity. The partial-wave expansion of the scattering ampltiudes between $SU(2)_L$ multiplets and gauge bosons leads to $I_2 \leq 7/2$ for complex scalars and $I_2 \leq 4$ for real scalars \cite{Hally:2012pu}. In the following, we will see that the criterion for a strong first-order phase transition (SFOPT) also puts an upper bound for representations with $I_2 \lesssim 4$.

%%%%%%%%%%%%%%%%%%%%%%%%%%%%%%%%%%%%%%%%%%%%%%%%%%%%%%%%%%%%
%%%%%%%%%%%%%%%%%%%%%%%%%%%%%%%%%%%%%%%%%%%%%%%%%%%%%%%%%%%%
%%%%%%%%%%%%%%%%%%%%%%%%%%%%%%%%%%%%%%%%%%%%%%%%%%%%%%%%%%%%
\section{The one-loop effective potential}
\label{sec:potential}
%%%%%%%%%%%%%%%%%%%%%%%%%%%%%%%%%%%%%%%%%%%%%%%%%%%%%%%%%%%%
%%%%%%%%%%%%%%%%%%%%%%%%%%%%%%%%%%%%%%%%%%%%%%%%%%%%%%%%%%%%
%%%%%%%%%%%%%%%%%%%%%%%%%%%%%%%%%%%%%%%%%%%%%%%%%%%%%%%%%%%%

To analyze the behavior of the effective potential as the temperature varieties, one needs to evaluate the effective potential at least up to the one-loop level. In our work, it is assumed that only the charge-neutral components of scalars generate VEVs, so we only consider the effective potential of these components. The $CP$-even neutral components of $\Phi_1$ and $\Phi_2$ are $h$ and $\operatorname{Re} \Phi_2^{\mathcal{I} = -Y_2}$ respectively, and we here define the corresponding classical background fields as $h_1$ and $h_2$. The effective potential with the finite temperature effects is given as
%%%%%%%
\begin{align}
  \label{eq:effpot}
	V_\text{eff}\left( h_1 , h_2 , T\right) = V_0(h_1, h_2) + V_{\text{CW}}(h_1, h_2) +  \Delta V_T(h_1, h_2, T) + V_T^{\rm ring}(h_1, h_2, T),
	\end{align}
%%%%%%%
where we have included the daisy resummation contributions to avoid IR divergence.\footnote{The inclusion of one-loop thermal potential and daisy resummation does not fully solve the problems of the dependence of the potential on gauge choices and renormalization scale, which may lead to large systematic errors \cite{Croon:2020cgk}. More rigorous approaches are required to alleviate these problems, such as the dimensional reduction technique \cite{Niemi:2018asa,Gould:2019qek,Kainulainen:2019kyp,Niemi:2021qvp,Ekstedt:2022bff}.}

Generally, each term in Eq.~\eqref{eq:effpot} is dependent on the field-dependent mass as
\begin{align}
  \label{eq:CWpot}
  V_{\text{CW}} & = \sum_i\frac{n_i}{64\pi^2} \, M^4_i\left( h_1 , h_2 \right)\,\left[  \ln\left( \frac{M^2_i\left( h_1 , h_2 \right)}{Q^2} \right) - c_i \right], \\
  \label{eq:finTpot}
  \Delta V_T &= 
  \frac{T^4}{2\pi^2} \sum_{i=\text{bosons}} 
  n_i  \int_0^\infty {\rm d} x \,x^2\ln 
  \left[ 1- \exp \left( -\sqrt{x^2+(M_i\left( h_1 , h_2 \right)/T)^2}\right) \right]\nonumber\\
  & \phantom{=} + \frac{T^4}{2\pi^2} \sum_{i = \text{fermions}} n_i  \int_0^\infty {\rm d} x \,x^2\ln 
  \left[ 1+ \exp \left( -\sqrt{x^2+(M_i\left( h_1 , h_2 \right)/T)^2}\right) \right], \\
  \label{eq:Daisypot}
  V_T^{\rm ring} & = \frac{T}{12\pi}\sum_{i = \text{bosons}}n_i \left( (M_i^2(h_1, h_2))^{3/2} - (M_{i,T}^2(h_1, h_2) )^{3/2}\right).
\end{align}
Here the Coleman-Weinberg potential Eq.~\eqref{eq:CWpot} is calculated under the $\overline{\text{MS}}$ scheme with $Q$ the renormalization scale, and $c_i = 3/2$ for scalars and fermions and $5/6$ for gauge bosons. The index $i$ in the summation runs over all the fields in the model, with the field-dependent masses given by square roots of the eigenvalues of the mass-squared matrix
\begin{align}
    M_{ij}^2(h_1, h_2) = \frac{\partial^2 V_0}{\partial \phi_i \partial \phi_j} \Bigg|_{h_1, h_2}
\end{align}
for bosons and similar for fermions. A general expression for the field-dependent masses of scalars is hard to obtain, and  in appendix~\ref{sec:appmass}, we present the results of the singlet, doublet, and triplet extended models. For gauge bosons and fermions, they are given as
\begin{align}
    M_W^2 =  \frac{1}{4}g^2\left(h_1 ^2  + I_W^2  h_2 ^2 \right),~ M_Z^2 =  \frac{1}{4}(g^2 + g'^2)\left( h_1 ^2 + 4Y_{2}^2 h_2 ^2 \right),~ M_t^2 = \frac{1}{2}y_t^2 h_1^2,
\end{align}
with $I_W^2 = 2I_{2} (I_{2}+1) - 2Y_{2}^2$.

The $M_{i,T}^2(h_1, h_2)$ in Eq.~\eqref{eq:Daisypot} are the eigenvalues of thermal-corrected field-dependent mass-squared matrix, which is given by adding the Debye mass term $\Pi_{ij}(T)$ to the mass-squared matrix. The Debye mass term can be regarded as an effective mass term induced by the plasma, whose leading order comes from the one-loop diagram. For example, the following shows the contribution from the $b_m$-terms to the Debye mass of $\Phi_2$:
\begin{align}
    (\Delta \Pi^{\Phi_2 \Phi_2^*})^{i_1\dots j_{2I_2}} &=
    \begin{gathered}
    	\begin{tikzpicture}[scale=1, transform shape]
            \begin{feynman}
                \vertex (a) at (0,0);
                \vertex (b) at (1,0);
                \vertex (c) at (2,0);
                \diagram*{
                (a) -- [scalar] (b) -- [scalar] (c)
                };
                \draw[scalar] (b) arc [start angle=-90, end angle=270, radius=0.7cm];
            \end{feynman}
        \end{tikzpicture}
    \end{gathered} \nonumber 
    \\
    & = -2i \sum_{m = 0}^{\lfloor I_2 \rfloor}b_m \times \left\{ \left[ \prod_{n = 1}^{2m} \epsilon^{i_nk_n}\epsilon^{j_nl_n}\prod_{n = 2m+1}^{2I_2}\epsilon^{i_nj_n}\epsilon^{k_nl_n} + (j \leftrightarrow l) \right] \right. \nonumber
     \\
    & \phantom{=} + {\text{all perms of } i,j,k,l} \Bigg\} 
    \left[\frac{1}{(2I_2)!}\right]^4 
    \times \prod_{n = 1}^{2I_2} \epsilon_{k_nl_n} \, 
    \sum\!\!\!\!\!\!\! \int_K 
    \frac{1}{k^2 + \omega_n^2}.
\end{align}
The ``sumint'' symbol stands for the integral over 3D space and the summation over all the Matsubara mode $\omega_n$ with a $T$ prefactor. The full result of Debye mass for scalars are
\begin{align}
\label{eq:debyemass1}
    \Pi^{\Phi_1 \Phi_1^*}_{ij} & = \delta_{ij}T^2\left[\frac{\lambda_1}{2} + \frac{3g^2}{16} + \frac{g'^2}{16} + \frac{y_t^2}{4}  + \frac{1}{6} (2I_2 + 1) \left(a_0 + \frac{1}{2} a_1 \right)\right], \\
    \label{eq:debyemassreal2}
    \Pi^{\Phi_2 \Phi_2}_{\mathcal{IJ}} & = \delta_{\mathcal{IJ}} T^2 \left[ \frac{g^2}{4} \left(I_{2}^2 + I_{2}\right) + \frac{g'^2}{4}Y_{2}^2 + \frac{1}{3}\sum_{m = 0}^{\lfloor I_2/2 \rfloor}f(I_2, m) b_m 
    + \frac{1}{6}a_0\right]
\end{align}
for real $\Phi_2$ and
\begin{align}
    \label{eq:debyemasscomplex2}
    \Pi^{\Phi_2 \Phi_2^*}_{\mathcal{IJ}} & = \delta_{\mathcal{I}\mathcal{J}}T^2 \left[ \frac{g^2}{4} \left(I_{2}^2 + I_{2}\right) + \frac{g'^2}{4}Y_{2}^2 + \frac{1}{6}\sum_{m = 0}^{\lfloor I_2 \rfloor} b_m \left( \frac{2I_2 + 1}{2m+1} + 1 \right) + \frac{1}{6}a_0 + \frac{1}{12}a_1 \right] 
\end{align}
for complex $\Phi_2$, while for even $2I_2$ and $Y_2 = 0$ extra terms contribute as
\begin{align}\label{eq:debyemasscomplex2extra}
    \Delta \Pi^{\Phi_2 \Phi_2^*}_{\mathcal{IJ}} & = \delta_{\mathcal{I}, -\mathcal{J}}(-1)^{j-\mathcal{I}}\frac{T^2}{3}\sum_{m = 0}^{\lfloor I_2/2 \rfloor} f(I_2, m) \left(\begin{array}{cc}
        \operatorname{Re}(b_m'') & -\operatorname{Im}(b_m'') \\
        -\operatorname{Im}(b_m'') & -\operatorname{Re}(b_m'')
    \end{array}\right).
\end{align}
The nondiagonal term in Eq.~\eqref{eq:debyemasscomplex2extra} indicates the mixing between $\operatorname{Re} \Phi_2^{\mathcal{I}}$ and $\operatorname{Im} \Phi_2^{\mathcal{I}}$ and
\begin{align}
    \label{eq:debyemassdeff}
    f(I_2, m) & = \frac{2I_2 + 1}{2m+1} + \frac{2I_2 + 1}{2I_2 - 2m+1} + 1.
\end{align}
For gauge bosons, only the longitudinal modes receive contributions from thermal loops at leading order, which leads to
\begin{align}
    \Pi^{WW}_{IJ} & = \delta_{IJ}g^2T^2\left[\frac{11}{6} + \frac{1}{9}I_{2}\left(1 + I_{2}\right)\left(1 + 2I_{2}\right)\right], \\
    \Pi^{BB} & = g'^2 T^2\left[\frac{11}{6} + \frac{1}{3}\left(1 + 2I_{2}\right)Y_{2}^2 \right].
\end{align}
If $\Phi_2$ is real, the terms in the bracket related to $I_2$ should be reduced by half.

%%%%%%%%%%%%%%%%%%%%%%%%%%%%%%%%%%%%%%%%%%%%%%%%%%%%%%%%%%%%
%%%%%%%%%%%%%%%%%%%%%%%%%%%%%%%%%%%%%%%%%%%%%%%%%%%%%%%%%%%%
%%%%%%%%%%%%%%%%%%%%%%%%%%%%%%%%%%%%%%%%%%%%%%%%%%%%%%%%%%%%
\section{Analysis of two-step phase transition}
\label{sec:twostep}
%%%%%%%%%%%%%%%%%%%%%%%%%%%%%%%%%%%%%%%%%%%%%%%%%%%%%%%%%%%%
%%%%%%%%%%%%%%%%%%%%%%%%%%%%%%%%%%%%%%%%%%%%%%%%%%%%%%%%%%%%
%%%%%%%%%%%%%%%%%%%%%%%%%%%%%%%%%%%%%%%%%%%%%%%%%%%%%%%%%%%%

Since the numerical simulation of the phase transition can be tedious and time consuming, it is better to have a brief analytic calculation on the possibility of a two-step phase transition prior to the numerical analysis.
For simplicity, we will take the leading terms in Eq.~\eqref{eq:effpot} only when we derive the conditions for two-step phase transitions analytically. The leading terms can be parametrized as
%%%%%%%%%%%%%%%%%%%%%%%%%%%%%%%%%%%%%%%%%%%%%%%%%%%%%%%%%%%%%%
\begin{align}\label{eq:leadingpot}
    V_{\text{eff}}\left(h_1, h_2, T\right) & = \frac{1}{2}\left(-\hat{\mu}_1^2 + D_1 T^2\right) h_1^2 + \frac{1}{2}\left(-\hat{\mu}_2^2 + D_2 T^2\right) h_2^2 + \frac{1}{4}\hat{\lambda}_{12} h_1^2 h_2^2 + \frac{1}{4}\hat{\lambda}_1 h_1^4 + \frac{1}{4}\hat{\lambda}_2 h_2^4,
\end{align}
%%%%%%%%%%%%%%%%%%%%%%%%%%%%%%%%%%%%%%%%%%%%%%%%%%%%%%%%%%%%%%
with the relation between original and hatted parameters given by
\begin{align}
    \hat{\mu_1}^2 = \mu_1^2,~ \hat{\mu_2}^2 = (-1)^{I_2}\mu_2^2,~ \hat{\lambda}_{12} = (-1)^{I_2}2a_0,~ \hat{\lambda}_1 = \lambda_1,~ \hat{\lambda}_2 = 4\sum_{m = 0}^{\lfloor I_2 / 2 \rfloor} b_m F(I_2, 0, m)
\end{align}
for real $\Phi_2$ and
\begin{align}
    \hat{\mu_1}^2 = \mu_1^2,~ \hat{\mu_2}^2 = \mu_2^2,~ \hat{\lambda}_{12} = a_0 + \frac{I_2 - Y_2}{2I_2}a_1,~ \hat{\lambda}_1 = \lambda_1,~ 
    \hat{\lambda}_2 =  \sum_{m = 0}^{\lfloor I_2 \rfloor} b_m F(I_2, Y_2, m)
\end{align}
for complex $\Phi_2$. The contributions from extra terms are
\begin{align}
    \Delta\hat{\lambda}_{12} = (-1)^{I_2} 2\operatorname{Re}(a_0'),~ \Delta\hat{\lambda}_2 = \sum_{m = 0}^{\lfloor I_2 / 2 \rfloor} 2 F(I_2, 0, m) \operatorname{Re}(b_m' + (-1)^{I_2}b_m'')
\end{align}
when $2I_2$ is even and $Y_2 = 0$,
and
\begin{align}
    \Delta\hat{\lambda}_{12} = (-1)^{I_2 - 1/2} \frac{I_2 + 1/2}{I_2}\operatorname{Re}(a_1')
\end{align}
when $2I_2$ is odd and $Y_2 = 1 / 2$.
The definition of the function $F$ is
\begin{align}
    F(I, Y, m) = & \sum_{s = 0}^{2m} \left[\frac{(I - Y)!(I + Y)!}{(2I)!}\right]^2\left[\frac{(2m)!}{s!(2m-s)!}\right]^2 \nonumber \\
    & \times \frac{[(2I - 2m)!]^2}{(I-Y-s)!(I+Y-s)!(I - Y - 2m + s)!(I + Y - 2m + s)!}
\end{align}
with all terms in factorials non-negative. $D$-terms can be extracted directly from the Debye mass terms
\begin{align}\label{eq:D1D2def}
D_1 = \Pi_{22}^{\Phi_1\Phi_1^*}/T^2, \quad D_2 = \Pi_{Y_2Y_2}^{\Phi_2\Phi_2^{(*)}}/T^2.
\end{align}

The five parameters in the potential $\hat{\mu}_1^2$, $\hat{\mu}_2^2$, $\hat{\lambda}_1$, $\hat{\lambda}_2$ and $\hat{\lambda}_{12}$ will be replaced by five input parameters $v$, $m_h$, $m_H$, $\hat{\lambda}_2$ and $\hat{\lambda}_{12}$ in this section, where $v=246$ GeV, $m_h=125$ GeV and $m_H$ is the mass of the additional $CP$-even neutral scalar. At the tree level, their relation is
\begin{align}\label{eq:trcon}
    \hat{\mu}_1^2 = \frac{1}{2}m_h^2,~ \hat{\mu}_2^2 = \frac{1}{2}\lambda_{12}v^2 - m_H^2,~ \hat{\lambda}_1 = \frac{m_h^2}{2v^2}.
\end{align}
In the following, we will discuss the conditions and the strength of the two-step phase transition with the hat marks implicit.

%%%%%%%%%%%%%%%%%%%%%%%%%%%%%%%%%%%%%%%%%%%%%%%%%%%%%%%%%%%%
%%%%%%%%%%%%%%%%%%%%%%%%%%%%%%%%%%%%%%%%%%%%%%%%%%%%%%%%%%%%
%%%%%%%%%%%%%%%%%%%%%%%%%%%%%%%%%%%%%%%%%%%%%%%%%%%%%%%%%%%%
\subsection{Conditions for two-step transition}
%%%%%%%%%%%%%%%%%%%%%%%%%%%%%%%%%%%%%%%%%%%%%%%%%%%%%%%%%%%%
%%%%%%%%%%%%%%%%%%%%%%%%%%%%%%%%%%%%%%%%%%%%%%%%%%%%%%%%%%%%
%%%%%%%%%%%%%%%%%%%%%%%%%%%%%%%%%%%%%%%%%%%%%%%%%%%%%%%%%%%%

The general form of the potential in \eqref{eq:leadingpot} does not imply a two-step EWPT during the cooling of the universe. At extremely high temperature, the origin is the only local minimum due to the thermal correction. As the universe cools, two local minima must appear successively and when $T = 0$ the SM vacuum $v = 246\,\GeV$ should be the global minimum. In this section, we discuss about three necessary conditions for a two-step phase transition. The tree-level potential is assumed to be bounded from below, which implies $\lambda_1, ~\lambda_2 > 0$ and $\lambda_{12} > -2\sqrt{\lambda_1\lambda_2}$.

\paragraph{Two different local minima}

%%%%%%%%%%%%%%%%%%%%%%%%%%%%%%%%%%%%
%%%%%%%%%%%%%%%%%%%%%%%%%%%%%%%%%%%%
\begin{figure}[t]
	\begin{center}
	\includegraphics[width=0.75\textwidth]{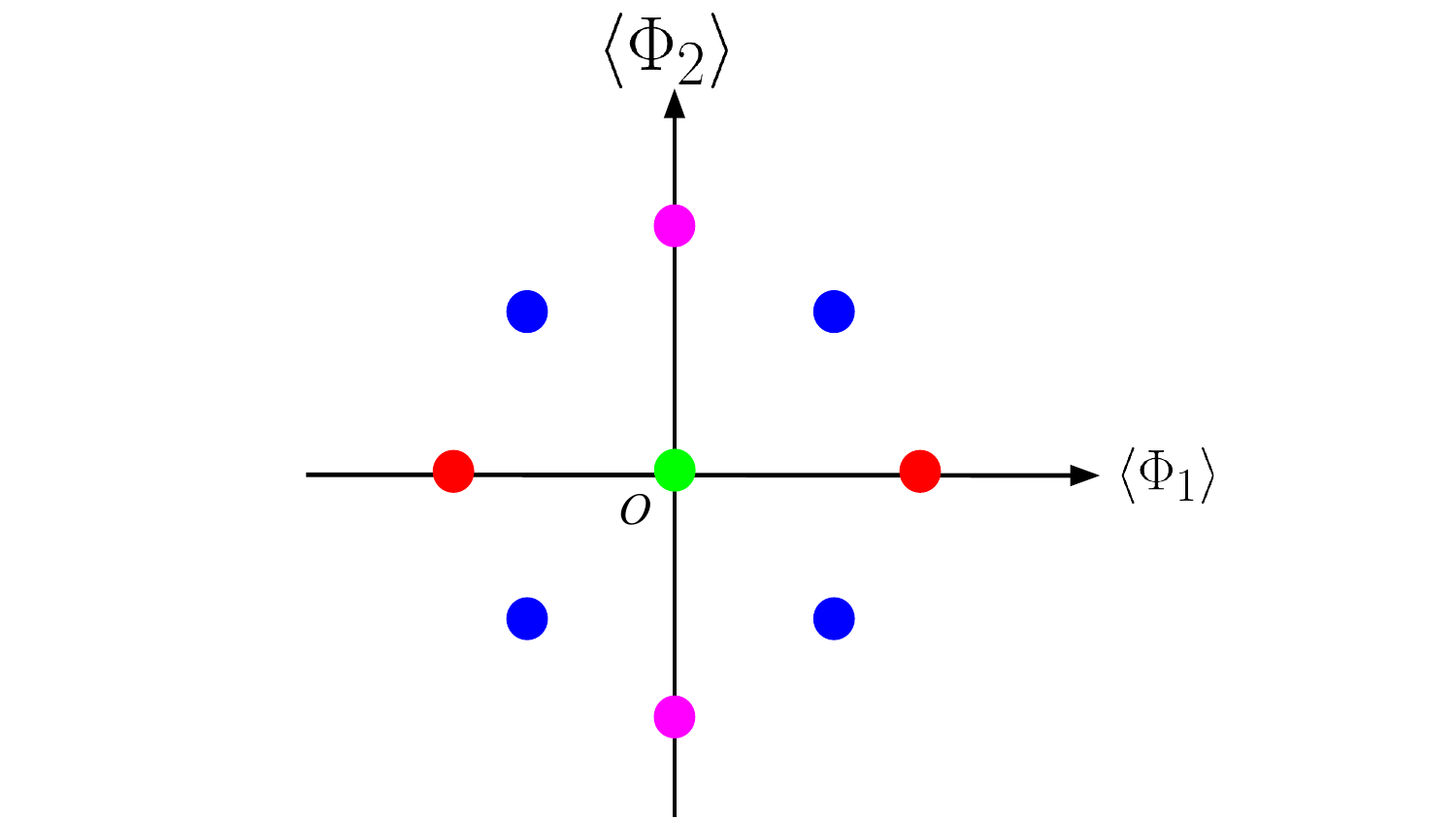}
	\caption{Extrema in the tree-level potential.}
	\label{Vtree}
	\end{center}
\end{figure}
%%%%%%%%%%%%%%%%%%%%%%%%%%%%%%%%%%%%
%%%%%%%%%%%%%%%%%%%%%%%%%%%%%%%%%%%%

There are nine potential local minima of the tree-level potential marked by green, red, magenta and blue color according to their behaviors under the $\mathbb{Z}_2 \times \mathbb{Z}_2$ symmetry in Fig.\,\ref{Vtree}. Only a transition from the magenta points to the red points can be a first-order phase transition since a barrier is necessary in the scenario. At zero temperature, the magenta and red points are given by $h_1^2 = \mu_1^2/\lambda_1,~ h_2 = 0$ and $h_1 = 0,~ h_2^2 = \mu_2^2/\lambda_2$ respectively. By further requiring that both of them are local minima of the tree-level potential, the positive-definiteness of the Hesse matrix
\begin{align}
	{\cal H} =  \begin{pmatrix}
	-\mu_1^2 + 3 \lambda_{1} h_1^2  + \lambda_{12} h_2^2/2  &  \lambda_{12}  h_1 h_2 \\ 
	\lambda_{12}  h_1 h_2 & -\mu_2^2 + 3 \lambda_{2} h_2^2  + \lambda_{12} h_1^2/2
	\end{pmatrix}
\end{align}
leads to~\cite{Cai:2022bcf}
\begin{align}\label{eq:con1}
	\lambda_{12}\mu_1^2 > 2\lambda_1 \mu_2^2 > 0, \quad \lambda_{12}\mu_2^2 > 2\lambda_2 \mu_1^2 > 0 \quad \Rightarrow \quad \lambda_{12} > 2 \sqrt{\lambda_1 \lambda_2}.
\end{align}
The last condition can also be obtained by requiring that the blue points are saddle points of the potential.

\paragraph{The global minimum}

As is discussed at the beginning of this subsection, the global minimum should lie at the SM-Higgs axis with $h_1 = 246 \GeV$, which requires~\cite{Cai:2022bcf}
\begin{align}\label{eq:con2}
	\frac{\mu_1^4}{\lambda_1} > \frac{\mu_2^4}{\lambda_2}.
\end{align}

\paragraph{$h_2$ vacuum appears first when cooling down}

As is indicated in Fig.~\ref{RM}-(b), a two-step phase transition requires that a local minimum along the $h_2$ axis appears first when cooling down. 
Along each axis ($h_2 = 0$ or $h_1 = 0$), it could be found that whether there is another extremum besides the origin depends on the quadratic term, $-\mu_i^2 + D_i T^2$. At high temperatures, all coefficients in the potential are positive, and the global minimum lies at the origin. When the temperature drops and the quadratic term becomes negative, the local minimum changes to $h_i^2 \sim (-\mu_i^2 + D_i T^2) / \lambda_i$ along each axis. We could define the turning temperature $T_i$ at which the quadratic coefficient changes sign, given by $T_i^2 = \mu_i^2 / D_i$.
Therefore, approximately the criterion for a two-step phase transition can be presented as
\begin{align}\label{eq:con3}
	T_2 > T_1,~ \text{i.e. } \frac{\mu_2^2}{D_2} > \frac{\mu_1^2}{D_1},
\end{align}
where $D_{1,2}$ are defined in Eq.~(\ref{eq:D1D2def}).

%%%%%%%%%%%%%%%%%%%%%%%%%%%%%%%%%%%%%%%%%%%%%%%%%%%%%%%%%%%%%%%%%
\begin{figure}[htb]
	\begin{center}
        ~~~~~~~~~~~~~~~
        \includegraphics[width=0.53\textwidth]{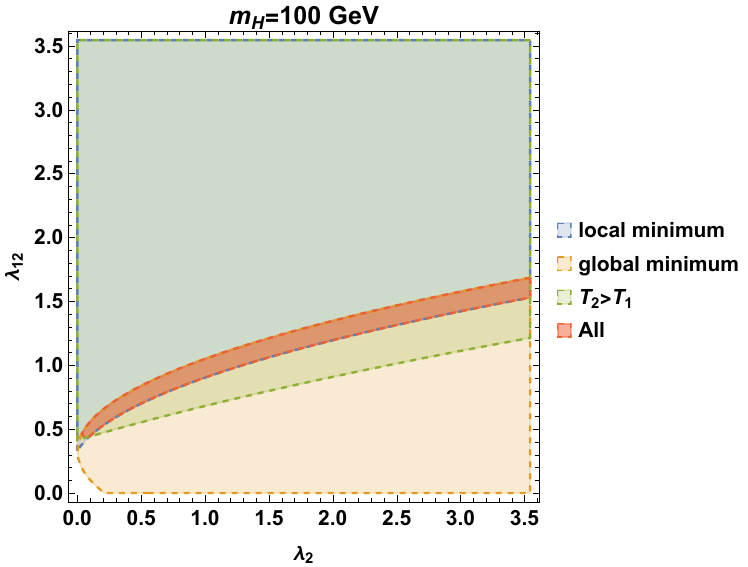}\\
        \includegraphics[width=0.4\textwidth]{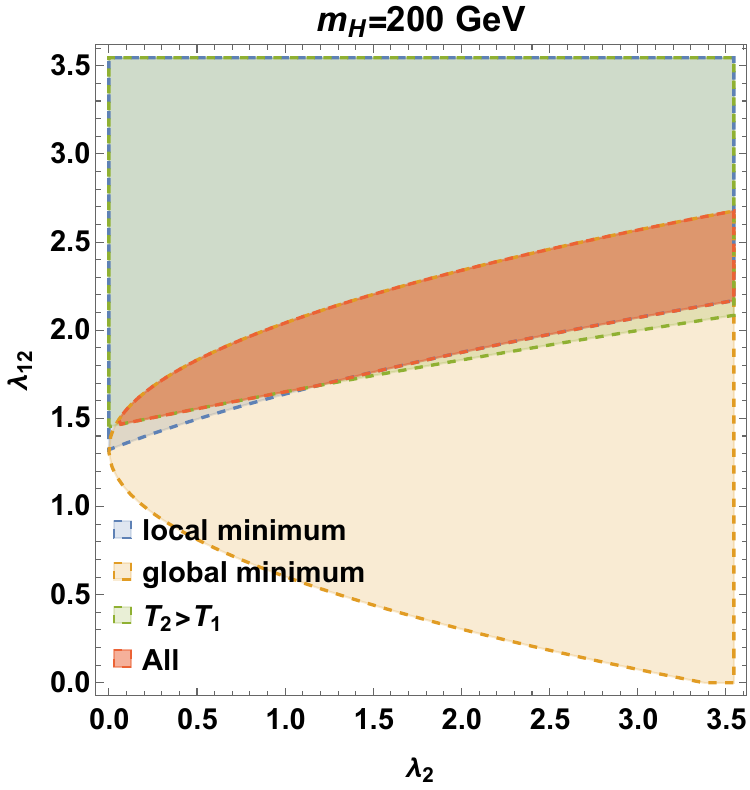}
        \includegraphics[width=0.4\textwidth]{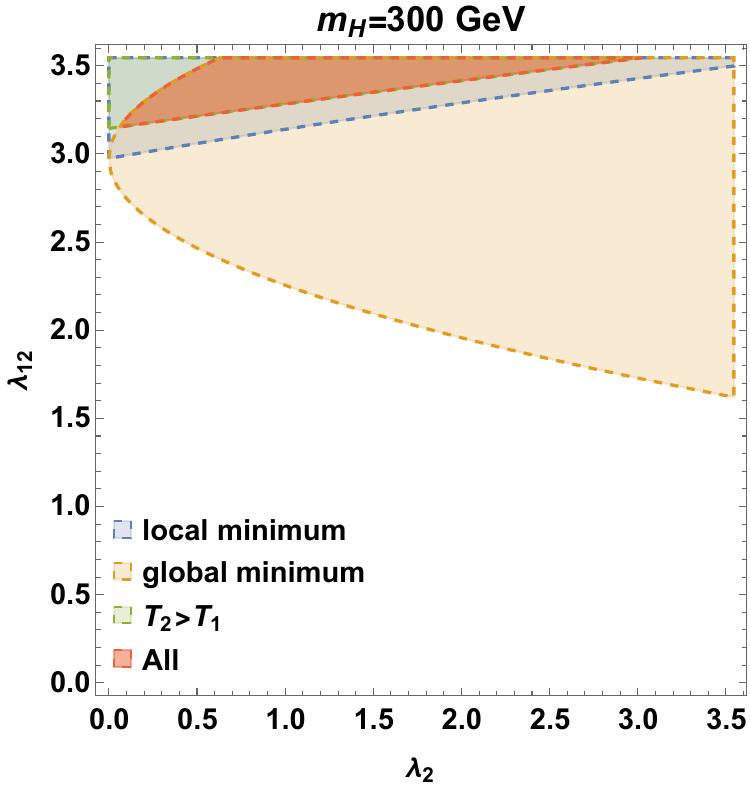}
        \caption{Parameter regions that satisfies the conditions of Eqs.~\eqref{eq:con1}, \eqref{eq:con2} and \eqref{eq:con3}, with $(I_2, Y_2) = (1/2, 1/2)$ and $m_H = 100, 200, 300 \,\GeV$. The region enclosed by blue, yellow, and green dashed lines corresponds to each condition, and the red is the union region. It could be found that the allowed regions for $\lambda_{12}$ are narrow belts scaling as $m_H$, and the upper bounds are set by the `global minimum' condition. Also, the lower bounds from `local minimum' and $T_2 > T_1$ are quite close, indicating that the phase transition goes through two steps given the condition that there are two local minima.
        }
        \label{fig:region}
    \end{center}
\end{figure}

\begin{figure}[htb]
    \centering
    \subfloat{
        \begin{tabular}{c|c}
        \centering
        \includegraphics[width=0.45\textwidth]{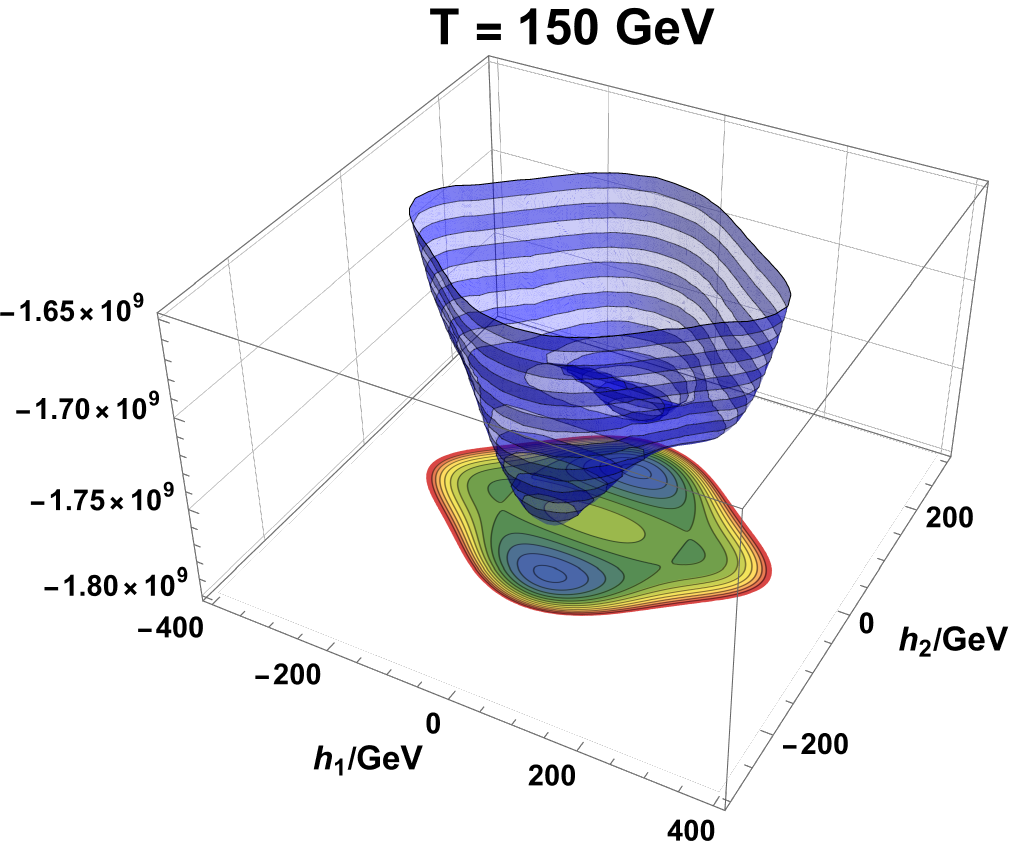}&
        \includegraphics[width=0.45\textwidth]{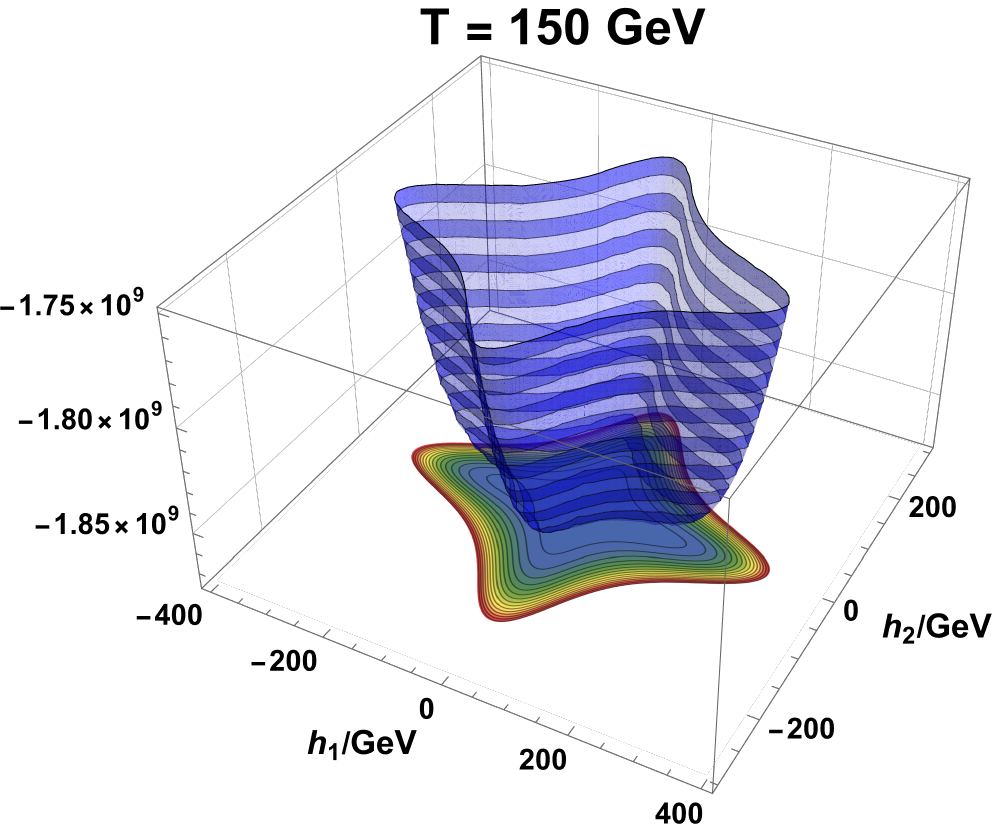}\\
        \includegraphics[width=0.45\textwidth]{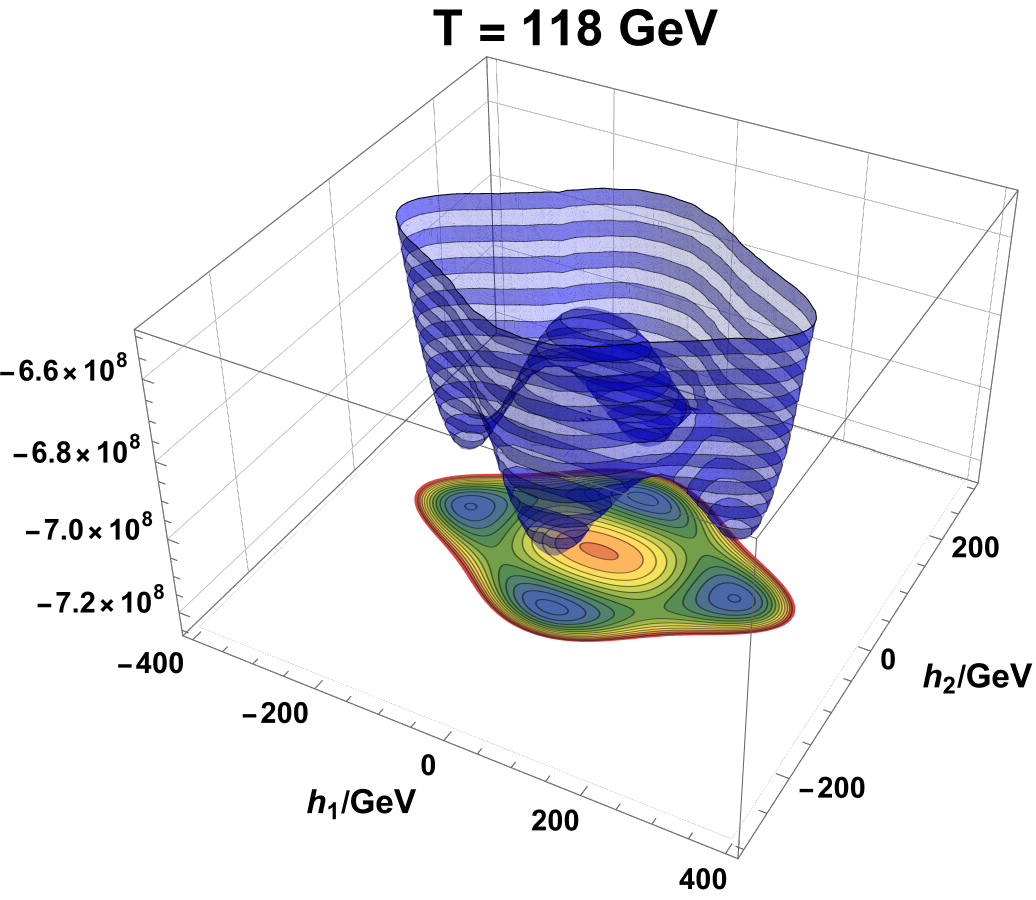}&
        \includegraphics[width=0.45\textwidth]{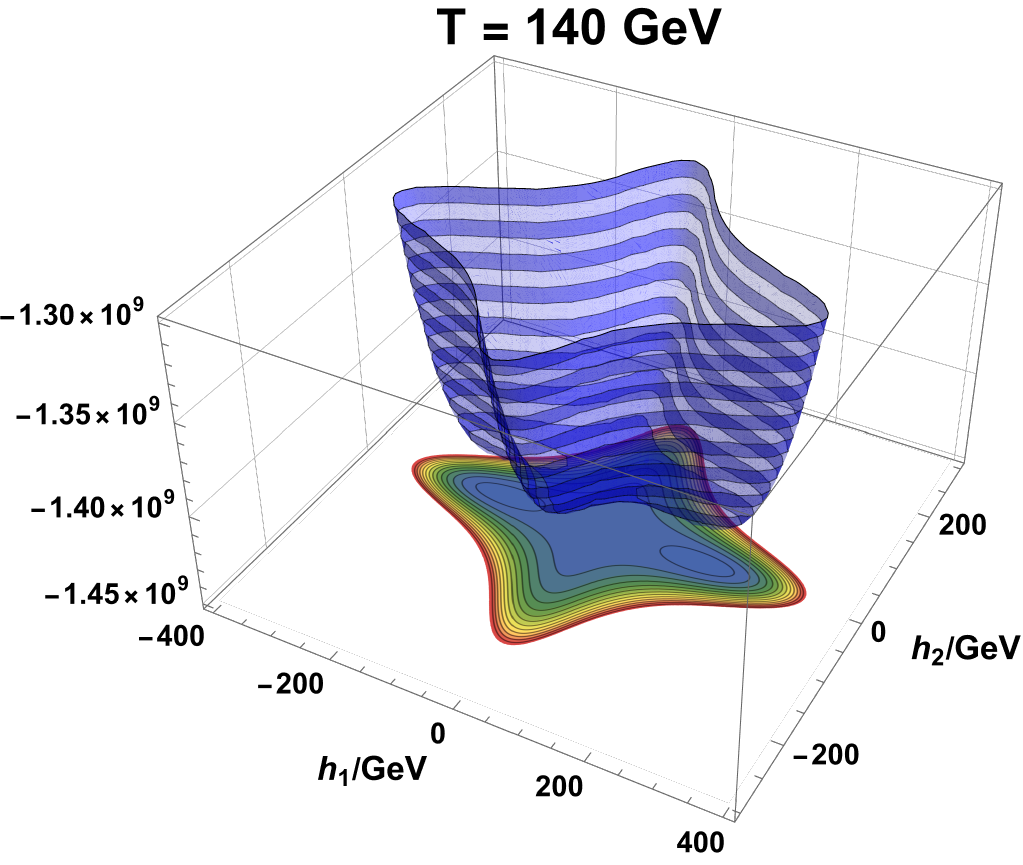}\\
        \includegraphics[width=0.45\textwidth]{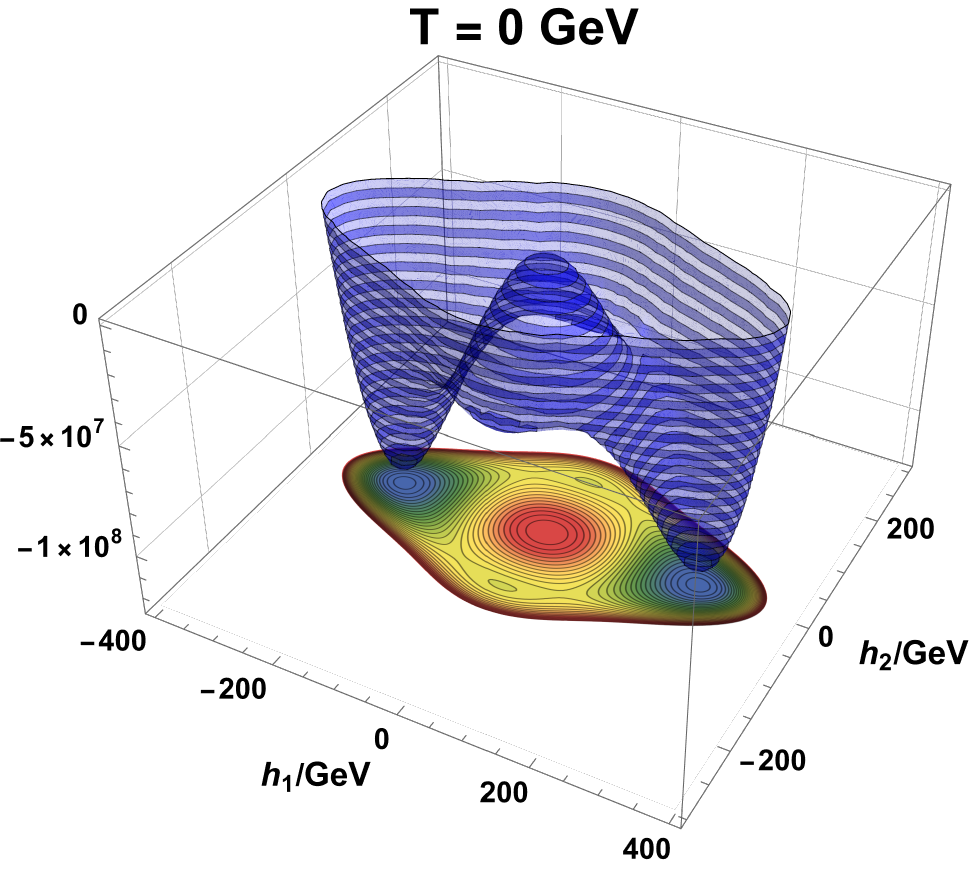}&
        \includegraphics[width=0.45\textwidth]{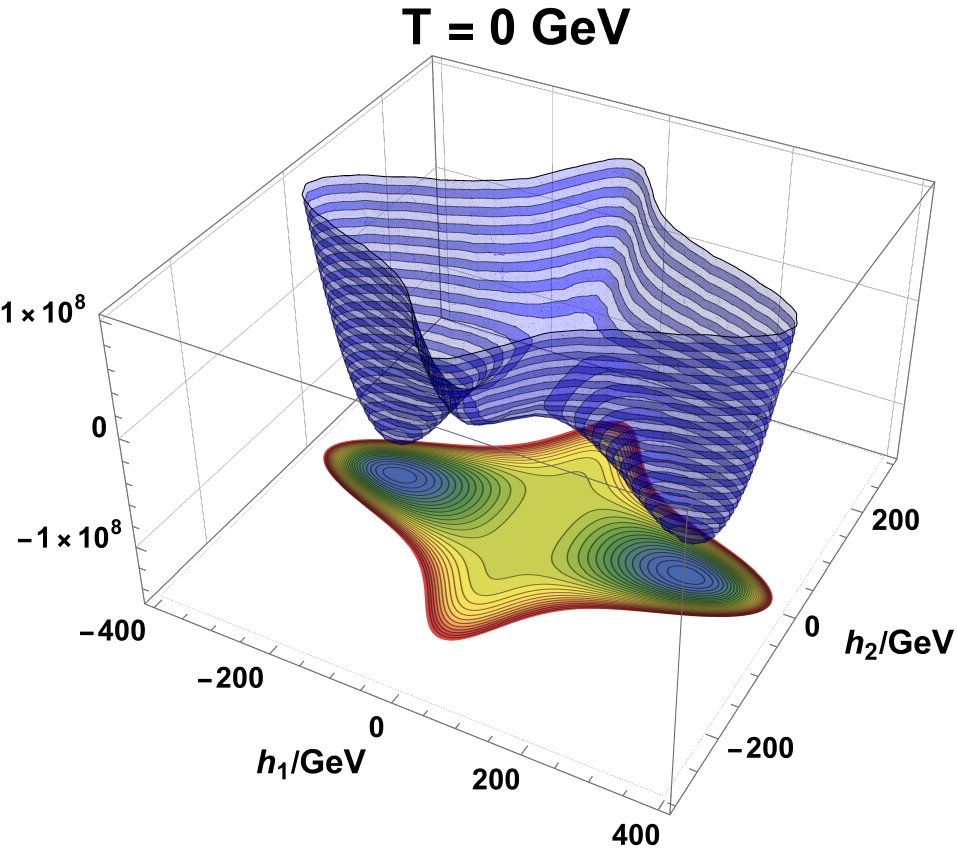}\\
        \end{tabular}
    }
    \caption{The potential as the temperature varieties. The left column shows the effective potential at $T = 150~\GeV$, $118~\GeV$ and $T = 0~\GeV$ respectively with $\lambda_2 = \lambda_{12} = 2.0,~ m_H = 200 ~\GeV$ and $(I_2, Y_2) = (1/2, 1/2)$, while the right is $T = 150 ~\GeV$, $140 \GeV$ and $T = 0 ~\GeV$ respectively with $\lambda_2 = 0.25,~ \lambda_{12} = 3.02,~ m_H = 300~ \GeV$ and $(I_2, Y_2) = (1, 1)$. Only the left case goes through a two-step phase transition.}
	\label{fig:6pot}
\end{figure}

We plot the parameter region that satisfies the three conditions above in Fig.~\ref{fig:region}. The region enclosed by blue, yellow, and green dashed lines corresponds to each condition, as shown in the legends, and the red one is the overlap region. It could be found that $\lambda_{12}$ is allowed in a narrow band in each plot, scaling as $m_H^2$. The upper bounds for $\lambda_{12}$ are set by the `global minimum' condition. In contrast, the lower bounds from the `local minimum' and $T_2 > T_1$ conditions are quite close, indicating that the phase transition goes through two steps given the condition that there are two local minima. 
We find that the two-step phase transition happens in a narrow region of the $\lambda_{12}$ couplings. This coupling should be moderately large and must be larger as the mass of the $\Phi_2$ increases. Furthermore, we find that as the isospin increases, the required  $\lambda_{12}$ coupling should be larger to realize a two-step phase transition.

Fig.~\ref{fig:6pot} shows how the potential evolves when the temperature drops. The numerical analysis of the effective potential includes the effect of tree-level, one-loop, and Daisy-resummed parts. The left column presents the case that $\lambda_2 = \lambda_{12} = 2.0,~ m_H = 200 ~\GeV$ and $(I_2, Y_2) = (1/2, 1/2)$. This set of parameters satisfies three conditions and does realize a two-step phase transition. By contrast, the right one only goes through one step, with the parameter $\lambda_2 = 0.25,~ \lambda_{12} = 3.02,~ m_H = 300 ~\GeV$ and $(I_2, Y_2) = (1, 1)$ lying outside the allowed region, even when there are two local minima at zero temperature.

%%%%%%%%%%%%%%%%%%%%%%%%%%%%%%%%%%%%%%%%%%%%%%%%%%%%%%%%%%%%
%%%%%%%%%%%%%%%%%%%%%%%%%%%%%%%%%%%%%%%%%%%%%%%%%%%%%%%%%%%%
%%%%%%%%%%%%%%%%%%%%%%%%%%%%%%%%%%%%%%%%%%%%%%%%%%%%%%%%%%%%
\subsection{Potential barriers for two-step transition}
%%%%%%%%%%%%%%%%%%%%%%%%%%%%%%%%%%%%%%%%%%%%%%%%%%%%%%%%%%%%
%%%%%%%%%%%%%%%%%%%%%%%%%%%%%%%%%%%%%%%%%%%%%%%%%%%%%%%%%%%%
%%%%%%%%%%%%%%%%%%%%%%%%%%%%%%%%%%%%%%%%%%%%%%%%%%%%%%%%%%%%

In the following, we will discuss the possibility of two SFOPTs in Fig.~\ref{RM}-(b) by the criterion that $v(T_C)/T_C \gtrsim 1$, where $T_C$ denotes the critical temperature at which the potential values at two local minima degenerate. For simplicity, we assume that the trajectory of the phase transition is straight in this subsection. A complete examination of the trajectory is instead performed in our numerical analysis in the following section.

To evaluate the criterion in the first step (green $\to $ magenta), we could analyze the potential along the $h_2$ axis. However, Eq.~\eqref{eq:leadingpot} cannot generate a barrier separating the origin and another local minimum because there only exists a quadratic and a quartic term. It is necessary to add the effects of the next-to-leading order, which is roughly
%%%%%%%%%%%%%%%%%%%%%%%%%%%%%%%%%%%%%%%%%%%%%%%%%%%%%%%%%%%
\begin{align}
    \label{eq:1stPT}
    V_{\text{eff}}\left(0, h_2, T\right) \sim \frac{1}{2} (-\mu_2^2 + D_2 T^2) h_2^2 - E_2 T h_2^3 +  \frac{1}{4} \lambda_2 h_2^4. 
\end{align}
%%%%%%%%%%%%%%%%%%%%%%%%%%%%%%%%%%%%%%%%%%%%%%%%%%%%%%%%%%%%
The cubic term $E_2$ comes from the thermal loops of bosons, and the contributions from gauge bosons and scalars are 
 %%%%%%%%%%%%%%%%%%%%%%%%%%%%%%%%%%%%%%%%%%%%%%%%%%%%%%%%%%%%
\begin{align}
 E_2^g &= \frac{1}{16\pi}g^3 I_{W}^{3} + \frac{1}{4\pi}(g^2+g'^2)^{3/2}Y_{2}^3,\nonumber\\
 E_2^\Phi &\sim \frac{2I_2 + 1}{6\pi}\sum_{m} b_m \times{\cal O}(1) \sim \frac{2I_2 + 1}{6\pi} \lambda_2.
\end{align}
%%%%%%%%%%%%%%%%%%%%%%%%%%%%%%%%%%%%%%%%%%%%%%%%%%%%%%%%%%%%
Here the explicit expression of $E_2^\Phi$ requires a specific model, and we only give a rough result with the approximation that $\lambda_2 \sim \sum_m b_m F(I_2, Y_2, m) \sim \sum_m b_m$.
The criterion for an SFOPT shows that
%%%%%%%%%%%%%%%%%%%%%%%%%%%%%%%%%%%%%%%%%%%%%%%%%%%%%%%%%%%%
\begin{align}
\label{}
   \frac{v_{2}(T_{C,1})}{T_{C,1}}=\frac{2E_2}{\lambda_2} \gtrsim 1 \quad \Rightarrow \quad \left(1 - \frac{2I_2 + 1}{3\pi}\right)\lambda_2 \lesssim 2E_2^g,
\end{align}
%%%%%%%%%%%%%%%%%%%%%%%%%%%%%%%%%%%%%%%%%%%%%%%%%%%%%%%%%%%%
where $v_2$ is the value of the local minimum and $T_{C,1}$ stands for the critical temperature at which $V_{\text{eff}}(0, 0, T_{C,1})=V_{\text{eff}}(0, v_{2}(T_{C,1}), T_{C,1})$. This estimation sets an upper bound for $\lambda_2$ under low representations $N \leq 9$, which indicates that an SFOPT in the first step may require a high representation of $SU(2)_L$.
In Sec.~\ref{sec:num} we will give a numerical analysis of the doublet model.

%----------- fig ---------------
\begin{figure}[t]
\begin{center}
    \includegraphics[width=0.8\textwidth]{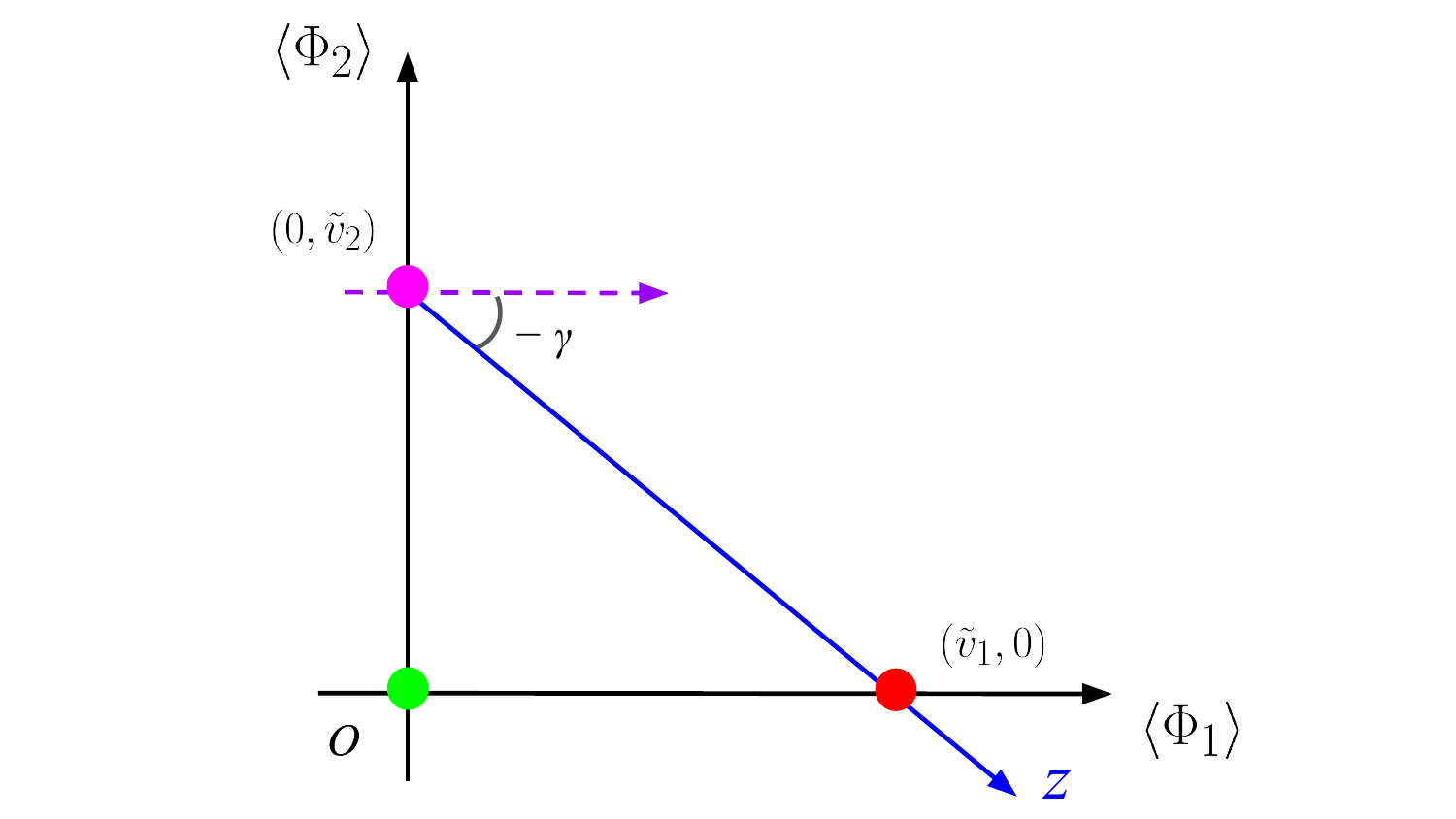}
    \caption{Polar coordinates in the effective potential.
    $z$ and $\gamma$ are a radius and an angle in the coordinates.}
    \label{fig:polar}
\end{center}
\end{figure}
%----------------------------------

\begin{figure}[t]
\begin{center}
    \includegraphics[width=0.4\textwidth]{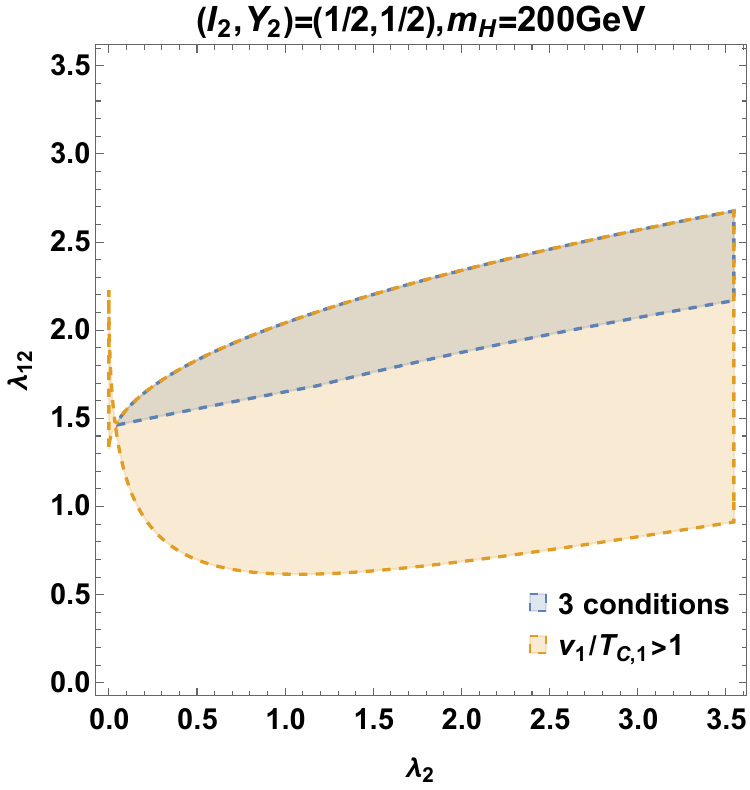}
    \includegraphics[width=0.4\textwidth]{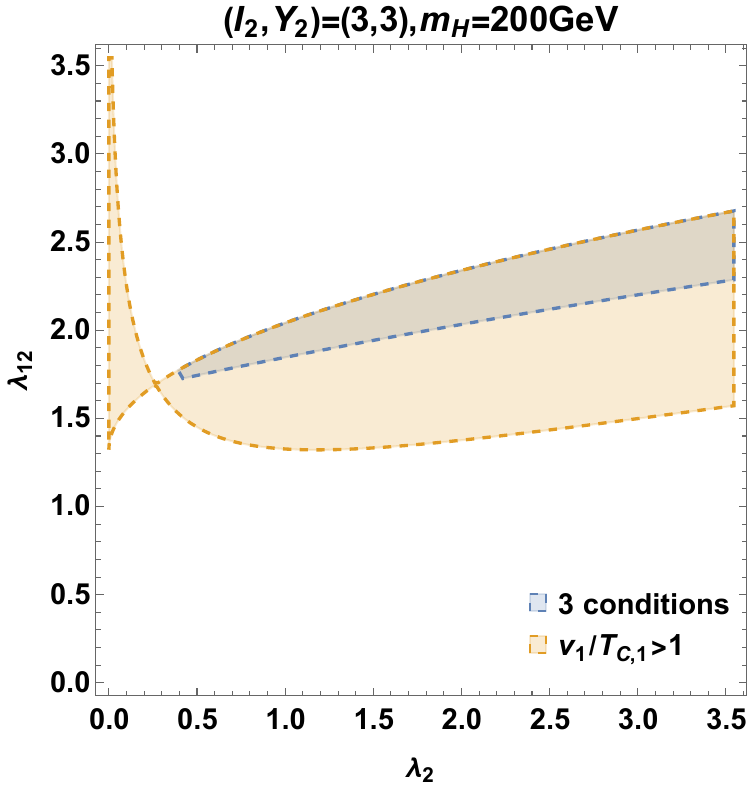}
    \caption{Comparison between the parameter region satisfying three conditions for two-step phase transition and SFOPT for the second step. The second region comprises the first one for both figures, with different isospins and hypercharges of $\Phi_2$.}
    \label{fig:criterion}
\end{center}
\end{figure}

For the second step of the EWPT (i.e., the path of magenta $ \to $ red) in Fig.~\ref{RM}-(b), we choose to reparametrize the field $h_1$ and $h_2$ by a polar coordinate $(z, \gamma)$ as is shown in Fig.~\ref{fig:polar}. The center of the polar coordinate is set to be the local minimum along the $h_2$ axis. The relation between two coordinates is given by $h_1 = z \cos \gamma$ and $h_2 = v_2(T) + z \sin \gamma$, and the coordinates of two minima are called $(0, 0)$ and $(\tilde{z}, \tilde{\gamma})$. Since a barrier already exists between two minima due to condition~\eqref{eq:con1}, it is sufficient to use the leading-order potential~\eqref{eq:effpot} to describe the phase transition and therefore
\begin{align}
    v_1(T)^2 = \frac{ \mu_1^2 - D_1T^2}{\lambda_1},~ v_2(T)^2 = \frac{\mu_2^2 - D_2T^2}{\lambda_2}, \nonumber \\
    \tilde{z}(T)^2 = v_1(T)^2 + ~v_2(T)^2,~ \tan\tilde{\gamma}(T) = -\frac{v_2(T)}{v_1(T)}.
\end{align} 
Under the polar coordinates, the effective potential Eq.~\eqref{eq:effpot} is given by
%%%%%%%%%%%%%%%%%%%%%%%%%%%%%%%%%%%%%%%%%%%%%%%%%%%%%%%%%%%%
\begin{align}
V_{\text{eff}}\left(z , \gamma , T\right) &= C_0+  C_2 z^2 +C_3 z^3 +C_4z^4,
\end{align}
%%%%%%%%%%%%%%%%%%%%%%%%%%%%%%%%%%%%%%%%%%%%%%%%%%%%%%%%%%%%
where
%%%%%%%%%%%%%%%%%%%%%%%%%%%%%%%%%%%%%%%%%%%
\begin{align}
    C_0 & = -\frac{1}{4}\lambda_2v_2(T)^4, \quad C_2 = \lambda_2v_2(T)^2  s_\gamma^2 + \frac{1}{4}(-2\lambda_1v^2 + 2D_1T^2 + \lambda_{12}v_2(T)^2) c_\gamma^2, \nonumber\\
    C_3 & = \frac{1}{2} v_2(T) s_\gamma (2 \lambda_{2}s_\gamma^2 + \lambda_{12} c_\gamma^2), \quad C_4 = \frac{1}{4}(\lambda_2s_\gamma^4 + \lambda_{12}s_\gamma^2 c_\gamma^2 + \lambda_1 c_\gamma^4).
\end{align}
%%%%%%%%%%%%%%%%%%%%%%%%%%%%%%%%%%%%%%%%%%
Here $\sin\gamma$ and $\cos\gamma$ are denoted $s_\gamma$ and $c_\gamma$, respectively. At the critical temperature $T_{C,2}$ the values of the potential at two local minima degenerate $V_{\text{eff}}(0, 0,  T_{C,2})=V_{\text{eff}}(\tilde{z}(T_{C,2}), \tilde{\gamma}(T_{C,2}), T_{C,2})$, which yields that
\begin{align}\label{eq:crittemp2}
    T_{C,2}^2 = \frac{\mu_2^2 - \sqrt{\lambda_2/\lambda_1}\mu_1^2}{D_2 - \sqrt{\lambda_2/\lambda_1}D_1}. 
\end{align}
Thus the criterion becomes
\begin{align}
    \frac{v_1(T_{C,2})^2}{T_{C,2}^2} = \frac{v^2}{T_{C,2}^2} - D_1 \sim \frac{\sqrt{\lambda_2/\lambda_1}D_1 - D_2}{\sqrt{\lambda_1\lambda_2} - \mu_2^2/v^2} > \frac{\sqrt{\lambda_2/\lambda_1}D_1 - D_2}{\sqrt{\lambda_1\lambda_2} - 2\lambda_1\lambda_2/\lambda_{12}},
\end{align}
where we use the conditions~\eqref{eq:con1} and~\eqref{eq:con2} that $\sqrt{\lambda_1\lambda_2} > \mu_2^2/v^2 > 2\lambda_1\lambda_2/\lambda_{12}$ and $\sqrt{\lambda_1} \sim \sqrt{1/8} \ll \sqrt{1/\lambda_1}$. The third condition~\eqref{eq:con3} guarantees that $\sqrt{\lambda_2/\lambda_1}D_1 > D_2$. Under the approximation $\lambda_2 \sim \lambda_{12} \sim 1$ as well as $D_1 \sim g^2/4 + y_t^2/4 + \lambda_2I_2/3$ and $D_2 \sim g^2I_2^2 + \lambda_{12}I_2/3$, we could estimate the value of the ratio to be
\begin{align}
    \frac{v_1(T_{C,2})^2}{T_{C,2}^2} \gtrsim \frac{g^2(\sqrt{2}/2 - I_2^2) + y_t^2\sqrt{2}/2 + (2\sqrt{2}\lambda_2 - \lambda_{12})I_2/3}{(\sqrt{2\lambda_2} - \lambda_2/\lambda_{12})/4} \sim \frac{-0.45I_2^2 + 0.6I_2 + 1}{0.1} > 1
\end{align}
for at least $I_2 \leq 2$. It indicates that the second step of the phase transition in Fig.~\ref{RM}-(b) is probable to be a SFOPT. Fig.~\ref{fig:criterion}, which shows the parameter region satisfying the criterion, verifies the validation of our approximation. In this section we only utilize the leading-order effective potential to describe the phase transition, and a complete numerical analysis will be presented in the following with specific models.

%----------- fig ---------------
%----------------------------------

%%%%%%%%%%%%%%%%%%%%%%%%%%%%%%%%%%%%%%%%%%%%%%%%%%%%%%%%%%%%
%%%%%%%%%%%%%%%%%%%%%%%%%%%%%%%%%%%%%%%%%%%%%%%%%%%%%%%%%%%%
%%%%%%%%%%%%%%%%%%%%%%%%%%%%%%%%%%%%%%%%%%%%%%%%%%%%%%%%%%%%
\section{Gravitational wave spectrum}
\label{sec:gw}
%%%%%%%%%%%%%%%%%%%%%%%%%%%%%%%%%%%%%%%%%%%%%%%%%%%%%%%%%%%%
%%%%%%%%%%%%%%%%%%%%%%%%%%%%%%%%%%%%%%%%%%%%%%%%%%%%%%%%%%%%
%%%%%%%%%%%%%%%%%%%%%%%%%%%%%%%%%%%%%%%%%%%%%%%%%%%%%%%%%%%%

Before doing numerical analysis, we briefly discuss the GW produced by the first-order phase transition.
Typically, the GW spectrum from the first-order phase transition could be characterized by phase transition parameters: $T_t$, $\alpha$, $\beta/H$, and $v_b$.

The first parameter $T_t$ is the transition temperature that is defined by an equation that a bubble nucleation probability $\Gamma$ per Hubble volume per Hubble time reaches the unit~\cite{Coleman:1977py}:
%%%%%%%
	\begin{align} 
	\label{Ttgh1}
	\Gamma/H^4|_{T=T_t}=1,
	\end{align}
%%%%%%%
where
%%%%%%%
	\begin{align} 
	\Gamma\simeq {\rm max}\left[T^4\left(\frac{S_3}{2\pi T} \right)^{3/2} \exp (-S_3/T),\,\, R_0^{-4}\left(\frac{S_4}{2\pi } \right)^{2} \exp (-S_4)\right].
	\end{align}
%%%%%%%
Here, $S_3$ ($S_4$) is a three- (four-) dimensional Euclidean action for O(3) [O(4)] symmetric bounce solution.
Also, $R_0$ is the size of the nucleating bubble.
Eq.~(\ref{Ttgh1}) means one critical bubble nucleates into the causality. In order to calculate the bounce solution, we used the public code {\tt AnyBubble}~\cite{Masoumi:2016wot}.

The second parameter $\alpha$ is the normalized released energy density by the radiation energy density at $T_t$: 
%%%%%%%
	\begin{align} 
	\alpha\equiv \epsilon(T_t)/ \rho_\text{rad}(T_t),
	\end{align}
%%%%%%%
where $\rho_{rad}=(\pi^2/30)g_* T^4$ and $\epsilon$ is released energy density:
%%%%%%%%%%%%%%%%%%%%%%%%%%%%%%%%%%%%%%%%%%
 \begin{align}
 \label{latenth}
  \epsilon(T)
  =  \Delta V_\text{eff} -T
  \frac{\partial  \Delta V_\text{eff} }{\partial T},\quad  \Delta V_\text{eff} =  V_\text{eff}(\varphi_-(T),T) - V_\text{eff}(\varphi_+(T),T),
\end{align}
%%%%%%%%%%%%%%%%%%%%%%%%%%%%%%%%%%%%%%%%%%
where $\varphi_{+(-)}$ is $\varphi$ at the broken (unbroken) phase.

The third parameter $\beta/H$ is the normalized $\beta$ parameter that is related to the time variation scale of bubble nucleation rate $\Gamma(t)=\Gamma_0\exp (\beta t)$.
This normalized $\beta$ parameter is defined as  
%%%%%%%
	\begin{align} 
	\frac{\beta}{H}\equiv\left.\frac{\mathrm{d}S_3}{\mathrm{d}T}\right|_{T=T_t}.
	\end{align}
%%%%%%%	
The last one is the bubble wall velocity $v_b$.
In our following analysis, we fix it by hand for simplicity.

By using these parameters, the GW spectrum from first-order phase transition could be estimated.
There are three sources of GW spectrum: bubble collision~\cite{Huber:2008hg}, plasma turbulence~\cite{Caprini:2009yp}, and compression wave of plasma~\cite{Caprini:2015zlo}.
The spectra could be calculated by the details of bubble dynamics, and the fitting formulas of them are provided in Ref.~\cite{Caprini:2015zlo}.
In our analysis, we will focus on the GW spectrum from compression wave of plasma, which typically is the largest contribution among them.
The formula from the compression wave of plasma~\cite{Caprini:2015zlo} is given by
%%%%%%%%%%%%%%%
	\begin{align}
	\label{CompGW}
  \Omega_{\rm comp} (f) h^2 
  =\widetilde{\Omega}_{\rm comp} h^2 \times
    (f/\tilde{f}_{\rm comp})^3
  \left(\frac{7}{4+3(f/\tilde{f}_{\rm comp})^2}\right)^{7/2},  
	  \end{align}
%%%%%%%%%%%%%%%
where the tilde parameters are the peak of the spectrum
%%%%%%%%%%%%%%%
	\begin{align}
		\label{CompGWpeak}
  \widetilde{\Omega}_{\rm comp} h^2
  \simeq 2.65 \times 10^{-6} v_b 
  \tilde{\beta}^{-1}
  \left(\frac{\kappa_v \alpha}{1+\alpha}\right)^2
  \left(\frac{100}{g_\ast}\right)^{1/3}, 
	\end{align}
%%%%%%%%%%%%%%%
and the peak frequency
%%%%%%%%%%%%%%%
	\begin{align}
  \tilde{f}_{\rm comp} \simeq 1.9 \times 10^{-5}~{\rm Hz} \frac{1}{v_b}
  \tilde{\beta}
  \left(\frac{T_t}{100~{\rm GeV}}\right)
  \left(\frac{g_\ast}{100}\right)^{1/6}. 
	\end{align}
%%%%%%%%%%%%%%%
Here, the $g_\ast$ is the number of degrees of freedom in the plasma.
The parameter $\kappa_v$ in Eq.~(\ref{CompGWpeak}) is an efficiency factor, which denotes the fraction of the vacuum energy transformed into the bulk motion of the plasma fluid~\cite{Espinosa:2010hh}:
%%%%%%%%%%%%%%%
\begin{align}
  \kappa_v(v_b, \alpha)\simeq
  \begin{cases}
    & \frac{ c_s^{11/5}\kappa_A \kappa_B }{(c_s^{11/5} -  v_b^{11/5} )\kappa_B
      +  v_b c_s^{6/5} \kappa_A}\quad (\text{for}~ v_b \lesssim c_s ) \\
    & \kappa_B + ( v_b - c_s) \delta\kappa 
    + \frac{( v_b - c_s)^3}{ (v_J - c_s)^3} [ \kappa_C - \kappa_B -(v_J - c_s) 
    \delta\kappa ]
    \quad (\text{for}~ c_s <  v_b < v_J) \\
    & \frac{ (v_J - 1)^3 v_J^{5/2}  v_b^{-5/2}
      \kappa_C \kappa_D }
    {[( v_J -1)^3 - ( v_b-1)^3] v_J^{5/2} \kappa_C
      + ( v_b - 1)^3 \kappa_D }
    \quad ( \text{for}~ v_J \lesssim v_b )
  \end{cases},    
\end{align}
%%%%%%%%%%%%%%%
where $c_s$ is the velocity of sound ($c_s=0.577$) and 
%%%%%%%%%%%%%%%
\begin{align}
  \kappa_A 
  &\simeq v_b^{6/5} \frac{6.9 \alpha}{1.36 - 0.037 \sqrt{\alpha} + \alpha},
  \quad
  \kappa_B 
  \simeq \frac{\alpha^{2/5}}{0.017+ (0.997 + \alpha)^{2/5} },
    \nonumber \\
  \kappa_C 
  &\simeq \frac{\sqrt{\alpha}}{0.135 + \sqrt{0.98 + \alpha}},
  \quad
  \kappa_D 
  \simeq \frac{\alpha}{0.73 + 0.083 \sqrt{\alpha} + \alpha}.
\end{align}
%%%%%%%%%%%%%%%

In order to discuss the testability of the parameter region for the multi-step EWPT, we estimate the signal-to-noise ratio (SNR) for measurements of GW spectrum~\cite{Thrane:2013oya, Caprini:2015zlo}.
The SNR is defined as 
%%%%%%%
	\begin{align} 
	\label{SNRGW}
	\text{SNR} \equiv \sqrt{\delta\times t_\text{obs} \int^{f_\text{max}}_{f_\text{min}}df \left[\frac{h^2 \Omega_\text{GW}(f)}{h^2\Omega_\text{Sens}(f)}\right]^2},
	\end{align}
%%%%%%%	
where $\delta$ is the number of independent channels for the experiment, $t_{obs}$ corresponds to the experimental observation period, $h^2 \Omega_\text{GW}(f)$ denotes the GW signal coming from the first-order phase transition and $h^2\Omega_\text{Sens}(f)$ denotes the sensitivity of experiments, such as LISA~\cite{Klein:2015hvg}, DECIGO and BBO~\cite{Yagi:2011wg}.
When the SNR at the LISA experiment is larger than 10, we could detect the GW spectrum~\cite{Caprini:2015zlo}.
In our analysis, we take $\text{SNR}>10$ as a criterion for whether the GW spectrum could be detected or not.

%%%%%%%%%%%%%%%%%%%%%%%%%%%%%%%%%%%%%%%%%%%%%%%%%%%%%%%%%%%%
%%%%%%%%%%%%%%%%%%%%%%%%%%%%%%%%%%%%%%%%%%%%%%%%%%%%%%%%%%%%
%%%%%%%%%%%%%%%%%%%%%%%%%%%%%%%%%%%%%%%%%%%%%%%%%%%%%%%%%%%%
\subsection{Supercooling phase transition}
%%%%%%%%%%%%%%%%%%%%%%%%%%%%%%%%%%%%%%%%%%%%%%%%%%%%%%%%%%%%
%%%%%%%%%%%%%%%%%%%%%%%%%%%%%%%%%%%%%%%%%%%%%%%%%%%%%%%%%%%%
%%%%%%%%%%%%%%%%%%%%%%%%%%%%%%%%%%%%%%%%%%%%%%%%%%%%%%%%%%%%

For the second step of the EWPT in Fig.~\ref{RM}-(b), a barrier between magenta and red points appears at zero temperature.
Then, the supercooling phase transition may be realized.
The condition of the supercooling phase transition is typically given as $\alpha\gg1$.
In the case of the supercooling phase transition, the broken phase may not fill with causality.
Therefore we should estimate the temperature for bubble percolation for the phase transition to complete successfully.
The percolation temperature $T_p$ means the broken phase is filled at least $34\%$ of the comoving volume at this temperature~\cite{Enqvist:1991xw, Ellis:2018mja}.
To evaluate the $T_p$, we use the probability of finding a point still in the false vacuum:
%%%%%%%
	\begin{align} 
	\label{Perco}
	P(T)=e^{-I(T)},\quad I(T)=\frac{4\pi}{3}\int^T_{T_c} \mathrm{d}T' \Gamma(T') a(T')^3 r(T, T')^3, 
	\end{align}
%%%%%%%	
where $a(T')$ is the Friedmann-Robertson-Walker scale factor, and $r(T,T')$ is the growing comoving size of a bubble from $T'$ to $T$:
%%%%%%%
	\begin{align}
	 r(T, T')=\int^t_{t'}\frac{\mathrm{d}\tilde{t}v_w}{a(\tilde{t})}.
	\end{align}
%%%%%%%	
The $P(T)$ in Eq.~(\ref{Perco}) denotes the amount of true vacuum volume per unit comoving volume.
By using it, the percolation temperature $T_p$ is defined as 
%%%%%%%
	\begin{align} 
	I(T_p)=0.34.
	\end{align}
%%%%%%%	
The physical volume of the false vacuum space ${\cal V}_{\rm false}$ decreases around percolation to complete the phase transition.
The condition is given by \cite{Ellis:2018mja}
%%%%%%%
	\begin{align}
	 \frac{1}{{\cal V}_{\rm false}}\frac{\mathrm{d}{\cal V}_{\rm false}}{\mathrm{d}t}<0\quad \Rightarrow \quad H(T)\left(3+ T\frac{\mathrm{d}I(T)}{\mathrm{d}T}\right)<0.
	\end{align}
%%%%%%%	
From this condition and equation, we can obtain the $T_p$.
Since the bubbles for the broken phase collide with each other at this temperature, the phase transition parameters are determined at this temperature $T_p$.
In the SNR phase transition, the GW spectrum is replaced as \cite{Ellis:2018mja}
%%%%%%%%%%%%%%%
	\begin{align}
	\label{CompGWSC}
  \Omega_{\rm comp} (f) h^2 
  =2.65 \times 10^{-6} v_b 
 \frac{H\bar{R}}{(8\pi)^{1/3}}
   \left(\frac{\kappa_v \alpha}{1+\alpha}\right)^2
  \left(\frac{100}{g_\ast}\right)^{1/3}
    (f/\tilde{f}_{\rm comp})^3
  \left(\frac{7}{4+3(f/\tilde{f}_{\rm comp})^2}\right)^{7/2},  
	  \end{align}
%%%%%%%%%%%%%%%
where 
%%%%%%%%%%%%%%%
	\begin{align}
  \tilde{f}_{\rm comp} \simeq 1.9 \times 10^{-5}  \frac{1}{v_b}
 \frac{(8\pi)^{1/3}}{H\bar{R}}
  \left(\frac{T_t}{100~{\rm GeV}}\right)
  \left(\frac{g_\ast}{100}\right)^{1/6}. 
	\end{align}
%%%%%%%%%%%%%%%
where $\bar{R}\sim (v_w-c_s)R_{\rm max}$ and $R_{\rm max}$ is the size of the bubble of maximum-energy-configuration of bubble distributions at percolation.

We will evaluate the GW spectrum from the first-order phase transition in our model by the above equations.
In the following numerical results, we will obtain the phase transition parameters in the model with the EWPT in Fig.~\ref{RM}-(b) and will discuss the testability of the model parameters by the SNR.
After that, we will show whether the two first-order EWPTs could be realized or not.

%%%%%%%%%%%%%%%%%%%%%%%%%%%%%%%%%%%%%%%%%%%%%%%%%%%%%%%%%%%%
%%%%%%%%%%%%%%%%%%%%%%%%%%%%%%%%%%%%%%%%%%%%%%%%%%%%%%%%%%%%
%%%%%%%%%%%%%%%%%%%%%%%%%%%%%%%%%%%%%%%%%%%%%%%%%%%%%%%%%%%%
\section{Numerical results}
\label{sec:num}
%%%%%%%%%%%%%%%%%%%%%%%%%%%%%%%%%%%%%%%%%%%%%%%%%%%%%%%%%%%%
%%%%%%%%%%%%%%%%%%%%%%%%%%%%%%%%%%%%%%%%%%%%%%%%%%%%%%%%%%%%
%%%%%%%%%%%%%%%%%%%%%%%%%%%%%%%%%%%%%%%%%%%%%%%%%%%%%%%%%%%%

\subsection{Patterns of EWPTs}

\begin{figure}[t]
  \begin{center}
\includegraphics[width=0.4\textwidth]{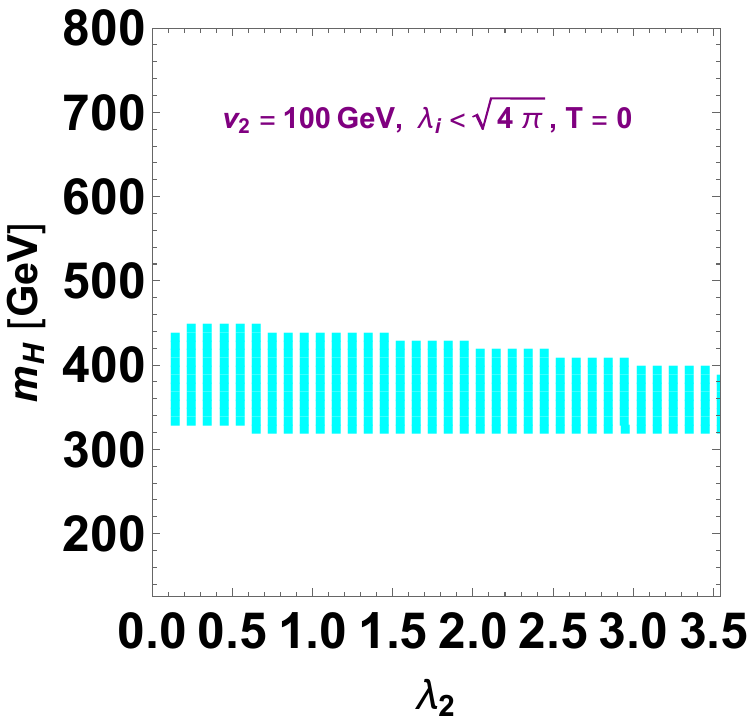}
\includegraphics[width=0.4\textwidth]{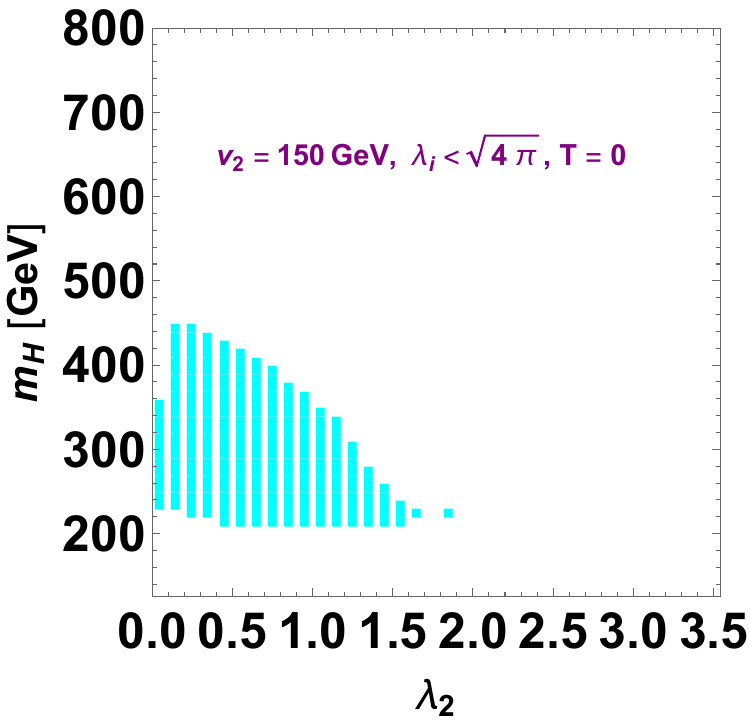}\\
\includegraphics[width=0.4\textwidth]{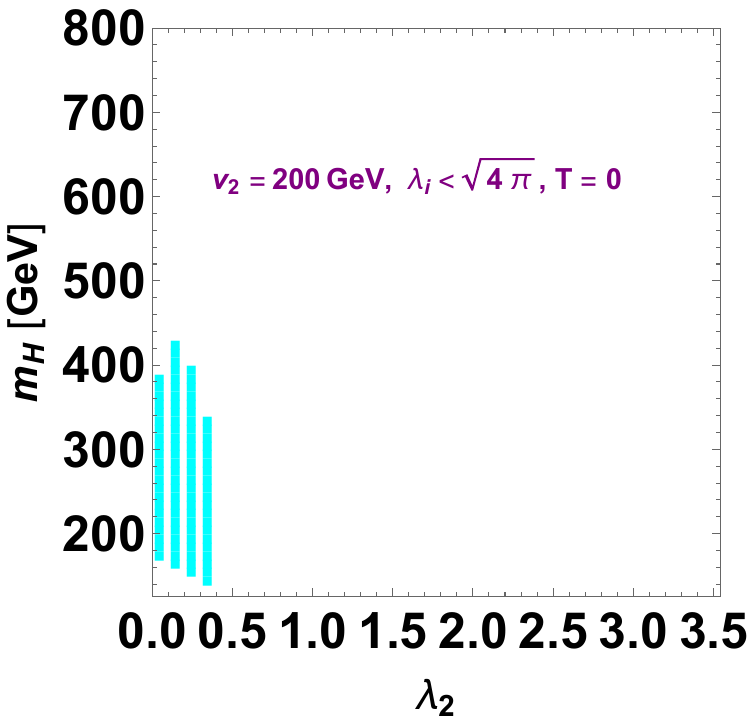}
\caption{Regions of realizing the EWPT in Figs.~\ref{RM}-(a) and -(b) by the potential with one-loop effects ($Y_{2}=1/2$ and $I_{2}=1/2$). 
These color regions can satisfy the conditions of ``two different local minima'' and ``the global minimum.''
}
\label{onres}
  \end{center}
\end{figure}

According to the results in Sec.~\ref{sec:twostep}, the EWPT in Figs.~\ref{RM}-(a) and -(b) could be realized in the model, especially, multi-step EWPT in Fig.~\ref{RM}-(b) can occur by the red parameter region in Fig.~\ref{fig:region}.
We will numerically analyze the detail of the possibility of the pattern of the EWPT in the model with $Y_{2}=1/2$ and $I_{2}=1/2$.
In the numerical analysis in this section, the four parameters in the potential, $ \mu_1^2$, $\mu_2^2$, $ \lambda_1$ and $\lambda_{12}$, will be taken as $v$, $v_2$, $m_h$ and $m_H$ by using the stationary conditions and second derivative of the potential of Eqs.~(\ref{redc1}), (\ref{redc2}), (\ref{redc3}) and (\ref{redc4}) in appendix B.
The conditions for the multi-step EWPT, which are shown in  Eqs.~(\ref{eq:con1}) and (\ref{eq:con2}), are given by Fig.~\ref{onres}.
%%%%%%%%%%%%%%%%%%%%%%%%%%%%%%%%%%%%
%%%%%%%%%%%%%%%%%%%%%%%%%%%%%%%%%%%%
%%%%%%%%%%%%%%%%%%%%%%%%%%%%%%%%%%%%
%%%%%%%%%%%%%%%%%%%%%%%%%%%%%%%%%%%%
These color regions can satisfy the conditions, including one-loop effects in the cases that $v_2=$ 100, 150, and 200 GeV.
In these cyan regions, the multi-step EWPT may be generated, however, we use the potential without finite temperature effects to describe these regions.
Therefore, we should consider the potential with finite temperature effects to get more detail on the possibility of the EWPTs, like the condition of ``the $h_2$ vacuum appears first when cooling down.''

\begin{figure}[t]
  \begin{center}
\includegraphics[width=0.4\textwidth]{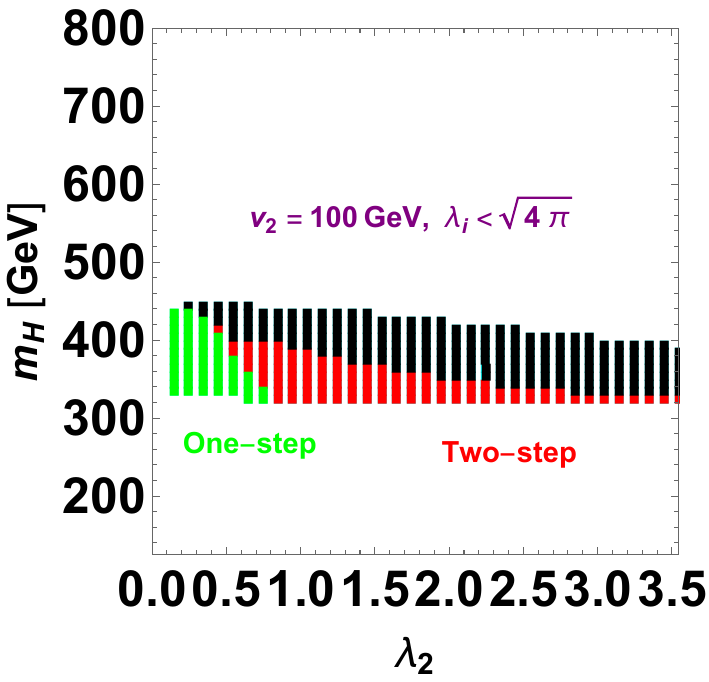}
\includegraphics[width=0.4\textwidth]{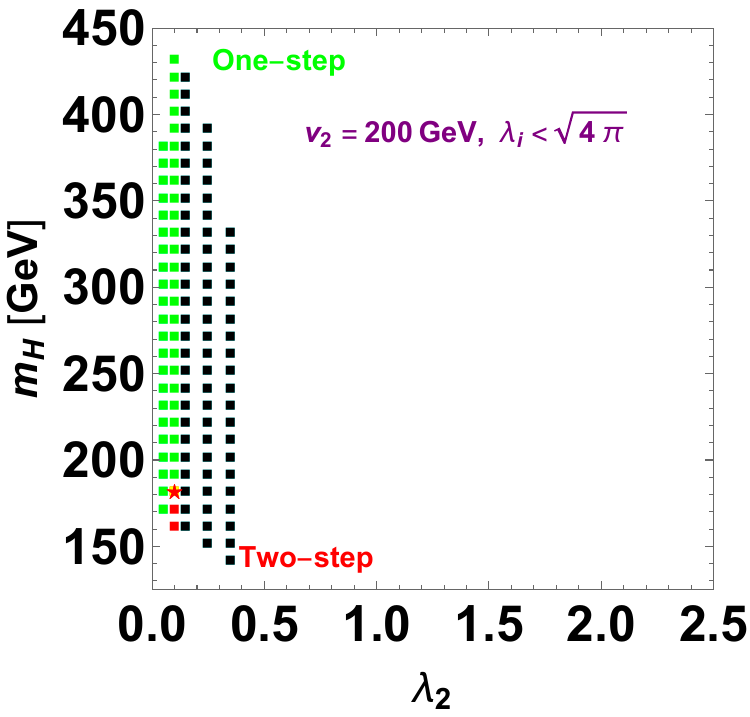}
\caption{EWPT pattern in the plane of ($\lambda_2$, $m_H$ [GeV]) for $v_2=$ 100 and 200 GeV.
The green squares represent the parameter region for the one-step phase transition of Fig.~\ref{RM}-(a).
The red squares represent the parameter region for the two-step phase transition of Fig.~\ref{RM}-(b).
In the black squares, the EWPT has not been completed in the current Universe.
A red star mark in the right panel could realize two first-order phase transitions in both steps of Fig.~\ref{RM}-(b).
}
\label{resgen}
  \end{center}
\end{figure}

The results, including the finite temperature effect, are given in Fig.~\ref{resgen}.
%%%%%%%%%%%%%%%%%%%%%%%%%%%%%%%%%%%%
%%%%%%%%%%%%%%%%%%%%%%%%%%%%%%%%%%%%
%%%%%%%%%%%%%%%%%%%%%%%%%%%%%%%%%%%%
%%%%%%%%%%%%%%%%%%%%%%%%%%%%%%%%%%%%
To clarify the one-step and multi-step EWPT in Figs.~\ref{RM}-(a) and -(b), we compare the temperatures starting the phase transition along $h_1$ and $h_2$ axes.
In the red (green) square marks, such a temperature along $h_2$ ($h_1$) is higher than $h_1$ ($h_2$).
Furthermore, the one-step EWPT of Fig.~\ref{RM}-(a) (green squares) and the second step of the EWPT in Fig.~\ref{RM}-(b) (red squares) are first order.
Since a barrier in the second path of the multi-step EWPT at red squares (one-step EWPT at green squares) occurs by the nonthermal effects (thermal effects), the frequency of GW may be different between the path of the EWPT of Figs.~\ref{RM}-(a) and -(b).
According to the numerical results, the peak frequency at red squares is less than about $10^{-1}$ Hz, while the frequency at green squares is ${\cal O} (1)$ Hz.
Therefore, we may distinguish the patterns of the EWPT of Figs.~\ref{RM}-(a) and -(b) by the GW observation.
At the black square marks, the EWPT cannot be completed at the current Universe because the barrier for the second step of the EWPT in Fig.~\ref{RM}-(b) does not disappear at zero temperature.
At the red star mark in the right panel of Fig.~\ref{resgen}, both steps of the EWPT in Fig.~\ref{RM}-(b) are first order.
It is difficult to realize such a phase transition because the first step relies on finite temperature effects.
The detail of that will be discussed in Sec.~\ref{sec:pos_two_ewpt}.
As the numerical results in Fig.~\ref{resgen}, the multi-step EWPT in Fig.~\ref{RM}-(b) could be realized in the case of the large $\lambda_2$ value, that can follow the condition in Eq.~(\ref{eq:con3}).
Also, when we take heavy $m_H$ and large $\lambda_2$ value, the EWPT has not been completed, because the height difference between the magenta and blue points $\Delta V$ becomes high by heavy $m_H$ and the large $\lambda_2$: $\Delta V\sim \frac{\lambda_2}{4 v^4}( 2v_2^2(m_H^2+\lambda_2v_2^2)-m_h^2v^2 )^2$. 
For example, the large $\lambda_2$ with $v_2=$ 200 GeV is prohibited by the condition of height potential in Eq.~(\ref{eq:con2}).
These behaviors could follow the discussion in Sec.~\ref{sec:twostep}.

%%%%%%%%%%%%%%%%%%%%%%%%%%%%%%%%%%%%%%%%%%%%%%%%%%%%%%%%%%%%
%%%%%%%%%%%%%%%%%%%%%%%%%%%%%%%%%%%%%%%%%%%%%%%%%%%%%%%%%%%%
\subsection{Isospin and hypercharge effects}
%%%%%%%%%%%%%%%%%%%%%%%%%%%%%%%%%%%%%%%%%%%%%%%%%%%%%%%%%%%%
%%%%%%%%%%%%%%%%%%%%%%%%%%%%%%%%%%%%%%%%%%%%%%%%%%%%%%%%%%%%

The isospin and hypercharge of the multiplet also contribute to the patterns of the EWPT.
First, we note that the hypercharge of additional scalar boson $Y_{2}$ does not contribute much to the order of step of the multi-step EWPT.
The effect of $Y_{2}$ appears in the denominator of $T_2$, and it is proportional to $(m_Z^2-m_W^2)$. 
Since the masses of these gauge bosons are almost the same, the $Y_{2}$ does not much affect the EWPT. 
On the other hand, the small $I_{2}$ makes the multi-step EWPT because the effect of $I_{2}^2$ in the denominator of $T_2$, which does not appear in $T_1$, can be neglected. 
Then, the $T_2^2>T_{1}^2$ condition can be realized. 
On the other hand, for the large $I_{2}$, the $T_{1}$ and $T_2$ are roughly given by 
 %%%%%%%
	\begin{align}
	T_{1}^2&\sim\frac{3m_h^2v^2}
	{2I_{2}(m_{H}^2 + \lambda_2 v_2^2 )} ,\quad T_2^2\sim  \frac{\lambda_2v_2^2v^2 + \frac{3I_{2} \lambda_2^2v^2v_2^2}{\pi^2}\left(  \ln \frac{T_2^2\alpha_B}{Q^2}  - \frac{3}{2}\right)}{ I_{2}^2m_W^2},
	\end{align}
%%%%%%% 
where the second term in the $T_2$ is the one-loop effect with the $I_{2}$ dependence.
One-loop effect in the $T_{1}$ could be neglected because such a term becomes small by using the condition of the height of potential in Eq.~(\ref{eq:con2}): $v_2^2<m_hv/\sqrt{2}$.
By using these terms $T_{1}$ and $T_2$, the ratio of these terms is
 %%%%%%%
	\begin{align}
	\label{TTIso}
	\frac{T_2^2}{T_{1}^2} \sim  \frac{2\lambda_2(m_{H}^2 + \lambda_2 v_2^2 )v_2^2\left(1 + \frac{3I_{2} \lambda_2}{\pi^2}\left(  \ln \frac{T_2^2\alpha_B}{Q^2}  - \frac{3}{2}\right)\right)}{ 3I_{2}m_W^2 m_h^2}.
	\end{align}
%%%%%%%
Becaused of this ratio, the denominator has the isospin effect, however, the $I_{2}$ dependence in this ratio disappears when the last term in the numerator is large. 
Then the value of $\lambda_2$ is an important source to realize the two-step phase transition.
For the heavy $m_H$ case, the $T_2^2/T_1^2$ ratio is roughly given by 
 %%%%%%%
	\begin{align}
	\label{ratheavymH}
	\frac{T_2^2}{T_{1}^2} \sim  \frac{ \lambda_2 v_2^2 (1+2I_{2}) + \frac{3m_h^2}{2\pi^2}\left(  \ln \frac{T_2^2\alpha_B}{Q^2}  - \frac{3}{2}\right) }{m_h^2}.
	\end{align}
%%%%%%%
In this case, the $I_2$ is important to realize the two-step phase transition.

\begin{figure}[t]
  \begin{center}
\includegraphics[width=0.4\textwidth]{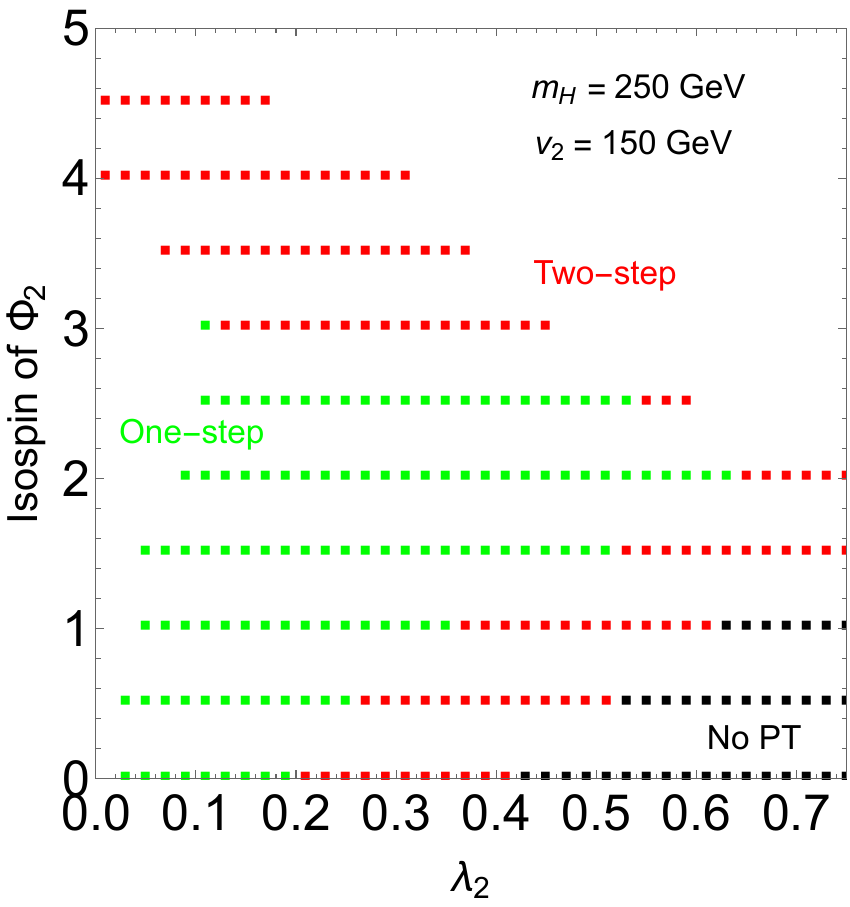}
\includegraphics[width=0.4\textwidth]{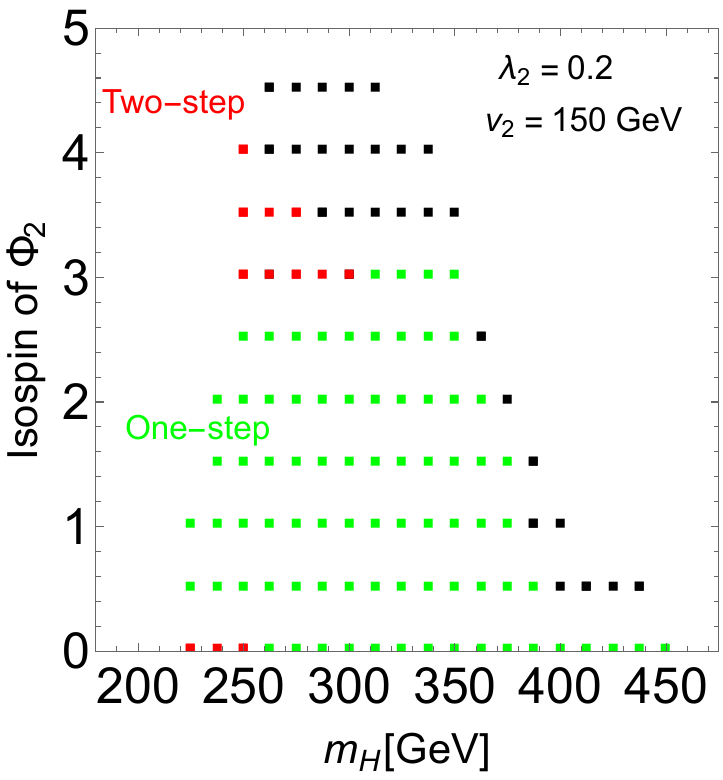}
\caption{The isospin dependence of the EWPT patterns.
The left figure shows the dependence for $\lambda_2$ with $m_H$ = 250 GeV and $v_2$ = 150 GeV.
The right figure shows the dependence for $m_H$ with $\lambda_2$ = 0.2 and $v_2$ = 150 GeV.
Other input parameters are the same as Fig.~\ref{resgen}. Note that high order $SU(2)_L$ multiplets receives constraints from perturbative unitarity, which yields $I_2 \leq 7/2$ for complex scalars and $I_2 \leq 4$ for real scalars \cite{Hally:2012pu}.}
\label{SNfigISO}
  \end{center}
\end{figure}

The numerical results of the isospin dependence are given by Fig.~\ref{SNfigISO}.
%%%%%%%%%%%%%%%%%%%%%%%%%%%%%%%%%%%%
%%%%%%%%%%%%%%%%%%%%%%%%%%%%%%%%%%%%
%%%%%%%%%%%%%%%%%%%%%%%%%%%%%%%%%%%%
%%%%%%%%%%%%%%%%%%%%%%%%%%%%%%%%%%%%
The horizontal axis in the left (right) figure is $\lambda_2$ ($m_H$), and the vertical axes are $I_{2}$.
In the black squares, the EWPT is multi-step and has not been completed in the current Universe.
According to these figures, the multi-step EWPT could be generated by the large $\lambda_2$.
Furthermore, the small or large $I_{2}$ value is the source of the multi-step EWPT.
These results could follow the above discussion. 
From that, we can find that the EWPT can assure the $\mathbb{Z}_2$ symmetry after the spontaneous EWSB in the model with $I_{2}<5$.

\subsection{The collider constraints}

%%%%%%%%%%%%%%%%%%%%%%%%%%%%%%%%%%%%
%%%%%%%%%%%%%%%%%%%%%%%%%%%%%%%%%%%%
\begin{figure}[t]
  \begin{center}
\includegraphics[width=0.4\textwidth]{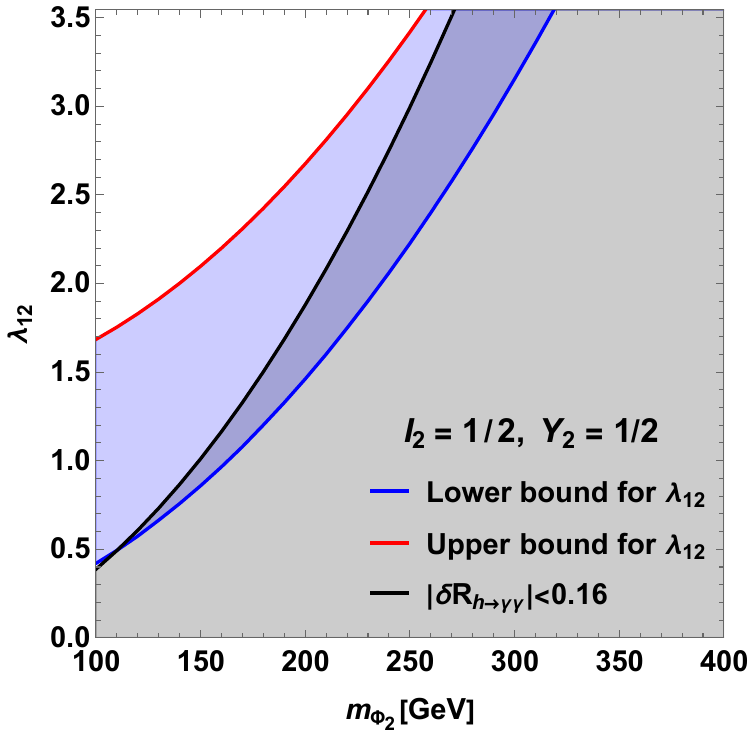}
\includegraphics[width=0.4\textwidth]{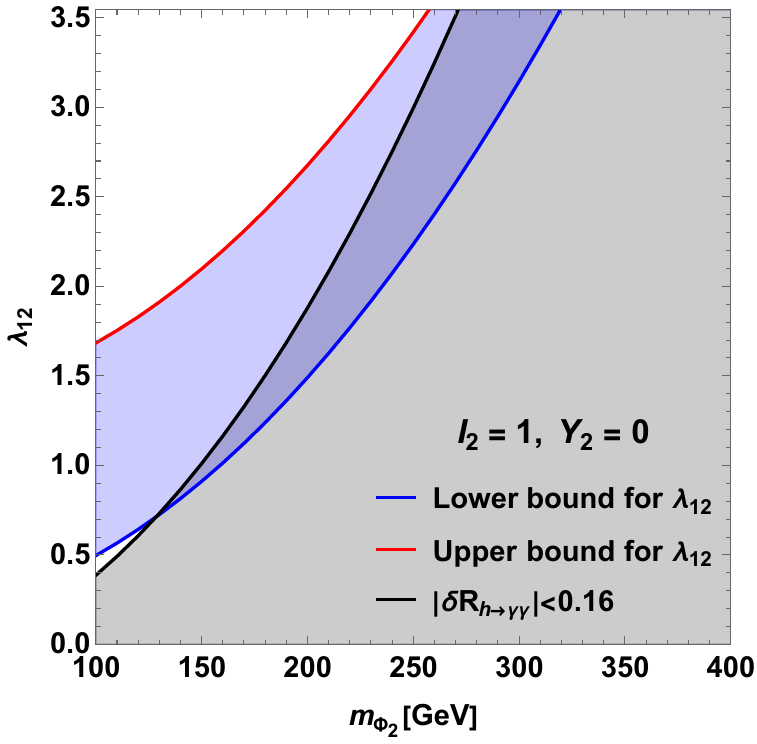}
\caption{The condition of two-step phase transition and $h\gamma\gamma$ measurement in the plane of ($m_{\Phi_2}$, $\lambda_{12}$) in the model with ($I_{2}$, $Y_{2}$) = ($1/2$, $1/2$) and ($1$, $0$).
The condition of a two-step phase transition can be realized in the blue region.
The black region is allowed by the $h\gamma\gamma$ measurements.
 }
\label{Fig:hphph}
  \end{center}
\end{figure}
%%%%%%%%%%%%%%%%%%%%%%%%%%%%%%%%%%%%
%%%%%%%%%%%%%%%%%%%%%%%%%%%%%%%%%%%%

Although the multiplet carries the $\mathbb{Z}_2$ symmetry, which forbids the mixing between the neutral scalar component and the SM Higgs boson, there are still constraints on the model parameters. At the LHC, the Drell-Yan process produces a pair of the charged components of the multiple, and the charged scalar subsequently decays into the SM weak gauge bosons and the neutral scalar. Thus, the charged scalars have been tightly constrained except that the charged scalars and neutral scalars have degenerate masses~\cite{ATLAS:2015eiz}. So in the following, we focus on the multiplet in which all the components have degenerate masses.\footnote{In fact, even the masses are degenerate for each component, there are still mass splittings among components due to radiative correction from the electroweak loop~\cite{Cirelli:2005uq}. However, we neglect such small corrections.} Furthermore, we also assume that the neutral scalar only contributes to part of the relic abundance, so the constraints from the relic density could be avoided.

One unavoidable constraint from the loop correction is the constraints from the $h\to\gamma\gamma$ measurements at the LHC. 
The deviation from the SM prediction in the $h\to\gamma\gamma$ is given by 
%%%%%%%
	\begin{align}
	\label{deltahpp}
	\delta R_{h\to\gamma\gamma} = \frac{\Gamma_{h\to\gamma\gamma}^{\rm SM+BSM}-\Gamma_{h\to\gamma\gamma}^{\rm SM}}{\Gamma_{h\to\gamma\gamma}^{\rm SM}},
	\end{align}
%%%%%%%
where the $\Gamma_{h\to\gamma\gamma}^{\rm SM+BSM}$ is the decay rate of $h\to\gamma\gamma$ in this model, which is given by~\cite{Ramsey-Musolf:2021ldh}
%%%%%%%
 \begin{align}
 \label{decay}
 \Gamma_{h\to\gamma\gamma}^{\rm SM+BSM}= \frac{v^2\alpha_e^2m_h^3}{256\pi^3}\left|\frac{4}{3}A_{1/2}\left(\frac{4m_t^2}{m_h^2}\right)+A_{1}\left(\frac{4m_W^2}{m_h^2}\right)+\sum_{n=-I_{2}}^{I_{2}}\frac{\lambda_{12}v^2}{2m_{\Phi_2}^2}\left(n+Y_{2}\right)A_{1}\left(\frac{4m_{\Phi_2}}{m_h^2}\right)\right|^2
 \end{align}
%%%%%%%
where $\alpha_e=e^2/4\pi$the loop function $A_{1/2}$, $A_1$ and $A_0$ is defined as~\cite{Chen:2013vi}
%%%%%%%
\begin{align}
  A_1(\tau) &= 2 + 3 \tau + 3 \tau(2 -\tau) f(\tau),\\
  A_{\frac{1}{2}}(\tau)   &= -2 \tau \left\{ 1 +(1 -\tau) f(\tau)\right\},\\
  A_0(\tau) &=\tau \left(1 -\tau f(\tau)\right)
\end{align}
%%%%%%%
with
%%%%%%%
\begin{align}
  f(\tau) = \left\{  
  \begin{array}{ll} 
    \arcsin^2\sqrt{1/\tau} & \quad\tau\geq 1 \\
    -\frac{1}{4} 
     \left( \ln\frac{1+\sqrt{1-\tau}}{1-\sqrt{1-\tau}} - i \pi \right)^2    & \quad\tau <1 \\ 
    \end{array}
    \right. .
\end{align}
%%%%%%%
The current bound for the effective coupling of the Higgs boson to the photon $\kappa_\gamma$ is $1.02^{+0.16}_{-0.14}$ at 95$\%$ confidence level~\cite{ATLAS:2022tnm}.
Fig.~\ref{Fig:hphph} represents the condition of two-step phase transition and the $h\gamma\gamma$ observation in the plane of ($m_{\Phi_2}$, $\lambda_{12}$).
$m_{\Phi_2}$ is the degenerate mass of additional charged-scalar bosons.
The left and right panels are the results in the model with $(I_{2}, Y_{2})=(1/2,1/2)$ and $(1,0)$, respectively.
The blue region can satisfy the conditions of a two-step phase transition.
The black region is allowed region with $|\delta R_{h\to\gamma\gamma}|<0.16$.
Because of this figure, the two-step phase transition can be realized in the allowed region for $h\gamma\gamma$ observation.
The numerical results in Figs.~\ref{resgen} and \ref{SNfigISO} can be avoided the current bound of $h\to\gamma\gamma$ observation.

%%%%%%%%%%%%%%%%%%%%%%%%%%%%%%%%%%%%%%%%%%%%%%%%%%%%%%%%%%%%
%%%%%%%%%%%%%%%%%%%%%%%%%%%%%%%%%%%%%%%%%%%%%%%%%%%%%%%%%%%%
\subsection{First-order phase transition and GW signatures}
%%%%%%%%%%%%%%%%%%%%%%%%%%%%%%%%%%%%%%%%%%%%%%%%%%%%%%%%%%%%
%%%%%%%%%%%%%%%%%%%%%%%%%%%%%%%%%%%%%%%%%%%%%%%%%%%%%%%%%%%%

\begin{figure}[t]
\begin{center}
\includegraphics[width=0.4\textwidth]{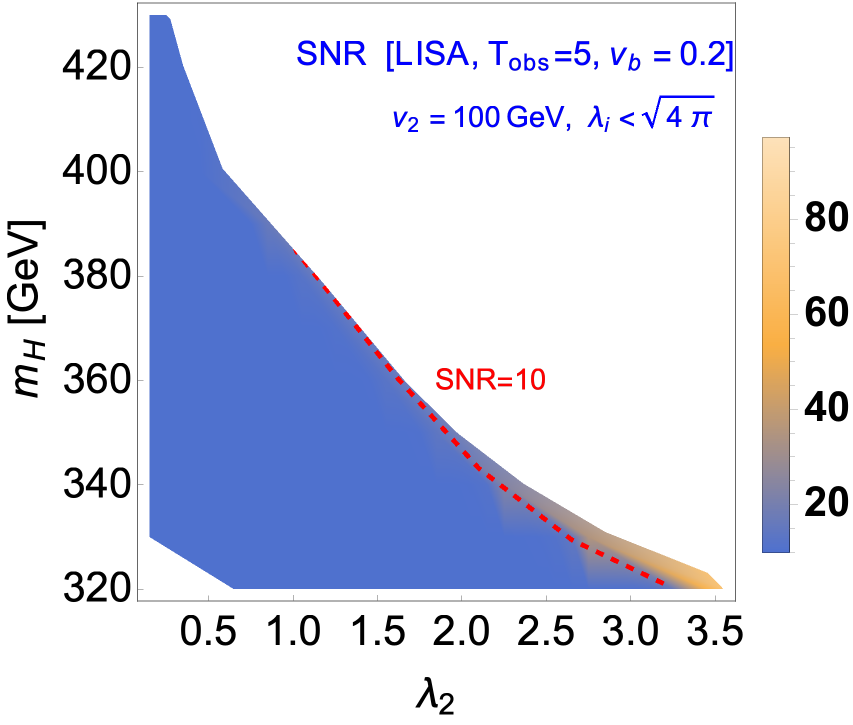}
\includegraphics[width=0.42\textwidth]{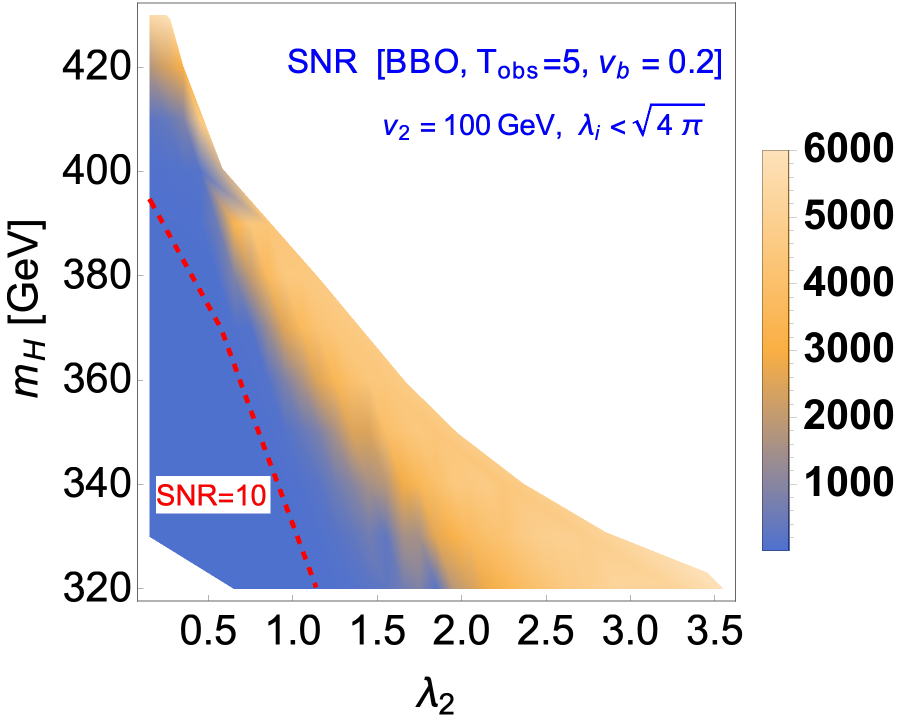}\\
\includegraphics[width=0.4\textwidth]{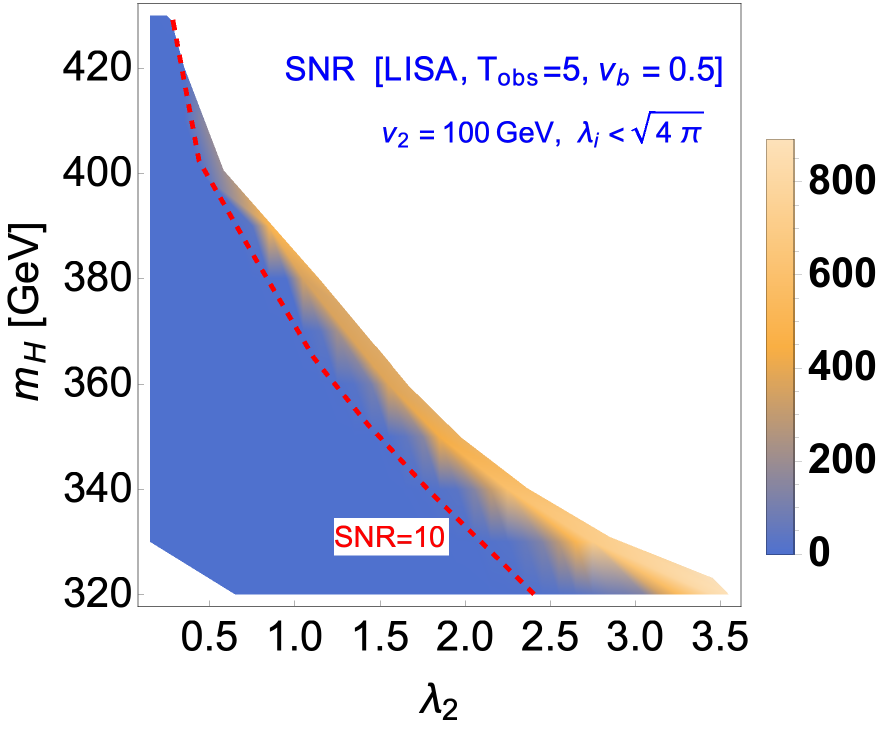}
\includegraphics[width=0.42\textwidth]{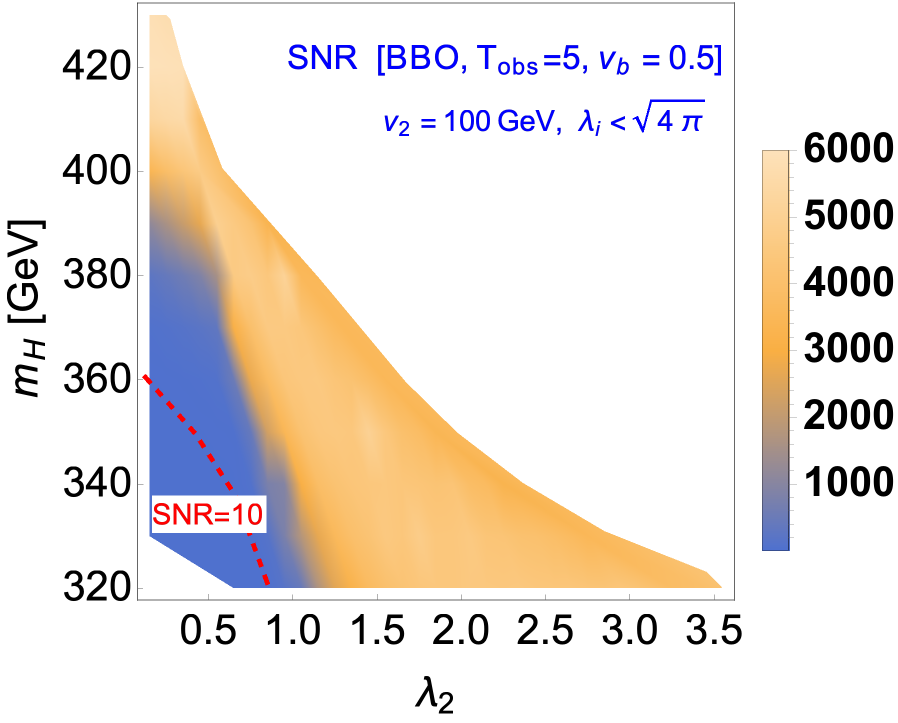}
\caption{The density plots of $S/N$ ratio at LISA (left two figures) and BBO (right two figures) experiments.
The experimental period is chosen as 5 years.
The top (bottom) two figures have $v_b$ = 0.2 (0.5), respectively. 
The red dashed lines in these figures represent $S/N=10$.
 }
\label{SNfig}
  \end{center}
\end{figure}

In this part, we will show the testability of the parameter region of the left panel of Fig.~\ref{resgen}. 
Fig.~\ref{SNfig} represents the results of the SNR, which is given by Eq.~(\ref{SNRGW}), of the green and red squares in the left panel of Fig.~\ref{resgen}.
In Fig.~\ref{SNfig}, the left (right) two figures represent the SNR for LISA~\cite{Klein:2015hvg} (BBO~\cite{Yagi:2011wg}).
The experimental period for the SNR is chosen as 5 years.
The red dashed line in these figures represents $\text{SNR}=10$.
The parameter regions above the red dashed line can realize $\text{SNR}>10$.
Therefore, such a parameter region can be tested by each experiment. 
%%%%%%%%%%%%%%%%%%%%%%%%%%%%%%%%%%%%
%%%%%%%%%%%%%%%%%%%%%%%%%%%%%%%%%%%%
%%%%%%%%%%%%%%%%%%%%%%%%%%%%%%%%%%%%
%%%%%%%%%%%%%%%%%%%%%%%%%%%%%%%%%%%%
The top (bottom) two figures have $v_b$ = 0.2 (0.5), respectively. 
The low speed of bubble wall velocity is required to generate the baryon asymmetry of the Universe~\cite{Joyce:1994fu}. 
The upper right parts of the white regions in these figures correspond to the black squares in Fig.~\ref{resgen}, which represent that the EWPT has not been completed in the current universe.
The parameter region close to the upper right white region can be tested at LISA.
On the other hand, the BBO can test the almost parameter region realizing, especially the multi-step EWPT in Fig.~\ref{RM}-(b).
Therefore, we can use the GW observation experiments to explore the possibility of the multi-step EWPT to restore the $\mathbb{Z}_2$ symmetry after the spontaneous EWSB.

The first-order EWPT in Fig.~\ref{RM}-(a) can be realized by the negative $\mu_2^2$ case. However, the first-order phase transition is not strong enough to generate the detectable GW spectrum because the $\lambda_{12}v^2$ is typically smaller for negative $\mu_2^2$ than for positive $\mu_2^2$ to determine the $m_H$.

%%%%%%%%%%%%%%%%%%%%%%%%%%%%%%%%%%%%%%%%%%%%%%%%%%%%%%%%%%%%
%%%%%%%%%%%%%%%%%%%%%%%%%%%%%%%%%%%%%%%%%%%%%%%%%%%%%%%%%%%%
\subsection{Possibility of two first-order EWPTs}
\label{sec:pos_two_ewpt}
%%%%%%%%%%%%%%%%%%%%%%%%%%%%%%%%%%%%%%%%%%%%%%%%%%%%%%%%%%%%
%%%%%%%%%%%%%%%%%%%%%%%%%%%%%%%%%%%%%%%%%%%%%%%%%%%%%%%%%%%%

\begin{figure}[t]
  \begin{center}
\includegraphics[width=0.4\textwidth]{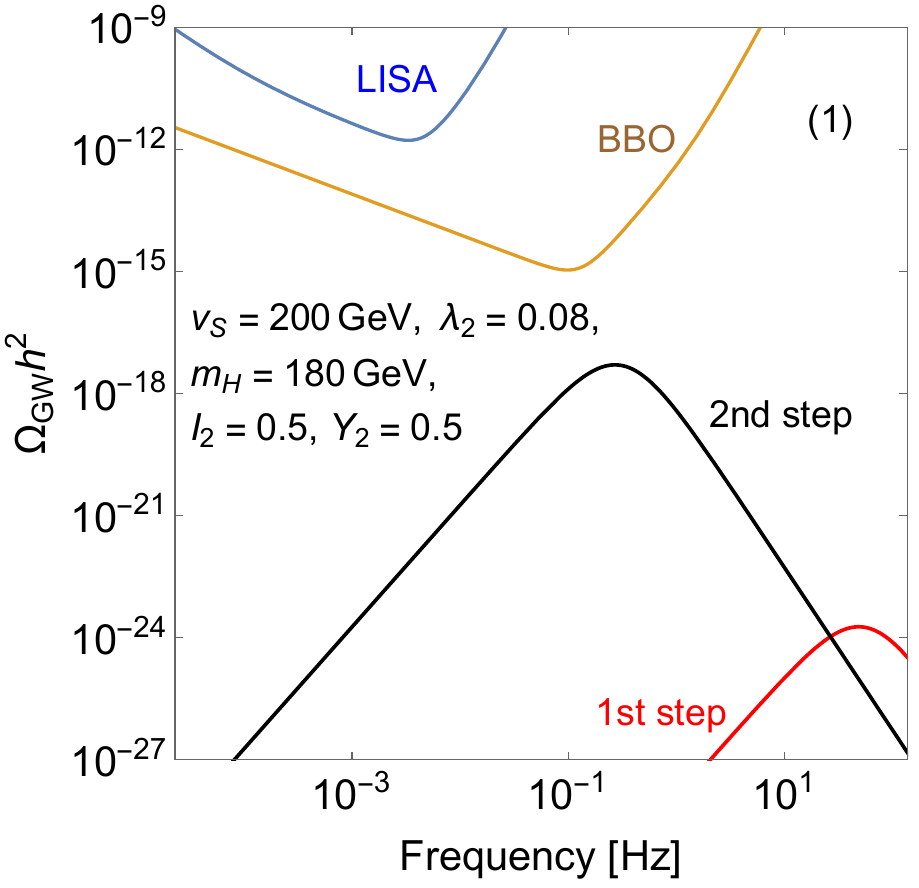}
\includegraphics[width=0.4\textwidth]{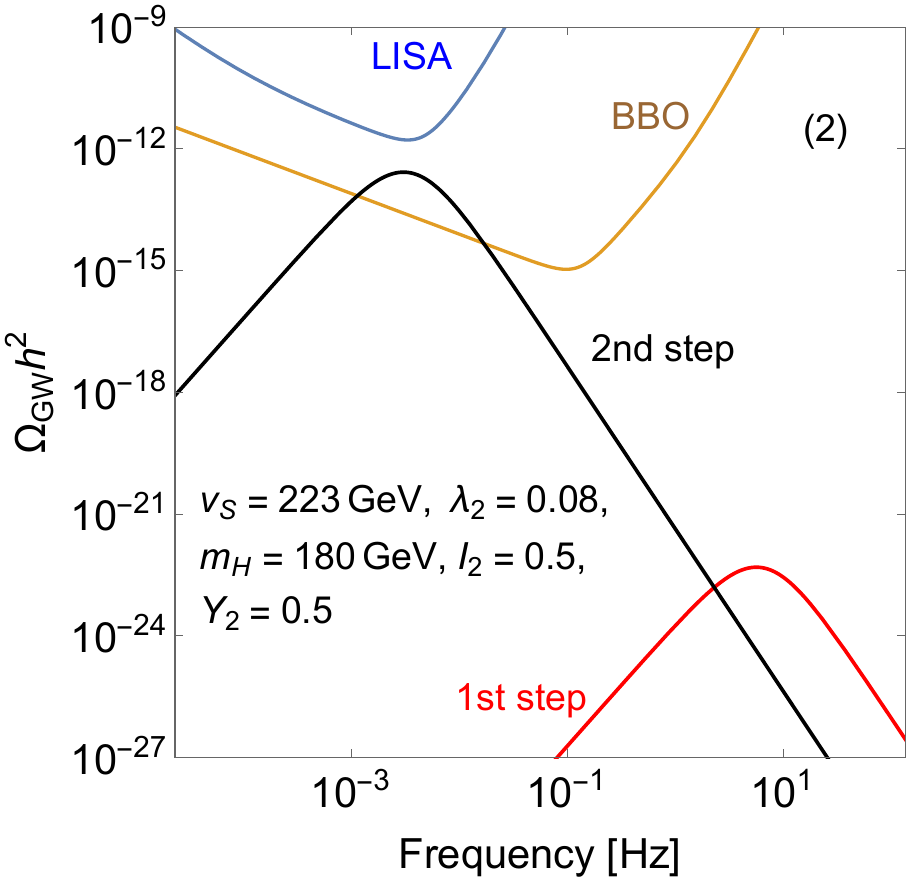}
\caption{Examples of two peaks of GW spectrum from compression wave of plasma with $v_b=0.5$ and $(m_H,\lambda_{2}, I_{2}, Y_{2})$ = (180 GeV, 0.08, 1/2, 1/2).
(1) Left figure is two peaks of GW spectra at the red star point in the right panel of Fig.~\ref{resgen}, which has $v_2$ = 200 GeV.
 (2) Right figure is the spectra about $v_2$ = 223 GeV.
The red (black) GW spectrum comes from the first-step (second-step) EWPT in Fig.~\ref{RM}-(b).
The blue and yellow lines are the sensitivity lines at LISA and BBO, respectively.
 }
\label{twopeakGW}
  \end{center}
\end{figure}

For the red star point in Fig.~\ref{resgen} with $v_2=200~{\rm GeV}$, 
the two peaks of GW spectra could be produced by the two first-order phase transitions in (green $\to$ magenta) and (magenta $\to$ red) paths,  depicted in Fig.~\ref{fig:polar}. 
 Parameters at the red star point are $(v_2,~ m_H,~ \lambda_{2},~ I_{2},~ Y_{2}) = (200~\mathrm{GeV},~ 180~\mathrm{GeV},~ 0.08,~ 1/2,~ 1/2)$.
 The GW spectra of this parameter are shown by Fig.~\ref{twopeakGW}-(a).
 The red (black) lines are the GW spectrum from the first-step (second-step) EWPT in Fig.~\ref{RM}-(b).
 The phase transition parameters are $(T_t,~ \alpha,~ \tilde{\beta},~ v_b)= (149~\mathrm{GeV},~ 3.79\times 10^{-4},~ 8.37\times 10^5,~ 0.5)$ (first step) and $(142~\mathrm{GeV},~ 4.70\times 10^{-3},~ 4.91\times 10^3,~ 0.5)$ (second step).
 The blue and yellow lines correspond to the sensitivity of the LISA and BBO experiments, respectively.
 The SNRs for these spectra are given as SNR (LISA) $\sim 10^{-16}$, SNR (BBO) $\sim 10^{-7}$ (first step) and SNR (LISA) $\sim 10^{-7}$, SNR (BBO) $\sim 0.565$ (second step).
 Therefore, it could not be easy to detect these GW spectra in Fig.~\ref{twopeakGW}-(a).
 On the other hand, we take the second benchmark point with larger $v_2$ than the star point: $v_2=223$ GeV.
%%%%%%%%%%%%%%%%%%%%%%%%%%%%%%%%%%%%
%%%%%%%%%%%%%%%%%%%%%%%%%%%%%%%%%%%%
%%%%%%%%%%%%%%%%%%%%%%%%%%%%%%%%%%%%
%%%%%%%%%%%%%%%%%%%%%%%%%%%%%%%%%%%%
The GW for the second benchmark point is described in Fig.~\ref{twopeakGW}-(b), and the phase transition parameters are $(T_t,~ \alpha,~ \tilde{\beta},~ v_b)= (153~\mathrm{GeV},~ 5.03\times 10^{-4},~ 9.58\times 10^4,~ 0.5)$ (first step) and $(109~\mathrm{GeV},~ 3.06\times 10^{-2},~ 72.2,~ 0.5)$ (second step). 
The SNRs for each path in Fig.~\ref{twopeakGW}-(b) are SNR (LISA) $\sim 10^{-16}$, SNR (BBO) $\sim 10^{-9}$ (first step) and SNR (LISA) $\sim$ 58.0, SNR (BBO) $\sim$ 238 (second step). 
Therefore, we could detect one peak of GW spectrum at least. However, these experiments are difficult to observe both GW spectra from the two first-order EWPTs in Fig.~\ref{RM}-(b). 
The reasons why the GW spectrum from the first step is so small are (i) first-step EWPT does not have nonthermal contributing to cubic term $h_2^3$, and (ii) $\lambda_2$ could not be small enough to generate the strongly first-order EWPT.
The detectable GW spectrum could require a large value of $v_C/T_C$; in other words, it could require a small value of $\lambda_2$.
However, the condition of $T_2>T_{1}$ in Eq.~(\ref{eq:con3}) cannot be satisfied by such a small value of $\lambda_2$.
In the case of large $v_2$ value, multi-step phase transitions are realized even for small values of $\lambda_2$, and two first-order EWPTs may be realized.
But the condition of the height of the potential between the magenta and red points in Fig.~\ref{Vtree} becomes strict by the large $v_2$: $ \lambda_2 v_2^4 < m_h^2v^2/2$.
 Therefore, the parameter region for producing the detectable two peaks of the GW spectrum is very narrow.

Furthermore, we study the supercooling phase transition ($\alpha\gg1$). 
The benchmark point of the supercooling phase transition is $(v_2, m_H,\lambda_{2}, I_{2}, Y_{2})$ = (225 GeV, 168.64 GeV, 0.08, 1/2, 1/2).
Then the percolation temperature of the second path is $T_p\sim 4$ GeV, and $\alpha\sim3880$. 
However, even in the supercooling case, the first step of the EWPT is not strong.
It is difficult to detect both peaks of the GW spectra coming from both steps of the EWPT in Fig.~\ref{RM}-(b); however, at least one peak of the spectra could be detected.
Suppose the multi-step EWPT is unable to restore the $\mathbb{Z}_2$ symmetry after the spontaneous EWSB, e.g., Fig.~\ref{RM}-(e) phase transition. In that case, we could detect both peaks of GW spectra produced by two first-order EWPTs~\cite{Aoki:2021oez}, because the condition of the height of the potential is not necessary to realize the multi-step EWPT.

%%%%%%%%%%%%%%%%%%%%%%%%%%%%%%%%%%%%%%%%%%%%%%%%%%%%%%%%%%%%
%%%%%%%%%%%%%%%%%%%%%%%%%%%%%%%%%%%%%%%%%%%%%%%%%%%%%%%%%%%%
%%%%%%%%%%%%%%%%%%%%%%%%%%%%%%%%%%%%%%%%%%%%%%%%%%%%%%%%%%%%
\section{Summary}
\label{sec:summary}
%%%%%%%%%%%%%%%%%%%%%%%%%%%%%%%%%%%%%%%%%%%%%%%%%%%%%%%%%%%%
%%%%%%%%%%%%%%%%%%%%%%%%%%%%%%%%%%%%%%%%%%%%%%%%%%%%%%%%%%%%
%%%%%%%%%%%%%%%%%%%%%%%%%%%%%%%%%%%%%%%%%%%%%%%%%%%%%%%%%%%%

We have investigated that the two-step phase transition can be naturally realized in the scalar potential with the discrete $\mathbb{Z}_2$ symmetry due to degenerate extreme along field directions. To lay out the general conditions on such a two-step phase transition, we consider the most general electroweak $N$-plet scalar extension. For the first time, we provide the general form of the Debye masses and obtain the two-step phase transition conditions for general isospin and hypercharge scalars. We found that there is a strong correlation between the coupling $\lambda_{12}$ and the heavy scalar mass, and the coupling $\lambda_{12}$ needs to be in a moderate region because of the global minimum and $T_2 > T_1$ conditions. As the heavy scalar mass increase, the coupling $\lambda_{12}$ is required to take a larger value, and similarly for larger isospin multiplets.

Although both steps of the EWPT could be first order, it is difficult to detect two peaks of GW spectra produced from these first-order EWPTs.
The first-order EWPT along the first step can be realized by thermal loop effects, and then typically, the GW spectrum frequency is ${\cal O} (1)$ Hz. The detectable GW spectrum from this step can be produced if there is a small $\lambda_2$ or large $v_2$ value. Since a barrier in the second step of the EWPT occurs by the tree-level effects, the detectable GW could be generated through the second step, of which the frequency is less than about $10^{-1}$ Hz. Therefore, we may distinguish different patterns of the EWPT by the GW observation.

Even in the case that the multi-step EWPT is able to restore the discrete symmetry, at least one peak of the spectrum could be detected by future GW observation experiments. According to the numerical results, the second step produces the GW spectrum that could be detected in the BBO experiment. Furthermore, it is possible to explore such parameter region of the moderately large $\lambda_{12}$ by the Higgs diphoton data in future hadron and lepton colliders. From these experiments, we would explore the possibility of multi-step phase transition comprehensively.

%====================================
\acknowledgments

This work is supported in part by the National Key Research and Development Program of China Grant No. 2020YFC2201501.  Q.H.C. is supported in part by the National Science Foundation of China under Grant No. 11675002, No. 11635001, No. 11725520, and No. 12235001. J.H.Y. is supported in part by the National Science Foundation of China under Grants No. 12022514, No. 11875003, and No. 12047503, and CAS Project for Young Scientists in Basic Research YSBR-006, and the Key Research Program of the CAS Grant No. XDPB15. K.H. is supported in part by Grant-in-Aid for Research Activity Start-up 23K19052. 

%%%%%%%%%%%%%%%%%%%%%%%%%%%%%%%%%%%%%%%%%%%%%%%%%%%%%%%%%%%%
%%%%%%%%%%%%%%%%%%%%%%%%%%%%%%%%%%%%%%%%%%%%%%%%%%%%%%%%%%%%
%%%%%%%%%%%%%%%%%%%%%%%%%%%%%%%%%%%%%%%%%%%%%%%%%%%%%%%%%%%%
\appendix
%%%%%%%%%%%%%%%%%%%%%%%%%%%%%%%%%%%%%%%%%%%%%%%%%%%%%%%%%%%%
%%%%%%%%%%%%%%%%%%%%%%%%%%%%%%%%%%%%%%%%%%%%%%%%%%%%%%%%%%%% %%%%%%%%%%%%%%%%%%%%%%%%%%%%%%%%%%%%%%%%%%%%%%%%%%%%%%%%%%%%
\section{Field-dependent masses in several multiplets}
%%%%%%%%%%%%%%%%%%%%%%%%%%%%%%%%%%%%%%%%%%%%%%%%%%%%%%%%%%%%
%%%%%%%%%%%%%%%%%%%%%%%%%%%%%%%%%%%%%%%%%%%%%%%%%%%%%%%%%%%%
%%%%%%%%%%%%%%%%%%%%%%%%%%%%%%%%%%%%%%%%%%%%%%%%%%%%%%%%%%%%
\label{sec:appmass}

The field-dependent mass (matrix) can be calculated by taking the second derivatives of the potential as
\begin{align}
    M_{ij}^2(h_1, h_2) = \frac{\partial^2 V_0}{\partial \phi_i \partial \phi_j} \Bigg|_{h_1, h_2},
\end{align}
where $\phi$ can be scalars, fermions, and gauge bosons. We list the mass matrix squared under various representations $(I_2, Y_2)$ as follows:
\paragraph{Notation}
\begin{itemize}
    \item NS, CS, and NCPC are abbreviations of neutral scalar, charged scalar, and neutral $CP$ conserving scalar. By default, $n_{\text{NS}} = n_{\text{NCPC}} = 1,~ n_{\text{CS}} = 2$.
    \item $\chi$ is components of $\Phi_1$ except $h_1$. By default, $n_\chi = 3$.
    \item $c^+ = 2\operatorname{Re}(c) = c + c^*,~ c^- = 2\operatorname{Im}(c) = -i(c - c^*)$ for parameter c.
\end{itemize}

\paragraph{Real singlet} $(I_2, Y_2) = (0, 0)$:
\begin{align}
    M_{\rm{NS}}^2 &= \left(\begin{array}{cc}
        3\lambda_{1}h_1^2 + a_0h_2^2 - \mu_1^2 & 2a_0h_1h_2 \\
        2a_0h_1h_2 & 12b_{0}h_2^2 + a_0h_1^2 - \mu_2^2
    \end{array}\right), \nonumber \\
    M_{\chi}^2 &= \lambda_{1}h_1^2 + a_0h_2^2 - \mu_1^2.
\end{align}

\paragraph{Complex singlet} $(I_2, Y_2) = (0, 0)$:
\begin{align}
    M_{\rm{NS}}^2 &= \left(\begin{array}{ccc}
        (M_{\rm{NS}}^{11})^2 & (a_0+a_1'^+)h_1h_2 & -a_0'^-h_1h_2 \\
        (a_0+a_1'^+)h_1h_2 & (M_{\rm{NS}}^{22})^2 & -\frac{1}{2}a_0'^-h_1^2 - 3\left(b_0'^- + \frac{1}{2}b_0''^-\right)h_2^2 \\
        -a_0'^-h_1h_2  & -\frac{1}{2}a_0'^-h_1^2 - 3\left(b_0'^- + \frac{1}{2}b_0''^-\right)h_2^2 & (M_{\rm{NS}}^{aa}))^2
    \end{array}\right), \nonumber \\
    M_{\chi}^2 &= \lambda_{1}h_1^2 + \frac{1}{2}a_0h_2^2 - \mu_1^2,
\end{align}
where
\begin{align}
    (M_{\rm{NS}}^{11})^2 & = 3\lambda_{1}h_1^2 + \frac{1}{2}(a_0+a_1'^+)h_2^2 - \mu_1^2, \nonumber \\
    (M_{\rm{NS}}^{22})^2 & = 3(b_{0}+b_0'^++b_0''^+)h_2^2 + \frac{1}{2}(a_0+a_1'^+)h_1^2 - \mu_2^2, \nonumber \\
    (M_{\rm{NS}}^{aa})^2 & = (b_0 - 3b_0'^+)h_2^2 + \frac{1}{2}(a_0 - a_0^+)h_1^2 - \mu_2^2.
\end{align}

\paragraph{Doublet}  $(I_2, Y_2) = (1/2, 1/2)$:
\begin{align}
    M_{\rm{NS}}^2 &= \left(\begin{array}{cccc}
        (M_{\rm{NS}}^{11})^2 & (a_0 + a_1'^+)h_1h_2 & -\frac{1}{2}a_1'^-h_2^2 & a_1'^-h_1h_2 \\
        (a_0 + a_1'^+)h_1h_2 & (M_{\rm{NS}}^{22})^2 & -a_1'^-h_1h_2 & \frac{1}{2}a_1'^-h_1^2 \\
        -\frac{1}{2}a_1^-h_2^2 & -a_1^-h_1h_2 & (M_{\rm{NS}}^{\phi_0\phi_0})^2 & a_1'^+h_1h_2 \\
        a_1'^-h_1h_2 & \frac{1}{2}a_1'^-h_1^2 & a_1'^+h_1h_2 & (M_{\rm{NS}}^{aa})^2
    \end{array}\right), \nonumber \\
    M_{\rm{CS}}^2 &= \left(\begin{array}{cc}
        \lambda_{1}h_1^2 + \frac{1}{2}(a_0 + a_1)h_2^2 - \mu_1^2 & -\frac{1}{2}(a_1 - 2a_1')h_1h_2 \nonumber \\
        -\frac{1}{2}(a_1 - 2a_1'^*)h_1h_2 & b_{0}h_2^2 + \frac{1}{2}(a_0 + a_1)h_1^2 - \mu_2^2
    \end{array}\right),
\end{align}
where 
\begin{align}
    (M_{\rm{NS}}^{11})^2 &= 3\lambda_{1}h_1^2 + \frac{1}{2}(a_0 + a_1'^+)h_2^2 - \mu_1^2, \nonumber\\
    (M_{\rm{NS}}^{22})^2 &= 3b_{0}h_2^2 + \frac{1}{2}(a_0 + a_1'^+)h_1^2 - \mu_2^2, \nonumber\\
    (M_{\rm{NS}}^{\phi_0\phi_0})^2 &= \lambda_{1}h_1^2 + \frac{1}{2}(a_0 - a_1'^+)h_2^2 - \mu_1^2, \nonumber\\
    (M_{\rm{NS}}^{aa})^2 &= b_{0}h_2^2 + \frac{1}{2}(a_0 - a_1'^+)h_1^2 - \mu_2^2.
\end{align}

\paragraph{Real triplet} $(I_2, Y_2) = (1, 0)$:
\begin{align}
    M_{\rm{NCPC}}^2 &= \left(\begin{array}{cc}
        3\lambda_{1}h_1^2 - a_0h_2^2 - \mu_1^2 & -2a_0h_1h_2 \\
        -2a_0h_1h_2 & 12b_{0}h_2^2 - a_0h_1^2 + \mu_2^2
    \end{array}\right), \nonumber \\
    M_{\chi}^2 &= \lambda_{1}h_1^2 - a_0h_2^2 - \mu_1^2, \nonumber \\
    M_{a}^2 &= 4b_{0}h_2^2 - a_0h_1^2 + \mu_2^2,~ n_{a} = 2.
\end{align}

\paragraph{Complex triplet} $(I_2, Y_2) = (1, 0)$:
\begin{align}
    M_{\rm{NS}}^2 &= \left(\begin{array}{ccc}
        (M_{\rm{NS}}^{11})^2 & \frac{1}{2}(2a_0 + a_1 - 2a_0'^+)h_1h_2 & a_0'^-h_1h_2 \\
        \frac{1}{2}(2a_0 + a_1 - 2a_0'^+)h_1h_2 & (M_{\rm{NS}}^{22})^2 & \frac{1}{2}a_0'^-h_1^2 - 3(b_0'^- - b_0''^-)h_2^2 \\
        a_0'^-h_1h_2 & \frac{1}{2}a_0'^-h_1^2 - 3(b_0'^- - b_0''^-)h_2^2 & (M_{\rm{NS}}^{aa})^2
    \end{array}\right), \nonumber \\
    M_{\rm{CS}}^2 &= \left(\begin{array}{ccc}
        (M_{\rm{CS}}^{G^\pm})^2 & -\frac{1}{2\sqrt{2}}a_1h_1h_2 & -\frac{1}{2\sqrt{2}}a_1h_1h_2 \\
        -\frac{1}{2\sqrt{2}}a_1h_1h_2 & (M_{\rm{CS}}^{s^+})^2 & a_0'h_1^2 - (b_{1} + 2b_{0}' - b_{0}'')h_2^2\\
        -\frac{1}{2\sqrt{2}}a_1h_1h_2 & a_0'h_1^2 - (b_{1} + 2b_{0}' - b_{0}'')h_2^2 & (M_{\rm{CS}}^{s^-})^2
    \end{array}\right), \nonumber \\
    M_{G^0}^2 &= \lambda_1h_1^2 + \frac{1}{4}(2a_0 + a_1 - 2a_1'^+)h_2^2 - \mu_1^2,~ n_{G^0} = 1,
\end{align}
where
\begin{align}
    (M_{\rm{NS}}^{11})^2 &= 3\lambda_{1}h_1^2 + \frac{1}{4}(2a_0 + a_1 - 2a_0'^+)h_2^2 - \mu_1^2, \nonumber\\
    (M_{\rm{NS}}^{22})^2 &= 3(b_{0} + b_{1} + b_0'^+ - b_0''^+)h_2^2 + \frac{1}{4}(2a_0 + a_1 - 2a_0'^+)h_1^2 - \mu_2^2, \nonumber\\
    (M_{\rm{NS}}^{aa})^2 &= (b_{0} + b_{1} - 3b_0'^+)h_2^2 + \frac{1}{4}(2a_0 + a_1 + 2a_0'^+)h_1^2 - \mu_2^2, \nonumber\\
    (M_{\rm{NS}}^{G^\pm})^2 &= M_{G^0}, \nonumber\\
    (M_{\rm{NS}}^{s^+})^2 &= \frac{1}{2}(2b_{0} - b_0''^+)h_2^2 + \frac{1}{2}(a_0 + a_1)h_1^2 - \mu_2^2, \nonumber\\
    (M_{\rm{NS}}^{s^-})^2 &= \frac{1}{2}(2b_{0} - b_0''^+)h_2^2 + \frac{1}{2}a_0h_1^2 - \mu_2^2.
\end{align}

\paragraph{Complex triplet} $(I_2, Y_2) = (1, 1)$:
\begin{align}
    M_{\rm{NCPC}}^2 &= \left(\begin{array}{cc}
        3\lambda_{1}h_1^2 + \frac{1}{2}a_0h_2^2 - \mu_1^2 & a_0h_1h_2 \\
        a_0h_1h_2 & 3b_{0}h_2^2 + \frac{1}{2}a_0h_1^2 - \mu_2^2
    \end{array}\right), \nonumber \\
    M_{G^0}^2 &= \lambda_{1}h_1^2 + \frac{1}{2}a_0h_2^2 - \mu_1^2,~ n_{G^0} = 1, \nonumber \\
    M_{a}^2 &= b_{0}h_2^2 + \frac{1}{2}a_0h_1^2 - \mu_2^2,~ n_{a} = 1, \nonumber \\
    M_{\rm{CS}}^2 &= \left(\begin{array}{cc}
        \lambda_{1}h_1^2 + \frac{1}{2}(a_0 + a_1)h_2^2 - \mu_1^2 & -\frac{1}{2\sqrt{2}}a_0h_1h_2 \nonumber \\
        -\frac{1}{2\sqrt{2}}a_0h_1h_2 & b_{0}h_2^2 + \frac{1}{4}(2a_0 + a_1)h_1^2 - \mu_2^2
    \end{array}\right), \nonumber \\
    M_{{s^{\pm\pm}}}^2 &= (b_{0} + 2b_{1})h_2^2 + \frac{1}{2}(a_0 + a_1)h_1^2 - \mu_2^2,~ n_{s^{\pm\pm}} = 2.
\end{align}

 %%%%%%%%%%%%%%%%%%%%%%%%%%%%%%%%%%%%%%%%%%%%%%%%%%%%%%%%%%%%
 %%%%%%%%%%%%%%%%%%%%%%%%%%%%%%%%%%%%%%%%%%%%%%%%%%%%%%%%%%%%
 %%%%%%%%%%%%%%%%%%%%%%%%%%%%%%%%%%%%%%%%%%%%%%%%%%%%%%%%%%%%
 \section{Stationary conditions and neutral scalar masses }
  %%%%%%%%%%%%%%%%%%%%%%%%%%%%%%%%%%%%%%%%%%%%%%%%%%%%%%%%%%%%
 
 %%%%%%%%%%%%%%%%%%%%%%%%%%%%%%%%%%%%%%%%%%%%%%%%%%%%%%%%%%%%
 We modify the tree-level conditions in Eq.~\ref{eq:trcon} by using the effective potential with the loop correction at zero temperature in Eq.~(\ref{eq:effpot}), with the nondiagonal terms neglected.
 We take $Q=v=246$ GeV and the tadpole conditions are defined as follows:
%%%%%%%%%%%%%%%%%%%%%%%%%%%%%%%%%%%%
 \begin{align}
  \label{redc1}
\left.\frac{\partial V_{\rm eff}}{\partial  \left\langle \Phi_1 \right\rangle} \right|_{ \left\langle \Phi_1 \right\rangle=v,  \left\langle \Phi_2 \right\rangle=0} &=
	-\mu_1^2+\lambda_1 v^2 +\frac{1}{16\pi^2} \Bigg[ \frac{6\lambda_1+\lambda_{12}}{4} f_+(m_{\Phi_{1,r}}^2,m_{\Phi_{2,r}}^2) + \frac{1}{4}   (6\lambda_1-\lambda_{12}) f_-(m_{\Phi_{1,r}}^2,m_{\Phi_{2,r}}^2) \nonumber\\
%%%
	& \quad +(2(2I_{2}-1)-1) \frac{\lambda_{12} m_{N_{2,r}}^2}{2}\left(\ln\frac{m_{N_{2,r}}^2}{Q^2}-1\right)  + \frac{6m_W^4}{ v^2}\left(\ln\frac{m_W^2}{Q^2}-\frac{1}{3}\right) \nonumber\\
%%%
	&\quad + \frac{3m_Z^4}{v^2}\left(\ln\frac{m_Z^2}{Q^2}-\frac{1}{3}\right) - \frac{12m_t^4}{v^2}\left(\ln\frac{m_t^2}{Q^2}-1\right)   \Bigg]= 0, 
 \end{align}
%  %%%%%%%%%%%%%%%%%%%%%%%%%%%%%%%%%%%% 
 \begin{align}
  \label{redc2}
\left.\frac{\partial V_{\rm eff}}{\partial  \left\langle \Phi_2 \right\rangle} \right|_{ \left\langle \Phi_1 \right\rangle=0,  \left\langle \Phi_2 \right\rangle=v_2}  &=  -\mu_2^2+\lambda_2  v_2^2 +\frac{1}{16\pi^2} \Bigg[\frac{6\lambda_2+\lambda_{12}}{4} f_+\left((m_{\Phi_{1,b}})^2,(m_{\Phi_{2,b}})^2\right)) \nonumber\\
%%%
	&\quad +\frac{1}{4}(\lambda_{12} - 6\lambda_2) f_-\left((m_{\Phi_{1,b}})^2,(m_{\Phi_{2,b}})^2\right) +3\frac{\lambda_{12}}{2} (m_{N_{1,b}}^s)^2\left(\ln\frac{(m_{N_{1,b}})^2}{Q^2}-1\right) \nonumber\\
%%%
	& \quad  +(2(2I_{2}-1)-1)\lambda_{2} (m_{N_{2,b}})^2\left(\ln\frac{(m_{N_{2,b}})^2}{Q^2}-1\right)   \nonumber\\
%%%
	&\quad + \frac{3Y_{2}^4m_Z^4v_2^2}{v^4}\left(\ln\frac{Y_{2}^2v_2^2m_Z^2}{Q^2v^2}-\frac{1}{3}\right)  + \frac{6I_W^4m_W^4v_2^2}{ v^4}\left(\ln\frac{I_W^2m_W^2v_2^2}{v^2Q^2}-\frac{1}{3}\right)  \Bigg]= 0,
\end{align}
  %%%%%%%%%%%%%%%%%%%%%%%%%%%%%%%%%%%%
where
  %%%%%%%%%%%%%%%%%%%%%%%%%%%%%%%%%%%%
\begin{align}
f_\pm(m_i^2, m_j^2) &= m_i^2\left(\ln\frac{m_i^2}{Q^2}-1\right)  \pm m_j^2\left(\ln\frac{m_j^2}{Q^2}-1\right), \\
\Delta m_\Phi^2 &= m_{\Phi_2}^2 - m_{\Phi_1}^2= \sqrt{\left({\cal M}_{\Phi_1\Phi_1}^2-{\cal M}_{\Phi_2\Phi_2}^2\right)^2+4{\cal M}_{\Phi_1\Phi_2}^4},\\
 \mathcal{M}_{h, H}^{2}&=
	\left(
	\begin{array}{ccc}
{\cal M}_{\Phi_1\Phi_1}^2  & {\cal M}_{\Phi_1\Phi_2}^2 \\ 
{\cal M}_{\Phi_2\Phi_1}^2 & {\cal M}_{\Phi_2\Phi_2}^2
	\end{array}
	\right),
\end{align}
%%%%%%%%%%%%%%%%%%%%%%%%%%%%%%%%%%%%
and 
%%%%%%%%%%%%%%%%%%%%%%%%%%%%%%%%%%%%
 \begin{align}
m_{\Phi_{1,r}}^2 &= -\mu_1^2 + 3\lambda_1 v^2  , &&m_{\Phi_{2,r}}^2 = -\mu_2^2 + \lambda_{12} v^2/2, \nonumber\\
m_{N_{1,r}}^2 &= -\mu_1^2 + \lambda_{1} v^2,  &&m_{N_{2,r}}^2 = -\mu_2^2 + \lambda_{12} v^2/2, \nonumber\\
 m_{\Phi_{1,b}}^2 &= -\mu_1^2 + \lambda_{12} v_2^2/2 ,  &&m_{\Phi_{2,b}}^2 = -\mu_2^2 + 3\lambda_{2} v_2^2 ,\nonumber\\
 m_{N_{1,b}}^2 &=-\mu_1^2 + \lambda_{12} v_2^2/2 , &&m_{N_{2,b}}^2 = -\mu_2^2 + \lambda_{2} v_2^2/2.
 \end{align}
%  %%%%%%%%%%%%%%%%%%%%%%%%%%%%%%%%%%%% 
To replace the Higgs boson mass $m_h$ and additional neutral $CP$-even boson mass $m_{ H}$, we use the following second derivative of the effective potentials: 
    %%%%%%%%%%%%%%%%%%%%%%%%%%%%%%%%%%%%
 \begin{align}
  \label{redc3}
m_h^2\equiv \left.\frac{\partial^2 V_{\rm eff}}{\partial  \left\langle \Phi_1 \right\rangle^2}  \right|_{ \left\langle \Phi_1 \right\rangle=v,  \left\langle \Phi_2 \right\rangle=0}  &=
	2 \lambda_1v^2 +\frac{v^2}{32\pi^2} \Bigg[A^r_2\ln\frac{m_{\Phi_{1,r}}^2m_{\Phi_{2,r}}^2}{Q^4} -A^r_3\ln\frac{m_{\Phi_{1,r}}^2}{m_{\Phi_{2,r}}^2}\nonumber\\
&\quad + 12\lambda_1^2 \ln\frac{m_{N_{1,r}}^2}{Q^2}+  4\lambda_{12}^2(2(2I_{2}-1)-1) \ln\frac{m_{N_{2,r}}^2}{Q^2} + 12  \frac{m_Z^4}{v^4}\left(\ln\frac{m_Z^2}{Q^2}+\frac{2}{3}\right)\nonumber\\
&\quad  +24  \frac{m_W^4}{ v^4}\left(\ln\frac{m_W^2}{Q^2}+\frac{2}{3}\right)  -48  \frac{m_t^4}{ v^4}\ln\frac{m_t^2}{Q^2} \Bigg] \\
%%%%%%%%%%
  \label{redc4}
m_{ H}^2 \equiv \left.\frac{\partial^2 V_{\rm eff}}{\partial  \left\langle \Phi_2 \right\rangle^2} \right|_{ \left\langle \Phi_1 \right\rangle=v,  \left\langle \Phi_2 \right\rangle=0} &= -\mu_2^2 +  \lambda_{12}v^2 +\frac{1}{64\pi^2} \Bigg[ B_1f_-(m_{\Phi_{2,r}}^2,m_{\Phi_{1,r}}^2)+ B_2f_+(m_{\Phi_{2,r}}^2,m_{\Phi_{1,r}}^2) \nonumber\\
&\quad %+ 2 n_1 \lambda_{12} m_{N_{1,r}}^2 \left(\ln\frac{m_{N_{1,r}}^2}{Q^2} - 1\right) 
 + 4(2(2I_{2}-1)-1) \lambda_2 m_{N_{2,r}}^2 \left(\ln\frac{m_{N_{2,r}}^2}{Q^2} - 1\right) + 12  \frac{m_Z^4Y_{2}^2}{v^2}\left(\ln\frac{m_Z^2}{Q^2}-\frac{1}{3}\right)  \nonumber\\
&\quad +24  \frac{m_W^4I_W^2}{ v^2}\left(\ln\frac{m_W^2}{Q^2}-\frac{1}{3}\right) \Bigg] ,  
 \end{align}
%%%%%%%%%%%%%%%%%%%%%%%%%%%%%%%%%%%%% 
 where
 %%%%%%%%%%%%%%%%%%%%%%%%%%%%%%%%%%%%%%%%%%%%%%%%%%%%%%%%%%%%
\begin{align}
  A^r_2 &=\frac{1}{4}\left( (6\lambda_1+\lambda_{12})^2 +(6\lambda_1-\lambda_{12})^2\right), \quad A^r_3 = -\frac{36\lambda_1^2-\lambda_{12}^2}{2}\\
   B_1&= \frac{2}{\Delta m^2}
\left[\left(-6\lambda_2+\lambda_{12}\right)  \left({\cal M}_{\Phi_1\Phi_1}^2-{\cal M}_{\Phi_2\Phi_2}^2\right)+ 4\lambda_{12}^2 v^2 \right]\\
B_2 &= 6\lambda_2+\lambda_{12} .
 \end{align}
%%%%%%%%%%%%%%%%%%%%%%%%%%%%%%%%%%%%%
The model parameters can be replaced by Eqs.~(\ref{redc1}), (\ref{redc2}) and (\ref{redc3}): 
 %%%%%%%%%%%%%%%%%%%%%%%%%%%%%%%%%%%%%%%%%%%%%%%%%%%%%%%%%%%%
\begin{align}
  (\mu_1^2, \mu_2^2,  \lambda_1,  \lambda_2,  \lambda_{12} )\to (v, v_2 , m_h ,  \lambda_2, m_{ \Phi_2}^2 ).
   \end{align}
%%%%%%%%%%%%%%%%%%%%%%%%%%%%%%%%%%%%% 


\begin{thebibliography}{99}




%%%%%%% 		EWBG 		%%%%%%%

%\cite{Kuzmin:1985mm}
\bibitem{Kuzmin:1985mm}
V.~Kuzmin, V.~Rubakov and M.~Shaposhnikov,
%``On the Anomalous Electroweak Baryon Number Nonconservation in the Early Universe,''
Phys. Lett. B \textbf{155} (1985), 36
doi:10.1016/0370-2693(85)91028-7
%2808 citations counted in INSPIRE as of 26 Jun 2020


\bibitem{Sakharov:1967dj} 
  A.~D.~Sakharov,
  %``Violation of CP Invariance, c Asymmetry, and Baryon Asymmetry of the Universe,''
  Pisma Zh.\ Eksp.\ Teor.\ Fiz.\  {\bf 5}, 32 (1967)
  [JETP Lett.\  {\bf 5}, 24 (1967)]
  [Sov.\ Phys.\ Usp.\  {\bf 34}, 392 (1991)]
  [Usp.\ Fiz.\ Nauk {\bf 161}, 61 (1991)].
  
  
%%%%%%%%% SM EWPT %%%%%%%%% 

%\cite{Dine:1992vs}
\bibitem{Dine:1992vs}
M.~Dine, R.~G.~Leigh, P.~Huet, A.~D.~Linde and D.~A.~Linde,
%``Comments on the electroweak phase transition,''
Phys. Lett. B \textbf{283} (1992), 319-325
[arXiv:hep-ph/9203201 [hep-ph]].
%\cite{Kajantie:1995kf}
%\bibitem{Kajantie:1995kf}
K.~Kajantie, M.~Laine, K.~Rummukainen and M.~E.~Shaposhnikov,
%``The Electroweak phase transition: A Nonperturbative analysis,''
Nucl. Phys. B \textbf{466} (1996), 189-258
[arXiv:hep-lat/9510020 [hep-lat]].





%%%%%%% 		LISA 		%%%%%%%
\bibitem{LISA}
H. Audley {\it et al.}, “Laser Interferometer Space Antenna,” [arXiv:1702.00786 [astro-ph.IM]].

%%%%%%% 		DECIGO 		%%%%%%%

%\cite{Seto:2001qf}
\bibitem{Seto:2001qf}
N.~Seto, S.~Kawamura and T.~Nakamura,
%``Possibility of direct measurement of the acceleration of the universe using 0.1-Hz band laser interferometer gravitational wave antenna in space,''
Phys. Rev. Lett. \textbf{87}, 221103 (2001)
%doi:10.1103/PhysRevLett.87.221103
[arXiv:astro-ph/0108011 [astro-ph]].


%%%%%%% 		BBO 		%%%%%%%
\bibitem{BBO} 
S.~Phinney {\it et al.}, {\it The Big Bang Observer: Direct Detection
of Gravitational Waves from the Birth of the Universe to the Present}, NASA Mission
Concept Study (2004).




%%%%%%%%%%%%%%%%%%%%%%%%%%%%%%%%%%%%%%%%%%%%%%%%%%%



\bibitem{Grojean:2006bp}
C.~Grojean and G.~Servant,
%``Gravitational Waves from Phase Transitions at the Electroweak Scale and Beyond,''
Phys. Rev. D \textbf{75}, 043507 (2007)
%doi:10.1103/PhysRevD.75.043507
[arXiv:hep-ph/0607107 [hep-ph]].



%%%%%%%%%%%%%%%%%%%%%%%%%%%%%%%%%%%%%%%%%%%%%%%%%%%
%%%%%%%%%%%%%%%%%%%%%%%%%%%%%%%%%%
%%%%%%%%%%%%%%%%%%%%%%%%%%%%%%%%%%
%\cite{Apreda:2001tj}
\bibitem{Apreda:2001tj}
R.~Apreda, M.~Maggiore, A.~Nicolis and A.~Riotto,
%``Supersymmetric phase transitions and gravitational waves at LISA,''
Class. Quant. Grav. \textbf{18}, L155-L162 (2001)
[arXiv:hep-ph/0102140 [hep-ph]].

%\cite{Apreda:2001us}
\bibitem{Apreda:2001us}
R.~Apreda, M.~Maggiore, A.~Nicolis and A.~Riotto,
%``Gravitational waves from electroweak phase transitions,''
Nucl. Phys. B \textbf{631}, 342-368 (2002)
[arXiv:gr-qc/0107033 [gr-qc]].


%\cite{Huber:2007vva}
\bibitem{Huber:2007vva}
S.~J.~Huber and T.~Konstandin,
%``Production of gravitational waves in the nMSSM,''
JCAP \textbf{05}, 017 (2008)
[arXiv:0709.2091 [hep-ph]].

%\cite{Espinosa:2008kw}
\bibitem{Espinosa:2008kw}
J.~R.~Espinosa, T.~Konstandin, J.~M.~No and M.~Quiros,
%``Some Cosmological Implications of Hidden Sectors,''
Phys. Rev. D \textbf{78}, 123528 (2008)
[arXiv:0809.3215 [hep-ph]].

%\cite{Ashoorioon:2009nf}
\bibitem{Ashoorioon:2009nf}
A.~Ashoorioon and T.~Konstandin,
%``Strong electroweak phase transitions without collider traces,''
JHEP \textbf{07}, 086 (2009)
[arXiv:0904.0353 [hep-ph]].

%\cite{Kang:2009rd}
\bibitem{Kang:2009rd}
J.~Kang, P.~Langacker, T.~Li and T.~Liu,
%``Electroweak Baryogenesis, CDM and Anomaly-free Supersymmetric U(1)' Models,''
JHEP \textbf{04}, 097 (2011)
[arXiv:0911.2939 [hep-ph]].

%\cite{Jarvinen:2009mh}
\bibitem{Jarvinen:2009mh}
M.~Jarvinen, C.~Kouvaris and F.~Sannino,
%``Gravitational Techniwaves,''
Phys. Rev. D \textbf{81}, 064027 (2010)
[arXiv:0911.4096 [hep-ph]].

%\cite{Konstandin:2010cd}
\bibitem{Konstandin:2010cd}
T.~Konstandin, G.~Nardini and M.~Quiros,
%``Gravitational Backreaction Effects on the Holographic Phase Transition,''
Phys. Rev. D \textbf{82}, 083513 (2010)
[arXiv:1007.1468 [hep-ph]].

%\cite{No:2011fi}
\bibitem{No:2011fi}
J.~M.~No,
%``Large Gravitational Wave Background Signals in Electroweak Baryogenesis Scenarios,''
Phys. Rev. D \textbf{84}, 124025 (2011)
[arXiv:1103.2159 [hep-ph]].

%\cite{Wainwright:2011qy}
\bibitem{Wainwright:2011qy}
C.~Wainwright, S.~Profumo and M.~J.~Ramsey-Musolf,
%``Gravity Waves from a Cosmological Phase Transition: Gauge Artifacts and Daisy Resummations,''
Phys. Rev. D \textbf{84}, 023521 (2011)
[arXiv:1104.5487 [hep-ph]].

%\cite{Barger:2011vm}
\bibitem{Barger:2011vm}
V.~Barger, D.~J.~H.~Chung, A.~J.~Long and L.~T.~Wang,
%``Strongly First Order Phase Transitions Near an Enhanced Discrete Symmetry Point,''
Phys. Lett. B \textbf{710}, 1-7 (2012)
[arXiv:1112.5460 [hep-ph]].

%\cite{Leitao:2012tx}
\bibitem{Leitao:2012tx}
L.~Leitao, A.~Megevand and A.~D.~Sanchez,
%``Gravitational waves from the electroweak phase transition,''
JCAP \textbf{10}, 024 (2012)
[arXiv:1205.3070 [astro-ph.CO]].

%\cite{Dorsch:2014qpa}
\bibitem{Dorsch:2014qpa}
G.~C.~Dorsch, S.~J.~Huber and J.~M.~No,
%``Cosmological Signatures of a UV-Conformal Standard Model,''
Phys. Rev. Lett. \textbf{113}, 121801 (2014)
[arXiv:1403.5583 [hep-ph]].

%\cite{Kozaczuk:2014kva}
\bibitem{Kozaczuk:2014kva}
J.~Kozaczuk, S.~Profumo, L.~S.~Haskins and C.~L.~Wainwright,
%``Cosmological Phase Transitions and their Properties in the NMSSM,''
JHEP \textbf{01}, 144 (2015)
[arXiv:1407.4134 [hep-ph]].

%\cite{Schwaller:2015tja}
\bibitem{Schwaller:2015tja}
P.~Schwaller,
%``Gravitational Waves from a Dark Phase Transition,''
Phys. Rev. Lett. \textbf{115}, no.18, 181101 (2015)
[arXiv:1504.07263 [hep-ph]].


%\cite{Xiao:2015tja}
\bibitem{Xiao:2015tja}
M.~L.~Xiao and J.~H.~Yu,
%``Electroweak baryogenesis in a scalar-assisted vectorlike fermion model,''
Phys. Rev. D \textbf{94}, no.1, 015011 (2016)
doi:10.1103/PhysRevD.94.015011
[arXiv:1509.02931 [hep-ph]].
%12 citations counted in INSPIRE as of 15 Dec 2022

%\cite{Kakizaki:2015wua}
\bibitem{Kakizaki:2015wua}
M.~Kakizaki, S.~Kanemura and T.~Matsui,
%``Gravitational waves as a probe of extended scalar sectors with the first order electroweak phase transition,''
Phys. Rev. D \textbf{92}, no.11, 115007 (2015)
[arXiv:1509.08394 [hep-ph]].

%\cite{Jinno:2015doa}
\bibitem{Jinno:2015doa}
R.~Jinno, K.~Nakayama and M.~Takimoto,
%``Gravitational waves from the first order phase transition of the Higgs field at high energy scales,''
Phys. Rev. D \textbf{93}, no.4, 045024 (2016)
[arXiv:1510.02697 [hep-ph]].

%\cite{Huber:2015znp}
\bibitem{Huber:2015znp}
S.~J.~Huber, T.~Konstandin, G.~Nardini and I.~Rues,
%``Detectable Gravitational Waves from Very Strong Phase Transitions in the General NMSSM,''
JCAP \textbf{03}, 036 (2016)
[arXiv:1512.06357 [hep-ph]].


%\cite{Leitao:2015fmj}
\bibitem{Leitao:2015fmj}
L.~Leitao and A.~Megevand,
%``Gravitational waves from a very strong electroweak phase transition,''
JCAP \textbf{05}, 037 (2016)
[arXiv:1512.08962 [astro-ph.CO]].

%\cite{Huang:2016odd}
\bibitem{Huang:2016odd}
F.~P.~Huang, Y.~Wan, D.~G.~Wang, Y.~F.~Cai and X.~Zhang,
%``Hearing the echoes of electroweak baryogenesis with gravitational wave detectors,''
Phys. Rev. D \textbf{94}, no.4, 041702 (2016)
[arXiv:1601.01640 [hep-ph]].

%\cite{Garcia-Pepin:2016hvs}
\bibitem{Garcia-Pepin:2016hvs}
M.~Garcia-Pepin and M.~Quiros,
%``Strong electroweak phase transition from Supersymmetric Custodial Triplets,''
JHEP \textbf{05}, 177 (2016)
[arXiv:1602.01351 [hep-ph]].

%\cite{Jaeckel:2016jlh}
\bibitem{Jaeckel:2016jlh}
J.~Jaeckel, V.~V.~Khoze and M.~Spannowsky,
%``Hearing the signal of dark sectors with gravitational wave detectors,''
Phys. Rev. D \textbf{94}, no.10, 103519 (2016)
[arXiv:1602.03901 [hep-ph]].

%\cite{Dev:2016feu}
\bibitem{Dev:2016feu}
P.~S.~B.~Dev and A.~Mazumdar,
%``Probing the Scale of New Physics by Advanced LIGO/VIRGO,''
Phys. Rev. D \textbf{93}, no.10, 104001 (2016)
[arXiv:1602.04203 [hep-ph]].

%\cite{Hashino:2016rvx}
\bibitem{Hashino:2016rvx}
K.~Hashino, M.~Kakizaki, S.~Kanemura and T.~Matsui,
%``Synergy between measurements of gravitational waves and the triple-Higgs coupling in probing the first-order electroweak phase transition,''
Phys. Rev. D \textbf{94}, no.1, 015005 (2016)
[arXiv:1604.02069 [hep-ph]].

%\cite{Jinno:2016knw}
\bibitem{Jinno:2016knw}
R.~Jinno and M.~Takimoto,
%``Probing a classically conformal B-L model with gravitational waves,''
Phys. Rev. D \textbf{95}, no.1, 015020 (2017)
[arXiv:1604.05035 [hep-ph]].

%\cite{Barenboim:2016mjm}
\bibitem{Barenboim:2016mjm}
G.~Barenboim and W.~I.~Park,
%``Gravitational waves from first order phase transitions as a probe of an early matter domination era and its inverse problem,''
Phys. Lett. B \textbf{759}, 430-438 (2016)
[arXiv:1605.03781 [astro-ph.CO]].

%\cite{Kobakhidze:2016mch}
\bibitem{Kobakhidze:2016mch}
A.~Kobakhidze, A.~Manning and J.~Yue,
%``Gravitational waves from the phase transition of a nonlinearly realized electroweak gauge symmetry,''
Int. J. Mod. Phys. D \textbf{26}, no.10, 1750114 (2017)
[arXiv:1607.00883 [hep-ph]].

%\cite{Hashino:2016xoj}
\bibitem{Hashino:2016xoj}
K.~Hashino, M.~Kakizaki, S.~Kanemura, P.~Ko and T.~Matsui,
%``Gravitational waves and Higgs boson couplings for exploring first order phase transition in the model with a singlet scalar field,''
Phys. Lett. B \textbf{766}, 49-54 (2017)
[arXiv:1609.00297 [hep-ph]].

%\cite{Artymowski:2016tme}
\bibitem{Artymowski:2016tme}
M.~Artymowski, M.~Lewicki and J.~D.~Wells,
%``Gravitational wave and collider implications of electroweak baryogenesis aided by non-standard cosmology,''
JHEP \textbf{03}, 066 (2017)
[arXiv:1609.07143 [hep-ph]].

%\cite{Kubo:2016kpb}
\bibitem{Kubo:2016kpb}
J.~Kubo and M.~Yamada,
%``Scale genesis and gravitational wave in a classically scale invariant extension of the standard model,''
JCAP \textbf{12}, 001 (2016)
[arXiv:1610.02241 [hep-ph]].

%\cite{Balazs:2016tbi}
\bibitem{Balazs:2016tbi}
C.~Balazs, A.~Fowlie, A.~Mazumdar and G.~White,
%``Gravitational waves at aLIGO and vacuum stability with a scalar singlet extension of the Standard Model,''
Phys. Rev. D \textbf{95}, no.4, 043505 (2017)
[arXiv:1611.01617 [hep-ph]].

%\cite{Vaskonen:2016yiu}
\bibitem{Vaskonen:2016yiu}
V.~Vaskonen,
%``Electroweak baryogenesis and gravitational waves from a real scalar singlet,''
Phys. Rev. D \textbf{95}, no.12, 123515 (2017)
[arXiv:1611.02073 [hep-ph]].

%\cite{Dorsch:2016nrg}
\bibitem{Dorsch:2016nrg}
G.~C.~Dorsch, S.~J.~Huber, T.~Konstandin and J.~M.~No,
%``A Second Higgs Doublet in the Early Universe: Baryogenesis and Gravitational Waves,''
JCAP \textbf{05}, 052 (2017)
[arXiv:1611.05874 [hep-ph]].

%\cite{Huang:2017laj}
\bibitem{Huang:2017laj}
F.~P.~Huang and X.~Zhang,
%``Probing the gauge symmetry breaking of the early universe in 3-3-1 models and beyond by gravitational waves,''
Phys. Lett. B \textbf{788}, 288-294 (2019)
[arXiv:1701.04338 [hep-ph]].

%\cite{Baldes:2017rcu}
\bibitem{Baldes:2017rcu}
I.~Baldes,
%``Gravitational waves from the asymmetric-dark-matter generating phase transition,''
JCAP \textbf{05}, 028 (2017)
[arXiv:1702.02117 [hep-ph]].

%\cite{Chao:2017vrq}
\bibitem{Chao:2017vrq}
W.~Chao, H.~K.~Guo and J.~Shu,
%``Gravitational Wave Signals of Electroweak Phase Transition Triggered by Dark Matter,''
JCAP \textbf{09}, 009 (2017)
[arXiv:1702.02698 [hep-ph]].

%\cite{Beniwal:2017eik}
\bibitem{Beniwal:2017eik}
A.~Beniwal, M.~Lewicki, J.~D.~Wells, M.~White and A.~G.~Williams,
%``Gravitational wave, collider and dark matter signals from a scalar singlet electroweak baryogenesis,''
JHEP \textbf{08}, 108 (2017)
[arXiv:1702.06124 [hep-ph]].

%\cite{Addazi:2017gpt}
\bibitem{Addazi:2017gpt}
A.~Addazi and A.~Marciano,
%``Gravitational waves from dark first order phase transitions and dark photons,''
Chin. Phys. C \textbf{42}, no.2, 023107 (2018)
[arXiv:1703.03248 [hep-ph]].

%\cite{Kobakhidze:2017mru}
\bibitem{Kobakhidze:2017mru}
A.~Kobakhidze, C.~Lagger, A.~Manning and J.~Yue,
%``Gravitational waves from a supercooled electroweak phase transition and their detection with pulsar timing arrays,''
Eur. Phys. J. C \textbf{77}, no.8, 570 (2017)
[arXiv:1703.06552 [hep-ph]].

%\cite{Tsumura:2017knk}
\bibitem{Tsumura:2017knk}
K.~Tsumura, M.~Yamada and Y.~Yamaguchi,
%``Gravitational wave from dark sector with dark pion,''
JCAP \textbf{07}, 044 (2017)
[arXiv:1704.00219 [hep-ph]].

%\cite{Marzola:2017jzl}
\bibitem{Marzola:2017jzl}
L.~Marzola, A.~Racioppi and V.~Vaskonen,
%``Phase transition and gravitational wave phenomenology of scalar conformal extensions of the Standard Model,''
Eur. Phys. J. C \textbf{77}, no.7, 484 (2017)
[arXiv:1704.01034 [hep-ph]].

%\cite{Bian:2017wfv}
\bibitem{Bian:2017wfv}
L.~Bian, H.~K.~Guo and J.~Shu,
%``Gravitational Waves, baryon asymmetry of the universe and electric dipole moment in the CP-violating NMSSM,''
Chin. Phys. C \textbf{42}, no.9, 093106 (2018)
[erratum: Chin. Phys. C \textbf{43}, no.12, 129101 (2019)]
[arXiv:1704.02488 [hep-ph]].

%\cite{Huang:2017rzf}
\bibitem{Huang:2017rzf}
F.~P.~Huang and J.~H.~Yu,
%``Exploring inert dark matter blind spots with gravitational wave signatures,''
Phys. Rev. D \textbf{98}, no.9, 095022 (2018)
[arXiv:1704.04201 [hep-ph]].

%\cite{Iso:2017uuu}
\bibitem{Iso:2017uuu}
S.~Iso, P.~D.~Serpico and K.~Shimada,
%``QCD-Electroweak First-Order Phase Transition in a Supercooled Universe,''
Phys. Rev. Lett. \textbf{119}, no.14, 141301 (2017)
[arXiv:1704.04955 [hep-ph]].

%\cite{Addazi:2017oge}
\bibitem{Addazi:2017oge}
A.~Addazi and A.~Marciano,
%``Limiting majoron self-interactions from gravitational wave experiments,''
Chin. Phys. C \textbf{42}, no.2, 023105 (2018)
[arXiv:1705.08346 [hep-ph]].

%\cite{Kang:2017mkl}
\bibitem{Kang:2017mkl}
Z.~Kang, P.~Ko and T.~Matsui,
%``Strong first order EWPT $\&$ strong gravitational waves in Z$_{3}$-symmetric singlet scalar extension,''
JHEP \textbf{02}, 115 (2018)
[arXiv:1706.09721 [hep-ph]].

%\cite{Cai:2017tmh}
\bibitem{Cai:2017tmh}
R.~G.~Cai, M.~Sasaki and S.~J.~Wang,
%``The gravitational waves from the first-order phase transition with a dimension-six operator,''
JCAP \textbf{08}, 004 (2017)
[arXiv:1707.03001 [astro-ph.CO]].

%\cite{Chao:2017ilw}
\bibitem{Chao:2017ilw}
W.~Chao, W.~F.~Cui, H.~K.~Guo and J.~Shu,
%``Gravitational wave imprint of new symmetry breaking,''
Chin. Phys. C \textbf{44}, no.12, 123102 (2020)
[arXiv:1707.09759 [hep-ph]].

%\cite{Aoki:2017aws}
\bibitem{Aoki:2017aws}
M.~Aoki, H.~Goto and J.~Kubo,
%``Gravitational Waves from Hidden QCD Phase Transition,''
Phys. Rev. D \textbf{96}, no.7, 075045 (2017)
[arXiv:1709.07572 [hep-ph]].

%\cite{Huang:2017kzu}
\bibitem{Huang:2017kzu}
F.~P.~Huang and C.~S.~Li,
%``Probing the baryogenesis and dark matter relaxed in phase transition by gravitational waves and colliders,''
Phys. Rev. D \textbf{96}, no.9, 095028 (2017)
[arXiv:1709.09691 [hep-ph]].

%\cite{Demidov:2017lzf}
\bibitem{Demidov:2017lzf}
S.~V.~Demidov, D.~S.~Gorbunov and D.~V.~Kirpichnikov,
%``Gravitational waves from phase transition in split NMSSM,''
Phys. Lett. B \textbf{779}, 191-194 (2018)
[arXiv:1712.00087 [hep-ph]].

%\cite{Chen:2017cyc}
\bibitem{Chen:2017cyc}
Y.~Chen, M.~Huang and Q.~S.~Yan,
%``Gravitation waves from QCD and electroweak phase transitions,''
JHEP \textbf{05}, 178 (2018)
[arXiv:1712.03470 [hep-ph]].

%\cite{Chala:2018ari}
\bibitem{Chala:2018ari}
M.~Chala, C.~Krause and G.~Nardini,
%``Signals of the electroweak phase transition at colliders and gravitational wave observatories,''
JHEP \textbf{07}, 062 (2018)
[arXiv:1802.02168 [hep-ph]].

%\cite{Hashino:2018zsi}
\bibitem{Hashino:2018zsi}
K.~Hashino, M.~Kakizaki, S.~Kanemura, P.~Ko and T.~Matsui,
%``Gravitational waves from first order electroweak phase transition in models with the U(1)$_{X}$ gauge symmetry,''
JHEP \textbf{06}, 088 (2018)
[arXiv:1802.02947 [hep-ph]].

%\cite{Morais:2018uou}
\bibitem{Morais:2018uou}
A.~P.~Morais, R.~Pasechnik and T.~Vieu,
%``Multi-peaked signatures of primordial gravitational waves from multi-step electroweak phase transition,''
PoS \textbf{EPS-HEP2019}, 054 (2020)
[arXiv:1802.10109 [hep-ph]].

%\cite{Croon:2018new}
\bibitem{Croon:2018new}
D.~Croon and G.~White,
%``Exotic Gravitational Wave Signatures from Simultaneous Phase Transitions,''
JHEP \textbf{05}, 210 (2018)
[arXiv:1803.05438 [hep-ph]].

%\cite{Bruggisser:2018mus}
\bibitem{Bruggisser:2018mus}
S.~Bruggisser, B.~Von Harling, O.~Matsedonskyi and G.~Servant,
%``Baryon Asymmetry from a Composite Higgs Boson,''
Phys. Rev. Lett. \textbf{121}, no.13, 131801 (2018)
[arXiv:1803.08546 [hep-ph]].

%\cite{Imtiaz:2018dfn}
\bibitem{Imtiaz:2018dfn}
B.~Imtiaz, Y.~F.~Cai and Y.~Wan,
%``Two-field cosmological phase transitions and gravitational waves in the singlet Majoron model,''
Eur. Phys. J. C \textbf{79}, no.1, 25 (2019)
[arXiv:1804.05835 [hep-ph]].

%\cite{Huang:2018aja}
\bibitem{Huang:2018aja}
F.~P.~Huang, Z.~Qian and M.~Zhang,
%``Exploring dynamical CP violation induced baryogenesis by gravitational waves and colliders,''
Phys. Rev. D \textbf{98}, no.1, 015014 (2018)
[arXiv:1804.06813 [hep-ph]].

%\cite{Bruggisser:2018mrt}
\bibitem{Bruggisser:2018mrt}
S.~Bruggisser, B.~Von Harling, O.~Matsedonskyi and G.~Servant,
%``Electroweak Phase Transition and Baryogenesis in Composite Higgs Models,''
JHEP \textbf{12}, 099 (2018)
[arXiv:1804.07314 [hep-ph]].

%\cite{FitzAxen:2018vdt}
\bibitem{FitzAxen:2018vdt}
M.~Fitz Axen, S.~Banagiri, A.~Matas, C.~Caprini and V.~Mandic,
%``Multiwavelength observations of cosmological phase transitions using LISA and Cosmic Explorer,''
Phys. Rev. D \textbf{98}, no.10, 103508 (2018)
[arXiv:1806.02500 [astro-ph.IM]].

%\cite{Megias:2018sxv}
\bibitem{Megias:2018sxv}
E.~Meg\'\i{}as, G.~Nardini and M.~Quir\'os,
%``Cosmological Phase Transitions in Warped Space: Gravitational Waves and Collider Signatures,''
JHEP \textbf{09}, 095 (2018)
[arXiv:1806.04877 [hep-ph]].

%\cite{Alves:2018oct}
\bibitem{Alves:2018oct}
A.~Alves, T.~Ghosh, H.~K.~Guo and K.~Sinha,
%``Resonant Di-Higgs Production at Gravitational Wave Benchmarks: A Collider Study using Machine Learning,''
JHEP \textbf{12}, 070 (2018)
[arXiv:1808.08974 [hep-ph]].


%\cite{Baldes:2018emh}
\bibitem{Baldes:2018emh}
I.~Baldes and C.~Garcia-Cely,
%``Strong gravitational radiation from a simple dark matter model,''
JHEP \textbf{05}, 190 (2019)
[arXiv:1809.01198 [hep-ph]].


%\cite{Ahriche:2018rao}
\bibitem{Ahriche:2018rao}
A.~Ahriche, K.~Hashino, S.~Kanemura and S.~Nasri,
%``Gravitational Waves from Phase Transitions in Models with Charged Singlets,''
Phys. Lett. B \textbf{789}, 119-126 (2019)
[arXiv:1809.09883 [hep-ph]].

%\cite{Prokopec:2018tnq}
\bibitem{Prokopec:2018tnq}
T.~Prokopec, J.~Rezacek and B.~\'Swie\.zewska,
%``Gravitational waves from conformal symmetry breaking,''
JCAP \textbf{02}, 009 (2019)
[arXiv:1809.11129 [hep-ph]].


%\cite{Fujikura:2018duw}
\bibitem{Fujikura:2018duw}
K.~Fujikura, K.~Kamada, Y.~Nakai and M.~Yamaguchi,
%``Phase Transitions in Twin Higgs Models,''
JHEP \textbf{12}, 018 (2018)
[arXiv:1810.00574 [hep-ph]].

%\cite{Beniwal:2018hyi}
\bibitem{Beniwal:2018hyi}
A.~Beniwal, M.~Lewicki, M.~White and A.~G.~Williams,
%``Gravitational waves and electroweak baryogenesis in a global study of the extended scalar singlet model,''
JHEP \textbf{02}, 183 (2019)
[arXiv:1810.02380 [hep-ph]].

%\cite{Brdar:2018num}
\bibitem{Brdar:2018num}
V.~Brdar, A.~J.~Helmboldt and J.~Kubo,
%``Gravitational Waves from First-Order Phase Transitions: LIGO as a Window to Unexplored Seesaw Scales,''
JCAP \textbf{02}, 021 (2019)
[arXiv:1810.12306 [hep-ph]].

%\cite{Mazumdar:2018dfl}
\bibitem{Mazumdar:2018dfl}
A.~Mazumdar and G.~White,
%``Review of cosmic phase transitions: their significance and experimental signatures,''
Rept. Prog. Phys. \textbf{82}, no.7, 076901 (2019)
[arXiv:1811.01948 [hep-ph]].

%\cite{Addazi:2018nzm}
\bibitem{Addazi:2018nzm}
A.~Addazi, A.~Marcian\`o and R.~Pasechnik,
%``Probing Trans-electroweak First Order Phase Transitions from Gravitational Waves,''
MDPI Physics \textbf{1}, no.1, 92-102 (2019)
[arXiv:1811.09074 [hep-ph]].

%\cite{Shajiee:2018jdq}
\bibitem{Shajiee:2018jdq}
V.~R.~Shajiee and A.~Tofighi,
%``Electroweak Phase Transition, Gravitational Waves and Dark Matter in Two Scalar Singlet Extension of The Standard Model,''
Eur. Phys. J. C \textbf{79}, no.4, 360 (2019)
[arXiv:1811.09807 [hep-ph]].

%\cite{Breitbach:2018ddu}
\bibitem{Breitbach:2018ddu}
M.~Breitbach, J.~Kopp, E.~Madge, T.~Opferkuch and P.~Schwaller,
%``Dark, Cold, and Noisy: Constraining Secluded Hidden Sectors with Gravitational Waves,''
JCAP \textbf{07}, 007 (2019)
[arXiv:1811.11175 [hep-ph]].

%\cite{Marzo:2018nov}
\bibitem{Marzo:2018nov}
C.~Marzo, L.~Marzola and V.~Vaskonen,
%``Phase transition and vacuum stability in the classically conformal B\textendash{}L model,''
Eur. Phys. J. C \textbf{79}, no.7, 601 (2019)
[arXiv:1811.11169 [hep-ph]].


%\cite{Megias:2018dki}
\bibitem{Megias:2018dki}
E.~Meg\'\i{}as, G.~Nardini and M.~Quir\'os,
%``Gravitational waves and collider signatures from holographic phase transitions in soft walls,''
PoS \textbf{Confinement2018}, 227 (2018)
[arXiv:1811.10891 [hep-ph]].

%\cite{Croon:2018kqn}
\bibitem{Croon:2018kqn}
D.~Croon, T.~E.~Gonzalo and G.~White,
%``Gravitational Waves from a Pati-Salam Phase Transition,''
JHEP \textbf{02}, 083 (2019)
[arXiv:1812.02747 [hep-ph]].

%\cite{Angelescu:2018dkk}
\bibitem{Angelescu:2018dkk}
A.~Angelescu and P.~Huang,
%``Multistep Strongly First Order Phase Transitions from New Fermions at the TeV Scale,''
Phys. Rev. D \textbf{99}, no.5, 055023 (2019)
[arXiv:1812.08293 [hep-ph]].

%\cite{Alves:2018jsw}
\bibitem{Alves:2018jsw}
A.~Alves, T.~Ghosh, H.~K.~Guo, K.~Sinha and D.~Vagie,
%``Collider and Gravitational Wave Complementarity in Exploring the Singlet Extension of the Standard Model,''
JHEP \textbf{04}, 052 (2019)
[arXiv:1812.09333 [hep-ph]].


%\cite{Abedi:2019msi}
\bibitem{Abedi:2019msi}
H.~Abedi, M.~Ahmadvand and S.~S.~Gousheh,
%``Electroweak phase transition in the presence of hypermagnetic field and the generation of gravitational waves,''
[arXiv:1901.05912 [hep-ph]].

%\cite{Fairbairn:2019xog}
\bibitem{Fairbairn:2019xog}
M.~Fairbairn, E.~Hardy and A.~Wickens,
%``Hearing without seeing: gravitational waves from hot and cold hidden sectors,''
JHEP \textbf{07}, 044 (2019)
[arXiv:1901.11038 [hep-ph]].

%\cite{Kainulainen:2019kyp}
\bibitem{Kainulainen:2019kyp}
K.~Kainulainen, V.~Keus, L.~Niemi, K.~Rummukainen, T.~V.~I.~Tenkanen and V.~Vaskonen,
%``On the validity of perturbative studies of the electroweak phase transition in the Two Higgs Doublet model,''
JHEP \textbf{06}, 075 (2019)
[arXiv:1904.01329 [hep-ph]].

%\cite{Hasegawa:2019amx}
\bibitem{Hasegawa:2019amx}
T.~Hasegawa, N.~Okada and O.~Seto,
%``Gravitational waves from the minimal gauged $U(1)_{B-L}$ model,''
Phys. Rev. D \textbf{99}, no.9, 095039 (2019)
[arXiv:1904.03020 [hep-ph]].

%\cite{Helmboldt:2019pan}
\bibitem{Helmboldt:2019pan}
A.~J.~Helmboldt, J.~Kubo and S.~van der Woude,
%``Observational prospects for gravitational waves from hidden or dark chiral phase transitions,''
Phys. Rev. D \textbf{100}, no.5, 055025 (2019)
[arXiv:1904.07891 [hep-ph]].

%\cite{Dev:2019njv}
\bibitem{Dev:2019njv}
P.~S.~B.~Dev, F.~Ferrer, Y.~Zhang and Y.~Zhang,
%``Gravitational Waves from First-Order Phase Transition in a Simple Axion-Like Particle Model,''
JCAP \textbf{11}, 006 (2019)
[arXiv:1905.00891 [hep-ph]].

%\cite{Ellis:2019flb}
\bibitem{Ellis:2019flb}
S.~A.~R.~Ellis, S.~Ipek and G.~White,
%``Electroweak Baryogenesis from Temperature-Varying Couplings,''
JHEP \textbf{08}, 002 (2019)
[arXiv:1905.11994 [hep-ph]].

%\cite{Cutting:2019zws}
\bibitem{Cutting:2019zws}
D.~Cutting, M.~Hindmarsh and D.~J.~Weir,
%``Vorticity, kinetic energy, and suppressed gravitational wave production in strong first order phase transitions,''
Phys. Rev. Lett. \textbf{125}, no.2, 021302 (2020)
[arXiv:1906.00480 [hep-ph]].

%\cite{Bian:2019bsn}
\bibitem{Bian:2019bsn}
L.~Bian, H.~K.~Guo, Y.~Wu and R.~Zhou,
%``Gravitational wave and collider searches for electroweak symmetry breaking patterns,''
Phys. Rev. D \textbf{101}, no.3, 035011 (2020)
[arXiv:1906.11664 [hep-ph]].

%\cite{Kannike:2019mzk}
\bibitem{Kannike:2019mzk}
K.~Kannike, K.~Loos and M.~Raidal,
%``Gravitational wave signals of pseudo-Goldstone dark matter in the $\mathbb{Z}_{3}$ complex singlet model,''
Phys. Rev. D \textbf{101}, no.3, 035001 (2020)
[arXiv:1907.13136 [hep-ph]].

%\cite{Bian:2019szo}
\bibitem{Bian:2019szo}
L.~Bian, W.~Cheng, H.~K.~Guo and Y.~Zhang,
%``Cosmological implications of a B \ensuremath{-} L charged hidden scalar: leptogenesis and gravitational waves,''
Chin. Phys. C \textbf{45}, no.11, 113104 (2021)
[arXiv:1907.13589 [hep-ph]].

%\cite{Dunsky:2019upk}
\bibitem{Dunsky:2019upk}
D.~Dunsky, L.~J.~Hall and K.~Harigaya,
%``Dark Matter, Dark Radiation and Gravitational Waves from Mirror Higgs Parity,''
JHEP \textbf{02}, 078 (2020)
[arXiv:1908.02756 [hep-ph]].

%\cite{Paul:2019pgt}
\bibitem{Paul:2019pgt}
A.~Paul, B.~Banerjee and D.~Majumdar,
%``Gravitational wave signatures from an extended inert doublet dark matter model,''
JCAP \textbf{10}, 062 (2019)
[arXiv:1908.00829 [hep-ph]].

%\cite{Brdar:2019fur}
\bibitem{Brdar:2019fur}
V.~Brdar, L.~Graf, A.~J.~Helmboldt and X.~J.~Xu,
%``Gravitational Waves as a Probe of Left-Right Symmetry Breaking,''
JCAP \textbf{12}, 027 (2019)
[arXiv:1909.02018 [hep-ph]].

%\cite{Wang:2019pet}
\bibitem{Wang:2019pet}
X.~Wang, F.~P.~Huang and X.~Zhang,
%``Gravitational wave and collider signals in complex two-Higgs doublet model with dynamical CP-violation at finite temperature,''
Phys. Rev. D \textbf{101}, no.1, 015015 (2020)
[arXiv:1909.02978 [hep-ph]].

%\cite{Alves:2019igs}
\bibitem{Alves:2019igs}
A.~Alves, D.~Gon\c{c}alves, T.~Ghosh, H.~K.~Guo and K.~Sinha,
%``Di-Higgs Production in the $4b$ Channel and Gravitational Wave Complementarity,''
JHEP \textbf{03}, 053 (2020)
[arXiv:1909.05268 [hep-ph]].


%\cite{DeCurtis:2019rxl}
\bibitem{DeCurtis:2019rxl}
S.~De Curtis, L.~Delle Rose and G.~Panico,
%``Composite Dynamics in the Early Universe,''
JHEP \textbf{12}, 149 (2019)
[arXiv:1909.07894 [hep-ph]].

%\cite{Hall:2019ank}
\bibitem{Hall:2019ank}
E.~Hall, T.~Konstandin, R.~McGehee, H.~Murayama and G.~Servant,
%``Baryogenesis From a Dark First-Order Phase Transition,''
JHEP \textbf{04}, 042 (2020)
%doi:10.1007/JHEP04(2020)042
[arXiv:1910.08068 [hep-ph]].


%\cite{Alanne:2019bsm}
\bibitem{Alanne:2019bsm}
T.~Alanne, T.~Hugle, M.~Platscher and K.~Schmitz,
%``A fresh look at the gravitational-wave signal from cosmological phase transitions,''
JHEP \textbf{03}, 004 (2020)
[arXiv:1909.11356 [hep-ph]].

%\cite{Zhou:2019uzq}
\bibitem{Zhou:2019uzq}
R.~Zhou, L.~Bian and H.~K.~Guo,
%``Connecting the electroweak sphaleron with gravitational waves,''
Phys. Rev. D \textbf{101}, no.9, 091903 (2020)
[arXiv:1910.00234 [hep-ph]].


%\cite{Morais:2019fnm}
\bibitem{Morais:2019fnm}
A.~P.~Morais and R.~Pasechnik,
%``Probing multi-step electroweak phase transition with multi-peaked primordial gravitational waves spectra,''
JCAP \textbf{04}, 036 (2020)
[arXiv:1910.00717 [hep-ph]].


%\cite{Greljo:2019xan}
\bibitem{Greljo:2019xan}
A.~Greljo, T.~Opferkuch and B.~A.~Stefanek,
%``Gravitational Imprints of Flavor Hierarchies,''
Phys. Rev. Lett. \textbf{124}, no.17, 171802 (2020)
[arXiv:1910.02014 [hep-ph]].



%\cite{Archer-Smith:2019gzq}
\bibitem{Archer-Smith:2019gzq}
P.~Archer-Smith, D.~Linthorne and D.~Stolarski,
%``Gravitational Wave Signals from Multiple Hidden Sectors,''
Phys. Rev. D \textbf{101}, no.9, 095016 (2020)
[arXiv:1910.02083 [hep-ph]].



%\cite{Aoki:2019mlt}
\bibitem{Aoki:2019mlt}
M.~Aoki and J.~Kubo,
%``Gravitational waves from chiral phase transition in a conformally extended standard model,''
JCAP \textbf{04}, 001 (2020)
[arXiv:1910.05025 [hep-ph]].


%\cite{Croon:2019rqu}
\bibitem{Croon:2019rqu}
D.~Croon, A.~Kusenko, A.~Mazumdar and G.~White,
%``Solitosynthesis and Gravitational Waves,''
Phys. Rev. D \textbf{101}, no.8, 085010 (2020)
[arXiv:1910.09562 [hep-ph]].

%\cite{Haba:2019qol}
\bibitem{Haba:2019qol}
N.~Haba and T.~Yamada,
%``Gravitational waves from phase transition in minimal SUSY $U(1)_{B-L}$  model,''
Phys. Rev. D \textbf{101}, no.7, 075027 (2020)
[arXiv:1911.01292 [hep-ph]].

%\cite{Carena:2019une}
\bibitem{Carena:2019une}
M.~Carena, Z.~Liu and Y.~Wang,
%``Electroweak phase transition with spontaneous Z$_{2}$-breaking,''
JHEP \textbf{08}, 107 (2020)
[arXiv:1911.10206 [hep-ph]].

%\cite{DelleRose:2019pgi}
\bibitem{DelleRose:2019pgi}
L.~Delle Rose, G.~Panico, M.~Redi and A.~Tesi,
%``Gravitational Waves from Supercool Axions,''
JHEP \textbf{04}, 025 (2020)
[arXiv:1912.06139 [hep-ph]].

%\cite{VonHarling:2019rgb}
\bibitem{VonHarling:2019rgb}
B.~Von Harling, A.~Pomarol, O.~Pujol\`as and F.~Rompineve,
%``Peccei-Quinn Phase Transition at LIGO,''
JHEP \textbf{04}, 195 (2020)
[arXiv:1912.07587 [hep-ph]].

%\cite{Chiang:2019oms}
\bibitem{Chiang:2019oms}
C.~W.~Chiang and B.~Q.~Lu,
%``First-order electroweak phase transition in a complex singlet model with $\mathbb{Z}_3$ symmetry,''
JHEP \textbf{07}, 082 (2020)
[arXiv:1912.12634 [hep-ph]].

%\cite{Barman:2019oda}
\bibitem{Barman:2019oda}
B.~Barman, A.~Dutta Banik and A.~Paul,
%``Singlet-doublet fermionic dark matter and gravitational waves in a two-Higgs-doublet extension of the Standard Model,''
Phys. Rev. D \textbf{101}, no.5, 055028 (2020)
[arXiv:1912.12899 [hep-ph]].

%\cite{Zhou:2020xqi}
\bibitem{Zhou:2020xqi}
R.~Zhou and L.~Bian,
%``Baryon asymmetry and detectable Gravitational Waves from Electroweak phase transition,''
[arXiv:2001.01237 [hep-ph]].

%\cite{Ellis:2020awk}
\bibitem{Ellis:2020awk}
J.~Ellis, M.~Lewicki and J.~M.~No,
%``Gravitational waves from first-order cosmological phase transitions: lifetime of the sound wave source,''
JCAP \textbf{07}, 050 (2020)
[arXiv:2003.07360 [hep-ph]].

%\cite{Wang:2020jrd}
\bibitem{Wang:2020jrd}
X.~Wang, F.~P.~Huang and X.~Zhang,
%``Phase transition dynamics and gravitational wave spectra of strong first-order phase transition in supercooled universe,''
JCAP \textbf{05}, 045 (2020)
[arXiv:2003.08892 [hep-ph]].



%\cite{Pandey:2020hoq}
\bibitem{Pandey:2020hoq}
M.~Pandey and A.~Paul,
%``Gravitational Wave Emissions from First Order Phase Transitions with Two Component FIMP Dark Matter,''
[arXiv:2003.08828 [hep-ph]].


%\cite{Zhou:2020stj}
\bibitem{Zhou:2020stj}
Z.~Zhou, J.~Yan, A.~Addazi, Y.~F.~Cai, A.~Marciano and R.~Pasechnik,
%``Probing new physics with multi-vacua quantum tunnelings beyond standard model through gravitational waves,''
Phys. Lett. B \textbf{812}, 136026 (2021)
[arXiv:2003.13244 [astro-ph.CO]].

%\cite{Blasi:2020wpy}
\bibitem{Blasi:2020wpy}
S.~Blasi, V.~Brdar and K.~Schmitz,
%``Fingerprint of low-scale leptogenesis in the primordial gravitational-wave spectrum,''
Phys. Rev. Res. \textbf{2}, no.4, 043321 (2020)
[arXiv:2004.02889 [hep-ph]].



%\cite{Lewicki:2020jiv}
\bibitem{Lewicki:2020jiv}
M.~Lewicki and V.~Vaskonen,
%``Gravitational wave spectra from strongly supercooled phase transitions,''
Eur. Phys. J. C \textbf{80}, no.11, 1003 (2020)
[arXiv:2007.04967 [astro-ph.CO]].

%\cite{Hindmarsh:2020hop}
\bibitem{Hindmarsh:2020hop}
M.~B.~Hindmarsh, M.~L\"uben, J.~Lumma and M.~Pauly,
%``Phase transitions in the early universe,''
SciPost Phys. Lect. Notes \textbf{24}, 1 (2021)
[arXiv:2008.09136 [astro-ph.CO]].



%\cite{Nakai:2020oit}
\bibitem{Nakai:2020oit}
Y.~Nakai, M.~Suzuki, F.~Takahashi and M.~Yamada,
%``Gravitational Waves and Dark Radiation from Dark Phase Transition: Connecting NANOGrav Pulsar Timing Data and Hubble Tension,''
Phys. Lett. B \textbf{816}, 136238 (2021)
[arXiv:2009.09754 [astro-ph.CO]].


%\cite{Eichhorn:2020upj}
\bibitem{Eichhorn:2020upj}
A.~Eichhorn, J.~Lumma, J.~M.~Pawlowski, M.~Reichert and M.~Yamada,
%``Universal gravitational-wave signatures from heavy new physics in the electroweak sector,''
JCAP \textbf{05}, 006 (2021)
[arXiv:2010.00017 [hep-ph]].



%\cite{Paul:2020wbz}
\bibitem{Paul:2020wbz}
A.~Paul, U.~Mukhopadhyay and D.~Majumdar,
%``Gravitational Wave Signatures from Domain Wall and Strong First-Order Phase Transitions in a Two Complex Scalar extension of the Standard Model,''
JHEP \textbf{05}, 223 (2021)
[arXiv:2010.03439 [hep-ph]].


%\cite{Han:2020ekm}
\bibitem{Han:2020ekm}
X.~F.~Han, L.~Wang and Y.~Zhang,
%``Dark matter, electroweak phase transition, and gravitational waves in the type II two-Higgs-doublet model with a singlet scalar field,''
Phys. Rev. D \textbf{103}, no.3, 035012 (2021)
[arXiv:2010.03730 [hep-ph]].


%\cite{Ares:2020lbt}
\bibitem{Ares:2020lbt}
F.~R.~Ares, M.~Hindmarsh, C.~Hoyos and N.~Jokela,
%``Gravitational waves from a holographic phase transition,''
JHEP \textbf{21}, 100 (2020)
doi:10.1007/JHEP04(2021)100
[arXiv:2011.12878 [hep-th]].



%\cite{Wang:2020wrk}
\bibitem{Wang:2020wrk}
Y.~Wang, C.~S.~Li and F.~P.~Huang,
%``Complementary probe of dark matter blind spots by lepton colliders and gravitational waves,''
Phys. Rev. D \textbf{104}, no.5, 053004 (2021)
[arXiv:2012.03920 [hep-ph]].

%\cite{Ghosh:2020ipy}
\bibitem{Ghosh:2020ipy}
T.~Ghosh, H.~K.~Guo, T.~Han and H.~Liu,
%``Electroweak phase transition with an SU(2) dark sector,''
JHEP \textbf{07}, 045 (2021)
[arXiv:2012.09758 [hep-ph]].

%\cite{Huang:2020crf}
\bibitem{Huang:2020crf}
W.~C.~Huang, M.~Reichert, F.~Sannino and Z.~W.~Wang,
%``Testing the dark SU(N) Yang-Mills theory confined landscape: From the lattice to gravitational waves,''
Phys. Rev. D \textbf{104}, no.3, 035005 (2021)
[arXiv:2012.11614 [hep-ph]].



%\cite{Chao:2020adk}
\bibitem{Chao:2020adk}
W.~Chao, X.~F.~Li and L.~Wang,
%``Filtered pseudo-scalar dark matter and gravitational waves from first order phase transition,''
JCAP \textbf{06}, 038 (2021)
[arXiv:2012.15113 [hep-ph]].

%\cite{Di:2020ivg}
\bibitem{Di:2020ivg}
Y.~Di, J.~Wang, R.~Zhou, L.~Bian, R.~G.~Cai and J.~Liu,
%``Magnetic Field and Gravitational Waves from the First-Order Phase Transition,''
Phys. Rev. Lett. \textbf{126}, no.25, 251102 (2021)
doi:10.1103/PhysRevLett.126.251102
[arXiv:2012.15625 [astro-ph.CO]].


%\cite{Liu:2021jyc}
\bibitem{Liu:2021jyc}
W.~Liu and K.~P.~Xie,
%``Probing electroweak phase transition with multi-TeV muon colliders and gravitational waves,''
JHEP \textbf{04}, 015 (2021)
[arXiv:2101.10469 [hep-ph]].


%\cite{Zhang:2021alu}
\bibitem{Zhang:2021alu}
Z.~Zhang, C.~Cai, X.~M.~Jiang, Y.~L.~Tang, Z.~H.~Yu and H.~H.~Zhang,
%``Phase transition gravitational waves from pseudo-Nambu-Goldstone dark matter and two Higgs doublets,''
JHEP \textbf{05}, 160 (2021)
[arXiv:2102.01588 [hep-ph]].


%\cite{Cline:2021iff}
\bibitem{Cline:2021iff}
J.~M.~Cline, A.~Friedlander, D.~M.~He, K.~Kainulainen, B.~Laurent and D.~Tucker-Smith,
%``Baryogenesis and gravity waves from a UV-completed electroweak phase transition,''
Phys. Rev. D \textbf{103}, no.12, 123529 (2021)
[arXiv:2102.12490 [hep-ph]].


%\cite{Cao:2021yau}
\bibitem{Cao:2021yau}
Q.~H.~Cao, K.~Hashino, X.~X.~Li, Z.~Ren and J.~H.~Yu,
%``Electroweak phase transition triggered by fermion sector,''
JHEP \textbf{01}, 001 (2022)
[arXiv:2103.05688 [hep-ph]].

%\cite{Niemi:2021qvp}
\bibitem{Niemi:2021qvp}
L.~Niemi, P.~Schicho and T.~V.~I.~Tenkanen,
%``Singlet-assisted electroweak phase transition at two loops,''
Phys. Rev. D \textbf{103}, no.11, 115035 (2021)
[arXiv:2103.07467 [hep-ph]].


%\cite{Funatsu:2021gnh}
\bibitem{Funatsu:2021gnh}
S.~Funatsu, H.~Hatanaka, Y.~Hosotani, Y.~Orikasa and N.~Yamatsu,
%``Electroweak and left-right phase transitions in SO(5)\texttimes{}U(1)\texttimes{}SU(3) gauge-Higgs unification,''
Phys. Rev. D \textbf{104}, no.11, 115018 (2021)
[arXiv:2104.02870 [hep-ph]].


%\cite{Zhou:2021cfu}
\bibitem{Zhou:2021cfu}
R.~Zhou, L.~Bian and J.~Shu,
%``Probing new physics for $(g-2)_\mu$ and gravitational waves,''
[arXiv:2104.03519 [hep-ph]].


%\cite{Liu:2021mhn}
\bibitem{Liu:2021mhn}
J.~Liu, X.~P.~Wang and K.~P.~Xie,
%``Searching for lepton portal dark matter with colliders and gravitational waves,''
JHEP \textbf{06}, 149 (2021)
[arXiv:2104.06421 [hep-ph]].


%\cite{Borah:2021ocu}
\bibitem{Borah:2021ocu}
D.~Borah, A.~Dasgupta and S.~K.~Kang,
%``Gravitational waves from a dark U(1)D phase transition in light of NANOGrav 12.5~yr data,''
Phys. Rev. D \textbf{104}, no.6, 063501 (2021)
[arXiv:2105.01007 [hep-ph]].


%\cite{Aoki:2021oez}
\bibitem{Aoki:2021oez}
M.~Aoki, T.~Komatsu and H.~Shibuya,
%``Possibility of multi-step electroweak phase transition in the two Higgs doublet models,''
[arXiv:2106.03439 [hep-ph]].



%\cite{Lewicki:2021xku}
\bibitem{Lewicki:2021xku}
M.~Lewicki, O.~Pujol\`as and V.~Vaskonen,
%``Escape from supercooling with or without bubbles: gravitational wave signatures,''
Eur. Phys. J. C \textbf{81}, no.9, 857 (2021)
[arXiv:2106.09706 [astro-ph.CO]].


%\cite{Dong:2021cxn}
\bibitem{Dong:2021cxn}
X.~X.~Dong, T.~F.~Feng, H.~B.~Zhang, S.~M.~Zhao and J.~L.~Yang,
%``Gravitational waves from the phase transition in the B-LSSM,''
JHEP \textbf{12}, 052 (2021)
[arXiv:2106.11084 [hep-ph]].

%\cite{Baldes:2021vyz}
\bibitem{Baldes:2021vyz}
I.~Baldes, S.~Blasi, A.~Mariotti, A.~Sevrin and K.~Turbang,
%``Baryogenesis via relativistic bubble expansion,''
Phys. Rev. D \textbf{104}, no.11, 115029 (2021)
[arXiv:2106.15602 [hep-ph]].

%\cite{Marfatia:2021twj}
\bibitem{Marfatia:2021twj}
D.~Marfatia and P.~Y.~Tseng,
%``Correlated gravitational wave and microlensing signals of macroscopic dark matter,''
JHEP \textbf{11}, 068 (2021)
[arXiv:2107.00859 [hep-ph]].


%\cite{Gould:2021ccf}
\bibitem{Gould:2021ccf}
O.~Gould and J.~Hirvonen,
%``Effective field theory approach to thermal bubble nucleation,''
Phys. Rev. D \textbf{104}, no.9, 9 (2021)
[arXiv:2108.04377 [hep-ph]].


%\cite{Lerambert-Potin:2021ohy}
\bibitem{Lerambert-Potin:2021ohy}
P.~Lerambert-Potin and J.~A.~de Freitas Pacheco,
%``Gravitational Waves from the Cosmological Quark-Hadron Phase Transition Revisited,''
Universe \textbf{7}, no.8, 304 (2021)
[arXiv:2108.10727 [hep-ph]].

%\cite{Freitas:2021yng}
\bibitem{Freitas:2021yng}
F.~F.~Freitas, G.~Louren\c{c}o, A.~P.~Morais, A.~Nunes, J.~Louren\c{c}o, R.~Pasechnik, R.~Santos and J.~Viana,
%``Impact of SM parameters and of the vacua of the Higgs potential in gravitational waves detection,''
[arXiv:2108.12810 [hep-ph]].

%\cite{Goncalves:2021egx}
\bibitem{Goncalves:2021egx}
D.~Gon\c{c}alves, A.~Kaladharan and Y.~Wu,
%``Electroweak phase transition in the 2HDM: collider and gravitational wave complementarity,''
[arXiv:2108.05356 [hep-ph]].

%\cite{Reichert:2021cvs}
\bibitem{Reichert:2021cvs}
M.~Reichert, F.~Sannino, Z.~W.~Wang and C.~Zhang,
%``Dark confinement and chiral phase transitions: gravitational waves vs matter representations,''
JHEP \textbf{01}, 003 (2022)
[arXiv:2109.11552 [hep-ph]].

%\cite{Borah:2021ftr}
\bibitem{Borah:2021ftr}
D.~Borah, A.~Dasgupta and S.~K.~Kang,
%``A first order dark SU(2)$_{D}$ phase transition with vector dark matter in the light of NANOGrav 12.5 yr data,''
JCAP \textbf{12}, no.12, 039 (2021)
[arXiv:2109.11558 [hep-ph]].


%\cite{Ares:2021ntv}
\bibitem{Ares:2021ntv}
F.~R.~Ares, O.~Henriksson, M.~Hindmarsh, C.~Hoyos and N.~Jokela,
%``Effective actions and bubble nucleation from holography,''
[arXiv:2109.13784 [hep-th]].

%\cite{Bai:2021ibt}
\bibitem{Bai:2021ibt}
Y.~Bai and M.~Korwar,
%``Cosmological Constraints on First-Order Phase Transitions,''
[arXiv:2109.14765 [hep-ph]].

%\cite{Ares:2021nap}
\bibitem{Ares:2021nap}
F.~R.~Ares, O.~Henriksson, M.~Hindmarsh, C.~Hoyos and N.~Jokela,
%``Gravitational Waves at Strong Coupling from an Effective Action,''
[arXiv:2110.14442 [hep-th]].

%\cite{Lewicki:2021pgr}
\bibitem{Lewicki:2021pgr}
M.~Lewicki, M.~Merchand and M.~Zych,
%``Electroweak bubble wall expansion: gravitational waves and baryogenesis in Standard Model-like thermal plasma,''
JHEP \textbf{02}, 017 (2022)
[arXiv:2111.02393 [astro-ph.CO]].


%\cite{Bandyopadhyay:2021ipw}
\bibitem{Bandyopadhyay:2021ipw}
P.~Bandyopadhyay and S.~Jangid,
%``Discerning Singlet and Triplet scalars at the electroweak phase transition and Gravitational Wave,''
[arXiv:2111.03866 [hep-ph]].

%\cite{Hashino:2021qoq}
\bibitem{Hashino:2021qoq}
K.~Hashino, S.~Kanemura and T.~Takahashi,
%``Primordial black holes as a probe of strongly first-order electroweak phase transition,''
[arXiv:2111.13099 [hep-ph]].


%\cite{Demidov:2021lyo}
\bibitem{Demidov:2021lyo}
S.~Demidov, D.~Gorbunov and E.~Kriukova,
%``Gravitational waves from first-order electroweak phase transition in a model with light sgoldstinos,''
[arXiv:2112.06083 [hep-ph]].


%\cite{Graf:2021xku}
\bibitem{Graf:2021xku}
L.~Gr\'af, S.~Jana, A.~Kaladharan and S.~Saad,
%``Gravitational Wave Imprints of Left-Right Symmetric Model with Minimal Higgs Sector,''
[arXiv:2112.12041 [hep-ph]].

%\cite{Kanemura:2022ozv}
\bibitem{Kanemura:2022ozv}
S.~Kanemura and M.~Tanaka,
%``Strongly first-order electroweak phase transition by relatively heavy additional Higgs bosons,''
[arXiv:2201.04791 [hep-ph]].

%\cite{Hamada:2022soj}
\bibitem{Hamada:2022soj}
Y.~Hamada, H.~Kawai, K.~Kawana, K.~y.~Oda and K.~Yagyu,
%``Gravitational waves in models with multicritical-point principle,''
[arXiv:2202.04221 [hep-ph]].




 %\cite{Blasi:2022woz}
\bibitem{Blasi:2022woz}
S.~Blasi and A.~Mariotti,
%``Domain walls seeding the electroweak phase transition,''
[arXiv:2203.16450 [hep-ph]].

%\cite{Benincasa:2022elt}
\bibitem{Benincasa:2022elt}
N.~Benincasa, L.~Delle Rose, K.~Kannike and L.~Marzola,
%``Multi-step phase transitions and gravitational waves in the inert doublet model,''
[arXiv:2205.06669 [hep-ph]].

%\cite{Azatov:2022tii}
\bibitem{Azatov:2022tii}
A.~Azatov, G.~Barni, S.~Chakraborty, M.~Vanvlasselaer and W.~Yin,
%``Ultra-relativistic bubbles from the simplest Higgs portal and their cosmological consequences,''
JHEP \textbf{10} (2022), 017
[arXiv:2207.02230 [hep-ph]].


 %\cite{Biekotter:2022kgf}
\bibitem{Biekotter:2022kgf}
T.~Biek\"otter, S.~Heinemeyer, J.~M.~No, M.~O.~Olea-Romacho and G.~Weiglein,
%``The trap in the early Universe: impact on the interplay between gravitational waves and LHC physics in the 2HDM,''
[arXiv:2208.14466 [hep-ph]].

%\cite{Phong:2022xpo}
\bibitem{Phong:2022xpo}
V.~Q.~Phong, N.~M.~Anh and H.~N.~Long,
%``Dual Electroweak Phase Transition in the Two-Higgs-doublet Model with the $S_3$ discrete symmetry,''
[arXiv:2209.14672 [hep-ph]].


%\cite{Croon:2020cgk}
\bibitem{Croon:2020cgk}
D.~Croon, O.~Gould, P.~Schicho, T.~V.~I.~Tenkanen and G.~White,
%``Theoretical uncertainties for cosmological first-order phase transitions,''
JHEP \textbf{04}, 055 (2021)
doi:10.1007/JHEP04(2021)055
[arXiv:2009.10080 [hep-ph]].




%\cite{Cai:2022bcf}
\bibitem{Cai:2022bcf}
R.~G.~Cai, K.~Hashino, S.~J.~Wang and J.~H.~Yu,
%``Gravitational waves from patterns of electroweak symmetry breaking: an effective perspective,''
Commun. Theor. Phys. \textbf{77} (2025) no.5, 055204
%doi:10.1088/1572-9494/ad9c3d
[arXiv:2202.08295 [hep-ph]].








%%%%%%%   multi-peaks   %%%%%%%





%%%%%%% 		Finite temperature effects cubic term 		%%%%%%%

  %\cite{Dolan:1973qd}
%\bibitem{Dolan:1973qd} 
%  L.~Dolan and R.~Jackiw,
  %``Symmetry Behavior at Finite Temperature,''
%  Phys.\ Rev.\ D {\bf 9}, 3320 (1974).
  
%  \bibitem{Pietroni:1992in}
%M.~Pietroni,
%``The Electroweak phase transition in a nonminimal supersymmetric model,''
%Nucl. Phys. B \textbf{402} (1993), 27-45
%[arXiv:hep-ph/9207227 [hep-ph]].
  
%\cite{Anderson:1991zb}
%\bibitem{Anderson:1991zb}
%G.~W.~Anderson and L.~J.~Hall,
%``The Electroweak phase transition and baryogenesis,''
%Phys. Rev. D \textbf{45} (1992), 2685-2698





%%%%%%% 		DM		%%%%%%%


%\cite{FileviezPerez:2008bj}
\bibitem{FileviezPerez:2008bj}
P.~Fileviez Perez, H.~H.~Patel, M.~J.~Ramsey-Musolf and K.~Wang,
%``Triplet Scalars and Dark Matter at the LHC,''
Phys. Rev. D \textbf{79}, 055024 (2009)
%doi:10.1103/PhysRevD.79.055024
[arXiv:0811.3957 [hep-ph]].

%\cite{Cui:2011qe}
\bibitem{Cui:2011qe}
Y.~Cui, L.~Randall and B.~Shuve,
%``Emergent Dark Matter, Baryon, and Lepton Numbers,''
JHEP \textbf{08}, 073 (2011)
%doi:10.1007/JHEP08(2011)073
[arXiv:1106.4834 [hep-ph]].

%\cite{Cline:2012hg}
\bibitem{Cline:2012hg}
J.~M.~Cline and K.~Kainulainen,
%``Electroweak baryogenesis and dark matter from a singlet Higgs,''
JCAP \textbf{01}, 012 (2013)
%doi:10.1088/1475-7516/2013/01/012
[arXiv:1210.4196 [hep-ph]].

%\cite{Patel:2012pi}
\bibitem{Patel:2012pi}
H.~H.~Patel and M.~J.~Ramsey-Musolf,
%``Stepping Into Electroweak Symmetry Breaking: Phase Transitions and Higgs Phenomenology,''
Phys. Rev. D \textbf{88}, 035013 (2013)
%doi:10.1103/PhysRevD.88.035013
[arXiv:1212.5652 [hep-ph]].

%\cite{Fairbairn:2013uta}
\bibitem{Fairbairn:2013uta}
M.~Fairbairn and R.~Hogan,
%``Singlet Fermionic Dark Matter and the Electroweak Phase Transition,''
JHEP \textbf{09}, 022 (2013)
%doi:10.1007/JHEP09(2013)022
[arXiv:1305.3452 [hep-ph]].

%\cite{Alanne:2014bra}
\bibitem{Alanne:2014bra}
T.~Alanne, K.~Tuominen and V.~Vaskonen,
%``Strong phase transition, dark matter and vacuum stability from simple hidden sectors,''
Nucl. Phys. B \textbf{889}, 692-711 (2014)
%doi:10.1016/j.nuclphysb.2014.11.001
[arXiv:1407.0688 [hep-ph]].

%\cite{Blinov:2015sna}
\bibitem{Blinov:2015sna}
N.~Blinov, J.~Kozaczuk, D.~E.~Morrissey and C.~Tamarit,
%``Electroweak Baryogenesis from Exotic Electroweak Symmetry Breaking,''
Phys. Rev. D \textbf{92}, no.3, 035012 (2015)
%doi:10.1103/PhysRevD.92.035012
[arXiv:1504.05195 [hep-ph]].

%\cite{Baker:2016xzo}
\bibitem{Baker:2016xzo}
M.~J.~Baker and J.~Kopp,
%``Dark Matter Decay between Phase Transitions at the Weak Scale,''
Phys. Rev. Lett. \textbf{119}, no.6, 061801 (2017)
%doi:10.1103/PhysRevLett.119.061801
[arXiv:1608.07578 [hep-ph]].

%\cite{Grzadkowski:2018nbc}
\bibitem{Grzadkowski:2018nbc}
B.~Grzadkowski and D.~Huang,
%``Spontaneous $CP$-Violating Electroweak Baryogenesis and Dark Matter from a Complex Singlet Scalar,''
JHEP \textbf{08}, 135 (2018)
%doi:10.1007/JHEP08(2018)135
[arXiv:1807.06987 [hep-ph]].

%\cite{Baker:2018vos}
\bibitem{Baker:2018vos}
M.~J.~Baker and L.~Mittnacht,
%``Variations on the Vev Flip-Flop: Instantaneous Freeze-out and Decaying Dark Matter,''
JHEP \textbf{05}, 070 (2019)
%doi:10.1007/JHEP05(2019)070
[arXiv:1811.03101 [hep-ph]].

%\cite{Bian:2018bxr}
\bibitem{Bian:2018bxr}
L.~Bian and X.~Liu,
%``Two-step strongly first-order electroweak phase transition modified FIMP dark matter, gravitational wave signals, and the neutrino mass,''
Phys. Rev. D \textbf{99}, no.5, 055003 (2019)
%doi:10.1103/PhysRevD.99.055003
[arXiv:1811.03279 [hep-ph]].



%\cite{Ghorbani:2019itr}
\bibitem{Ghorbani:2019itr}
K.~Ghorbani and P.~H.~Ghorbani,
%``A Simultaneous Study of Dark Matter and Phase Transition: Two-Scalar Scenario,''
JHEP \textbf{12}, 077 (2019)
%doi:10.1007/JHEP12(2019)077
[arXiv:1906.01823 [hep-ph]].

%\cite{Robens:2019kga}
\bibitem{Robens:2019kga}
T.~Robens, T.~Stefaniak and J.~Wittbrodt,
%``Two-real-scalar-singlet extension of the SM: LHC phenomenology and benchmark scenarios,''
Eur. Phys. J. C \textbf{80}, no.2, 151 (2020)
%doi:10.1140/epjc/s10052-020-7655-x
[arXiv:1908.08554 [hep-ph]].

%\cite{Chen:2019ebq}
\bibitem{Chen:2019ebq}
N.~Chen, T.~Li, Y.~Wu and L.~Bian,
%``Complementarity of the future $e^+ e^-$ colliders and gravitational waves in the probe of complex singlet extension to the standard model,''
Phys. Rev. D \textbf{101}, no.7, 075047 (2020)
%doi:10.1103/PhysRevD.101.075047
[arXiv:1911.05579 [hep-ph]].



%\cite{Bell:2020gug}
\bibitem{Bell:2020gug}
N.~F.~Bell, M.~J.~Dolan, L.~S.~Friedrich, M.~J.~Ramsey-Musolf and R.~R.~Volkas,
%``Two-Step Electroweak Symmetry-Breaking: Theory Meets Experiment,''
JHEP \textbf{05}, 050 (2020)
%doi:10.1007/JHEP05(2020)050
[arXiv:2001.05335 [hep-ph]].

%\cite{Chiang:2020yym}
\bibitem{Chiang:2020yym}
C.~W.~Chiang, D.~Huang and B.~Q.~Lu,
%``Electroweak phase transition confronted with dark matter detection constraints,''
JCAP \textbf{01}, 035 (2021)
%doi:10.1088/1475-7516/2021/01/035
[arXiv:2009.08635 [hep-ph]].


%%%%%%% 		BAU		%%%%%%%


%\cite{Profumo:2007wc}
\bibitem{Profumo:2007wc}
S.~Profumo, M.~J.~Ramsey-Musolf and G.~Shaughnessy,
%``Singlet Higgs phenomenology and the electroweak phase transition,''
JHEP \textbf{08}, 010 (2007)
%doi:10.1088/1126-6708/2007/08/010
[arXiv:0705.2425 [hep-ph]].





%\cite{Espinosa:2011ax}
\bibitem{Espinosa:2011ax}
J.~R.~Espinosa, T.~Konstandin and F.~Riva,
%``Strong Electroweak Phase Transitions in the Standard Model with a Singlet,''
Nucl. Phys. B \textbf{854}, 592-630 (2012)
%doi:10.1016/j.nuclphysb.2011.09.010
[arXiv:1107.5441 [hep-ph]].


%\cite{Espinosa:2011eu}
\bibitem{Espinosa:2011eu}
J.~R.~Espinosa, B.~Gripaios, T.~Konstandin and F.~Riva,
%``Electroweak Baryogenesis in Non-minimal Composite Higgs Models,''
JCAP \textbf{01}, 012 (2012)
%doi:10.1088/1475-7516/2012/01/012
[arXiv:1110.2876 [hep-ph]].

%\cite{Profumo:2014opa}
\bibitem{Profumo:2014opa}
S.~Profumo, M.~J.~Ramsey-Musolf, C.~L.~Wainwright and P.~Winslow,
%``Singlet-catalyzed electroweak phase transitions and precision Higgs boson studies,''
Phys. Rev. D \textbf{91}, no.3, 035018 (2015)
doi:10.1103/PhysRevD.91.035018
[arXiv:1407.5342 [hep-ph]].

%\cite{Inoue:2015pza}
\bibitem{Inoue:2015pza}
S.~Inoue, G.~Ovanesyan and M.~J.~Ramsey-Musolf,
%``Two-Step Electroweak Baryogenesis,''
Phys. Rev. D \textbf{93}, 015013 (2016)
%doi:10.1103/PhysRevD.93.015013
[arXiv:1508.05404 [hep-ph]].


%\cite{Tenkanen:2016idg}
\bibitem{Tenkanen:2016idg}
T.~Tenkanen, K.~Tuominen and V.~Vaskonen,
%``A Strong Electroweak Phase Transition from the Inflaton Field,''
JCAP \textbf{09}, 037 (2016)
%doi:10.1088/1475-7516/2016/09/037
[arXiv:1606.06063 [hep-ph]].


%\cite{Kurup:2017dzf}
\bibitem{Kurup:2017dzf}
G.~Kurup and M.~Perelstein,
%``Dynamics of Electroweak Phase Transition In Singlet-Scalar Extension of the Standard Model,''
Phys. Rev. D \textbf{96}, no.1, 015036 (2017)
%doi:10.1103/PhysRevD.96.015036
[arXiv:1704.03381 [hep-ph]].

%\cite{Chen:2017qcz}
\bibitem{Chen:2017qcz}
C.~Y.~Chen, J.~Kozaczuk and I.~M.~Lewis,
%``Non-resonant Collider Signatures of a Singlet-Driven Electroweak Phase Transition,''
JHEP \textbf{08}, 096 (2017)
%doi:10.1007/JHEP08(2017)096
[arXiv:1704.05844 [hep-ph]].

%\cite{Ramsey-Musolf:2017tgh}
\bibitem{Ramsey-Musolf:2017tgh}
M.~J.~Ramsey-Musolf, P.~Winslow and G.~White,
%``Color Breaking Baryogenesis,''
Phys. Rev. D \textbf{97}, no.12, 123509 (2018)
%doi:10.1103/PhysRevD.97.123509
[arXiv:1708.07511 [hep-ph]].

%\cite{Chala:2018opy}
\bibitem{Chala:2018opy}
M.~Chala, M.~Ramos and M.~Spannowsky,
%``Gravitational wave and collider probes of a triplet Higgs sector with a low cutoff,''
Eur. Phys. J. C \textbf{79}, no.2, 156 (2019)
%doi:10.1140/epjc/s10052-019-6655-1
[arXiv:1812.01901 [hep-ph]].

%\cite{Gould:2019qek}
\bibitem{Gould:2019qek}
O.~Gould, J.~Kozaczuk, L.~Niemi, M.~J.~Ramsey-Musolf, T.~V.~I.~Tenkanen and D.~J.~Weir,
%``Nonperturbative analysis of the gravitational waves from a first-order electroweak phase transition,''
Phys. Rev. D \textbf{100}, no.11, 115024 (2019)
%doi:10.1103/PhysRevD.100.115024
[arXiv:1903.11604 [hep-ph]].

%\cite{Caprini:2019egz}
\bibitem{Caprini:2019egz}
C.~Caprini, M.~Chala, G.~C.~Dorsch, M.~Hindmarsh, S.~J.~Huber, T.~Konstandin, J.~Kozaczuk, G.~Nardini, J.~M.~No and K.~Rummukainen, \textit{et al.}
%``Detecting gravitational waves from cosmological phase transitions with LISA: an update,''
JCAP \textbf{03}, 024 (2020)
%doi:10.1088/1475-7516/2020/03/024
[arXiv:1910.13125 [astro-ph.CO]].


%\cite{Kozaczuk:2019pet}
\bibitem{Kozaczuk:2019pet}
J.~Kozaczuk, M.~J.~Ramsey-Musolf and J.~Shelton,
%``Exotic Higgs boson decays and the electroweak phase transition,''
Phys. Rev. D \textbf{101}, no.11, 115035 (2020)
%doi:10.1103/PhysRevD.101.115035
[arXiv:1911.10210 [hep-ph]].


%\cite{Ramsey-Musolf:2019lsf}
\bibitem{Ramsey-Musolf:2019lsf}
M.~J.~Ramsey-Musolf,
%``The electroweak phase transition: a collider target,''
JHEP \textbf{09}, 179 (2020)
%doi:10.1007/JHEP09(2020)179
[arXiv:1912.07189 [hep-ph]].

%\cite{Senaha:2020mop}
\bibitem{Senaha:2020mop}
E.~Senaha,
%``Symmetry Restoration and Breaking at Finite Temperature: An Introductory Review,''
Symmetry \textbf{12}, no.5, 733 (2020)


%\cite{Ramsey-Musolf:2021ldh}
\bibitem{Ramsey-Musolf:2021ldh}
M.~J.~Ramsey-Musolf, J.~H.~Yu and J.~Zhou,
%``Probing extended scalar sectors with precision e$^{+}$e$^{−}$\textrightarrow{} Zh and Higgs diphoton studies,''
JHEP \textbf{10} (2021), 155
doi:10.1007/JHEP10(2021)155
[arXiv:2104.10709 [hep-ph]].




%\cite{Hally:2012pu}
\bibitem{Hally:2012pu}
K.~Hally, H.~E.~Logan and T.~Pilkington,
%``Constraints on large scalar multiplets from perturbative unitarity,''
Phys. Rev. D \textbf{85}, 095017 (2012)
doi:10.1103/PhysRevD.85.095017
[arXiv:1202.5073 [hep-ph]].



%%%%%%% 		Dimensional Reduction 		%%%%%%%

%\cite{Niemi:2018asa}
\bibitem{Niemi:2018asa}
L.~Niemi, H.~H.~Patel, M.~J.~Ramsey-Musolf, T.~V.~I.~Tenkanen and D.~J.~Weir,
%``Electroweak phase transition in the real triplet extension of the SM: Dimensional reduction,''
Phys. Rev. D \textbf{100}, no.3, 035002 (2019)
doi:10.1103/PhysRevD.100.035002
[arXiv:1802.10500 [hep-ph]].

% %\cite{Gould:2019qek}
% \bibitem{Gould:2019qek}
% O.~Gould, J.~Kozaczuk, L.~Niemi, M.~J.~Ramsey-Musolf, T.~V.~I.~Tenkanen and D.~J.~Weir,
% %``Nonperturbative analysis of the gravitational waves from a first-order electroweak phase transition,''
% Phys. Rev. D \textbf{100}, no.11, 115024 (2019)
% doi:10.1103/PhysRevD.100.115024
% [arXiv:1903.11604 [hep-ph]].
% %92 citations counted in INSPIRE as of 26 Mar 2025

% % \cite{Kainulainen:2019kyp}
% \bibitem{Kainulainen:2019kyp}
% K.~Kainulainen, V.~Keus, L.~Niemi, K.~Rummukainen, T.~V.~I.~Tenkanen and V.~Vaskonen,
% %``On the validity of perturbative studies of the electroweak phase transition in the Two Higgs Doublet model,''
% JHEP \textbf{06}, 075 (2019)
% doi:10.1007/JHEP06(2019)075
% [arXiv:1904.01329 [hep-ph]].
% %102 citations counted in INSPIRE as of 26 Mar 2025

% %\cite{Niemi:2021qvp}
% \bibitem{Niemi:2021qvp}
% L.~Niemi, P.~Schicho and T.~V.~I.~Tenkanen,
% %``Singlet-assisted electroweak phase transition at two loops,''
% Phys. Rev. D \textbf{103}, no.11, 115035 (2021)
% [erratum: Phys. Rev. D \textbf{109}, no.3, 039902 (2024)]
% doi:10.1103/PhysRevD.103.115035
% [arXiv:2103.07467 [hep-ph]].
% %92 citations counted in INSPIRE as of 26 Mar 2025

%\cite{Ekstedt:2022bff}
\bibitem{Ekstedt:2022bff}
A.~Ekstedt, P.~Schicho and T.~V.~I.~Tenkanen,
%``DRalgo: A package for effective field theory approach for thermal phase transitions,''
Comput. Phys. Commun. \textbf{288}, 108725 (2023)
doi:10.1016/j.cpc.2023.108725
[arXiv:2205.08815 [hep-ph]].




%\cite{Land:1992sm}
%\bibitem{Land:1992sm}
%D.~Land and E.~D.~Carlson,
%``Two stage phase transition in two Higgs models,''
%Phys. Lett. B \textbf{292} (1992), 107-112
%doi:10.1016/0370-2693(92)90616-C
%[arXiv:hep-ph/9208227 [hep-ph]].


%%%%%%% 		Gamma 		%%%%%%%

%\cite{Coleman:1977py}
\bibitem{Coleman:1977py}
S.~R.~Coleman,
%``The Fate of the False Vacuum. 1. Semiclassical Theory,''
Phys. Rev. D \textbf{15}, 2929-2936 (1977)
[erratum: Phys. Rev. D \textbf{16}, 1248 (1977)]


%%%%%%% 		Anybubbles 		%%%%%%%


%\cite{Masoumi:2016wot}
\bibitem{Masoumi:2016wot}
A.~Masoumi, K.~D.~Olum and B.~Shlaer,
%``Efficient numerical solution to vacuum decay with many fields,''
JCAP \textbf{01} (2017), 051
[arXiv:1610.06594 [gr-qc]].


%\cite{Kosowsky:1991ua}
%\bibitem{Kosowsky:1991ua}
%A.~Kosowsky, M.~S.~Turner and R.~Watkins,
%``Gravitational radiation from colliding vacuum bubbles,''
%Phys. Rev. D \textbf{45} (1992), 4514-4535
%doi:10.1103/PhysRevD.45.4514
%\cite{Kosowsky:1992rz}
%\bibitem{Kosowsky:1992rz}
%A.~Kosowsky, M.~S.~Turner and R.~Watkins,
%``Gravitational waves from first order cosmological phase transitions,''
%Phys. Rev. Lett. \textbf{69} (1992), 2026-2029
%doi:10.1103/PhysRevLett.69.2026


%%%%%%%%  GW collision %%%%%%%%
  
%\cite{Huber:2008hg}
\bibitem{Huber:2008hg} 
  S.~J.~Huber and T.~Konstandin,
  %``Gravitational Wave Production by Collisions: More Bubbles,''
  JCAP {\bf 0809}, 022 (2008). 




%%%%%%%%  GW turb %%%%%%%%
  %\cite{Caprini:2009yp}
\bibitem{Caprini:2009yp} 
  C.~Caprini, R.~Durrer and G.~Servant,
  %``The stochastic gravitational wave background from turbulence and magnetic fields generated by a first-order phase transition,''
  JCAP {\bf 0912}, 024 (2009), 
  %\cite{Binetruy:2012ze}
%\bibitem{Binetruy:2012ze} 
  P.~Binetruy, A.~Bohe, C.~Caprini and J.~F.~Dufaux,
  %``Cosmological Backgrounds of Gravitational Waves and eLISA/NGO: Phase Transitions, Cosmic Strings and Other Sources,''
  JCAP {\bf 1206}, 027 (2012)

  

%%%%%%%%  GW SW %%%%%%%%

%\cite{Caprini:2015zlo}
\bibitem{Caprini:2015zlo} 
  C.~Caprini {\it et al.},
  %``Science with the space-based interferometer eLISA. II: Gravitational waves from cosmological phase transitions,''
  arXiv:1512.06239 [astro-ph.CO].




%%%%%%%%  GW effi %%%%%%%%

%\cite{Espinosa:2010hh}
\bibitem{Espinosa:2010hh} 
  J.~R.~Espinosa, T.~Konstandin, J.~M.~No and G.~Servant,
  %``Energy Budget of Cosmological First-order Phase Transitions,''
  JCAP {\bf 1006}, 028 (2010).





%%%%%%%%  S/N ratio %%%%%%%%

%\cite{Thrane:2013oya}
\bibitem{Thrane:2013oya}
E.~Thrane and J.~D.~Romano,
%``Sensitivity curves for searches for gravitational-wave backgrounds,''
Phys. Rev. D \textbf{88} (2013) no.12, 124032
%[arXiv:1310.5300 [astro-ph.IM]].



% Sensitivity in experiments %

%\cite{Klein:2015hvg}
\bibitem{Klein:2015hvg}
A.~Klein, E.~Barausse, A.~Sesana, A.~Petiteau, E.~Berti, S.~Babak, J.~Gair, S.~Aoudia, I.~Hinder and F.~Ohme, \textit{et al.}
%``Science with the space-based interferometer eLISA: Supermassive black hole binaries,''
Phys. Rev. D \textbf{93}, no.2, 024003 (2016)
doi:10.1103/PhysRevD.93.024003
[arXiv:1511.05581 [gr-qc]].

%\cite{Yagi:2011wg}
\bibitem{Yagi:2011wg}
K.~Yagi and N.~Seto,
%``Detector configuration of DECIGO/BBO and identification of cosmological neutron-star binaries,''
Phys. Rev. D \textbf{83}, 044011 (2011)
[erratum: Phys. Rev. D \textbf{95}, no.10, 109901 (2017)]
doi:10.1103/PhysRevD.83.044011
[arXiv:1101.3940 [astro-ph.CO]].






%%%%%% Tp %%%%%%%

%\cite{Enqvist:1991xw}
\bibitem{Enqvist:1991xw}
K.~Enqvist, J.~Ignatius, K.~Kajantie and K.~Rummukainen,
%``Nucleation and bubble growth in a first order cosmological electroweak phase transition,''
Phys. Rev. D \textbf{45} (1992), 3415-3428

%\cite{Ellis:2018mja}
\bibitem{Ellis:2018mja}
J.~Ellis, M.~Lewicki and J.~M.~No,
%``On the Maximal Strength of a First-Order Electroweak Phase Transition and its Gravitational Wave Signal,''
JCAP \textbf{04} (2019), 003


%%% low vb %%%

\bibitem{Joyce:1994fu}
M.~Joyce, T.~Prokopec and N.~Turok,
%``Electroweak baryogenesis from a classical force,''
Phys. Rev. Lett. \textbf{75}, 1695-1698 (1995)
[erratum: Phys. Rev. Lett. \textbf{75}, 3375 (1995)]
%doi:10.1103/PhysRevLett.75.1695
[arXiv:hep-ph/9408339 [hep-ph]].




%\cite{Arnold:1992rz}
%\bibitem{Arnold:1992rz}
%P.~B.~Arnold and O.~Espinosa,
%``The Effective potential and first order phase transitions: Beyond leading-order,''
%Phys. Rev. D \textbf{47} (1993), 3546
%[erratum: Phys. Rev. D \textbf{50} (1994), 6662]
%doi:10.1103/PhysRevD.47.3546
%[arXiv:hep-ph/9212235 [hep-ph]].


%\cite{ATLAS:2015eiz}
\bibitem{ATLAS:2015eiz}
G.~Aad \textit{et al.} [ATLAS],
%``Search for the electroweak production of supersymmetric particles in $\sqrt{s}$=8 TeV $pp$ collisions with the ATLAS detector,''
Phys. Rev. D \textbf{93}, no.5, 052002 (2016)
doi:10.1103/PhysRevD.93.052002
[arXiv:1509.07152 [hep-ex]].
%153 citations counted in INSPIRE as of 13 Dec 2022


%\cite{Cirelli:2005uq}
\bibitem{Cirelli:2005uq}
M.~Cirelli, N.~Fornengo and A.~Strumia,
%``Minimal dark matter,''
Nucl. Phys. B \textbf{753}, 178-194 (2006)
doi:10.1016/j.nuclphysb.2006.07.012
[arXiv:hep-ph/0512090 [hep-ph]].
%925 citations counted in INSPIRE as of 15 Dec 2022


%\cite{Chen:2013vi}
\bibitem{Chen:2013vi}
C.~S.~Chen, C.~Q.~Geng, D.~Huang and L.~H.~Tsai,
%``New Scalar Contributions to $h\to Z\gamma$,''
Phys. Rev. D \textbf{87} (2013), 075019
doi:10.1103/PhysRevD.87.075019
[arXiv:1301.4694 [hep-ph]].


%\cite{ATLAS:2022tnm}
\bibitem{ATLAS:2022tnm}
ATLAS Collaboration,
%``Measurement of the properties of Higgs boson production at $\sqrt{s} = 13$ TeV in the $H\to\gamma\gamma$ channel using $139$ fb$^{-1}$ of $pp$ collision data with the ATLAS experiment,''
JHEP \textbf{07}, 088 (2023)
doi:10.1007/JHEP07(2023)088
[arXiv:2207.00348 [hep-ex]].



 
\end{thebibliography}
\end{document}